\documentclass[twoside,11pt]{article}
\usepackage{jmlr2e}

\usepackage{tikz}
\usetikzlibrary{arrows,positioning}

\usepackage{dsfont}    
\usepackage{hyperref}  
\usepackage{bm}        
\usepackage{mathtools} 
\usepackage{enumerate} 
\usepackage{lastpage}

\usepackage{algpseudocode}
\usepackage{algorithm}
\renewcommand{\algorithmicrequire}{\textbf{Input:}}
\renewcommand{\algorithmicensure}{\textbf{Output:}}

\providecommand{\gray}{\color{gray}}

\newcommand{\where}{\quad\text{where}\quad}
\newcommand{\andwhere}{\quad\text{and}\quad}

\DeclareMathOperator*{\argmax}{arg\,max} 
\DeclareMathOperator*{\argmin}{arg\,min} 
\DeclareMathOperator*{\tr}{tr}           
\DeclareMathOperator{\vect}{vec}         
\DeclareMathOperator{\diag}{diag}         
\DeclareMathOperator{\Cov}{Cov}          
\DeclareMathOperator{\SLOOCV}{SLOOCV}      
\DeclareMathOperator{\pa}{pa}            

\renewcommand{\vec}[1]{\bm{\mathbf{#1}}}

\DeclarePairedDelimiter\abs{\lvert}{\rvert}         
\DeclarePairedDelimiter\deter{\lvert}{\rvert}       
\DeclarePairedDelimiter\sqfnorm{\lVert}{\rVert_F^2} 

\newcommand\independent{\protect\mathpalette{\protect\independenT}{\perp}}
\def\independenT#1#2{\mathrel{\rlap{$#1#2$}\mkern2mu{#1#2}}}

\newcommand{\missingarg}{{\;\cdot\;}} 
\newcommand{\summed}{\bullet}         
\newcommand{\svdot}{\raisebox{3pt}{$\scalebox{.75}{$\vdots$}$}}
\newcommand{\sddot}{\raisebox{3pt}{$\scalebox{.75}{$\ddots$}$}}

\newcommand{\bbE}{\mathbb{E}}
\newcommand{\bbR}{\mathbb{R}}

\newcommand{\bbOne}{\mathds{1}}
\newcommand{\calN}{\mathcal{N}}
\newcommand{\calS}{\mathcal{S}}
\newcommand{\calP}{\mathcal{P}}
\newcommand{\calL}{\mathcal{L}}
\newcommand{\calG}{\mathcal{G}}

\newcommand{\vSigma}{{\vec{\Sigma}}}
\newcommand{\hvSigma}{{\hat{\vec{\Sigma}}}{}}
\newcommand{\vOmega}{{\vec{\Omega}}}
\newcommand{\hvOmega}{{\hat{\vec{\Omega}}}{}}

\newcommand{\hvX}{{\hat{\vec{X}}}{}}
\newcommand{\vPsi}{{\vec{\Psi}}}
\newcommand{\vPhi}{{\vec{\Phi}}}

\newcommand{\vTheta}{{\vec{\Theta}}}

\newcommand{\vy}{{\boldsymbol{\mathrm{y}}}}

\newcommand{\vLambda}{{\vec{\Lambda}}}
\newcommand{\vX}{\vec{X}}
\newcommand{\vY}{\vec{Y}}
\newcommand{\vS}{\vec{S}}

\newcommand{\vI}{\vec{I}}
\newcommand{\vJ}{\vec{J}}
\newcommand{\vT}{\vec{T}}

\newcommand{\vA}{\vec{A}}
\newcommand{\vB}{\vec{B}}
\newcommand{\vP}{\vec{P}}

\newcommand{\hvP}{\hat{\vP}{}}

\newcommand{\OmT}[1]{(\vOmega_{g_{#1}} \!- \vT_{g_{#1}})}
\newcommand{\OmTmOmT}{\OmT{1} - \OmT{2}}
\newcommand{\OmTpOmT}{\OmT{1} + \OmT{2}}

\newcommand{\lamDS}{{\lambda_{\mathrm{DS}}}}
\newcommand{\lamST}{{\lambda_{\mathrm{ST}}}}
\newcommand{\ABC}{{\mathrm{ABC}}}
\newcommand{\GCB}{{\mathrm{GCB}}}

\newcommand{\nth}[1]{$#1$th}

\newcommand{\R}{\textsf{R}}

\makeatletter
\renewcommand*{\@fnsymbol}[1]{\ensuremath{\ifcase#1\or \bigstar\or \dagger\or \ddagger\or
    \mathsection\or \mathparagraph\or \|\or **\or \dagger\dagger
    \or \ddagger\ddagger \else\@ctrerr\fi}}
\makeatother

\jmlrheading{21}{2020}{1-\pageref{LastPage}}{10/15; Revised 8/19}{3/20}{15-509}{Anders E. Bilgrau, Carel F.W. Peeters, Poul Svante Eriksen, Martin B{\o}gsted, and Wessel N.\ van Wieringen}

\ShortHeadings{Targeted Fused Ridge Precision Estimation}{Bilgrau \& Peeters et al.}
\firstpageno{1}

\begin{document}

\title{Targeted Fused Ridge Estimation of Inverse Covariance Matrices from Multiple High-Dimensional Data Classes}

\author{\name Anders Ellern Bilgrau\thanks{Shared first authorship.} \email anders.ellern.bilgrau@gmail.com\\
        \addr Department of Mathematical Sciences, \\
        Aalborg University\\
        9220 Aalborg \O, Denmark ~\&\\
        Department of Haematology, \\
        Aalborg University Hospital\\
        9000 Aalborg, Denmark
        \AND
        \name Carel F.W. Peeters$^\bigstar$ \email cf.peeters@amsterdamumc.nl \\
        \addr Department of Epidemiology \& Biostatistics, \\
        Amsterdam University medical centers, location VUmc\\
        Postbus 7057, 1007 MB Amsterdam, The Netherlands
        \AND
        \name Poul Svante Eriksen \email svante@math.aau.dk\\
        \addr Department of Mathematical Sciences, \\
        Aalborg University\\
        9220 Aalborg \O, Denmark
        \AND
        \name Martin B{\o}gsted \email m\_boegsted@dcm.aau.dk\\
        \addr Department of Haematology, \\
        Aalborg University Hospital\\
        9000 Aalborg, Denmark ~\&\\
        Department of Clinical Medicine, \\
        Aalborg University \\
        9000 Aalborg, Denmark
        \AND
        \name Wessel N. van Wieringen \email w.vanwieringen@amsterdamumc.nl \\
        \addr Department of Epidemiology \& Biostatistics, \\
        Amsterdam University medical centers, location VUmc\\
        Postbus 7057, 1007 MB Amsterdam, The Netherlands ~\&\\
        Department of Mathematics, \\
        VU University Amsterdam\\
        1081 HV Amsterdam, The Netherlands}

\editor{Francis Bach}

\maketitle
\phantomsection
\addcontentsline{toc}{section}{Abstract}
\begin{abstract}%
\noindent We consider the problem of jointly estimating multiple inverse covariance matrices from high-dimensional data consisting of distinct classes.
An $\ell_2$-penalized maximum likelihood approach is employed.
The suggested approach is flexible and generic, incorporating several other $\ell_2$-penalized estimators as special cases.
In addition, the approach allows specification of target matrices through which prior knowledge may be incorporated and which can stabilize the estimation procedure in high-dimensional settings.
The result is a targeted fused ridge estimator that is of use when the precision matrices of the constituent classes are believed to chiefly share the same structure while potentially differing in a number of locations of interest.
It has many applications in (multi)factorial study designs.
We focus on the graphical interpretation of precision matrices with the proposed estimator then serving as a basis for integrative or meta-analytic Gaussian graphical modeling.
Situations are considered in which the classes are defined by data sets and subtypes of diseases.
The performance of the proposed estimator in the graphical modeling setting is assessed through extensive simulation experiments.
Its practical usability is illustrated by the differential network modeling of 12 large-scale gene expression data sets of diffuse large B-cell lymphoma subtypes.
The estimator and its related procedures are incorporated into the {\R}-package \texttt{rags2ridges}.
\end{abstract}



\section{Introduction}
High-dimensional data are ubiquitous in modern statistics. Consequently, the fundamental problem of estimating the covariance matrix or its inverse (the precision matrix) has received renewed attention.
Suppose we have $n$ i.i.d.\ observations of a $p$-dimensional variate distributed as $\calN_p(\vec{\mu}, \vSigma)$. The Gaussian log-likelihood parameterized in terms of the precision matrix $\vOmega = \vSigma^{-1}$ is then given by:
\begin{equation}
  \label{eq:OLL}
  \calL(\vOmega; \vS)
  \propto \ln\deter{\vOmega} - \tr(\vS\vOmega),
\end{equation}
where $\vS$ is the sample covariance matrix.
When $n > p$ the maximum of \eqref{eq:OLL} is attained at the maximum likelihood estimate (MLE) $\hvOmega^\text{ML} = \vS^{-1}$.
However, in the high-dimensional case, i.e., when $p > n$, the sample covariance matrix $\vS$ is singular and its inverse ceases to exist.
Furthermore, when $p \approx n$, the sample covariance matrix may be ill-conditioned and the inversion becomes numerically unstable.
Hence, these situations necessitate usage of regularization techniques.

Here, we study the simultaneous estimation of numerous precision matrices when multiple classes of high-dimensional data are present.
Suppose $\vy_{ig}$ is a realization of a $p$-dimensional Gaussian random vector for $i = 1, \ldots, n_g$ independent observations nested within $g = 1, \ldots, G$ classes, each with class-dependent covariance $\vSigma_g$, i.e., $\vy_{ig} \sim \calN_p(\vec{\mu}_g, \vSigma_g)$ for each designated class $g$.
Hence, for each class a data set consisting of the $n_g \times p$ matrix
$\vY_g = [\vy_{1 g}, \ldots, \vy_{n_g g}]^\top$ is observed.
Without loss of generality $\vec{\mu}_g = \vec{0}$ can be assumed as each data set $\vY_g$ can be centered around its column means.
The class-specific sample covariance matrix is given by
\begin{equation*}
  \vS_g
  = \frac{1}{n_g} \sum_{i = 1}^{n_g} \vy_{ig}\vy_{ig}^\top
  = \frac{1}{n_g} \vY_g^\top\vY_g,
\end{equation*}
which constitutes the well-known MLE of $\vSigma_g$ as discussed above.
The closely related \emph{pooled} sample covariance matrix
\begin{equation}
  \label{eq:pooledcovar}
  \vS_\summed
   = \frac{1}{n_\summed} \sum_{g = 1}^G \sum_{i = 1}^{n_g} \vy_{ig}\vy_{ig}^\top
   = \frac{1}{n_\summed} \sum_{g = 1}^G n_g \vS_g,
\end{equation}
where $n_\summed = \sum_{g = 1}^G n_g$, is an oft-used estimate of the common covariance matrix across classes.
In the high-dimensional setting, in which $p > n_\summed$ (implying $p > n_g$), the $\vS_g$ and $\vS_\summed$ are singular and their inverses do not exist.
Our primary interest thus lies in estimating the precision matrices $\vOmega_1 = \vSigma_1^{-1}, \ldots, \vOmega_G = \vSigma_G^{-1}$, as well as their commonalities and differences, when $p > n_\summed$.
We will develop a general $\ell_2$-penalized ML framework to this end which we designate \emph{targeted fused ridge estimation}.

The estimation of multiple precision matrices from high-dimensional data classes is of interest in many applications. The field of oncogenomics, for example, often deals with high-dimensional data from high-throughput experiments.
Class membership may have different connotations in such settings.
It may refer to certain sub-classes within a single data set such as cancer subtypes (cancer is a very heterogeneous disease, even when present in a single organ).
It may also designate different data sets or studies.
Likewise, the class indicator may also refer to a conjunction of both subclass and study membership to form a two-way design of factors of interest (e.g., breast cancer subtypes present in a batch of study-specific data sets), as is often the case in oncogenomics.
Our approach is thus motivated by the meta-analytic setting, where we aim for an integrative analysis in terms of simultaneously considering multiple data (sub-)classes, data sets, or both.
Its desire is to borrow statistical power across classes by effectively increasing the sample size in order to improve sensitivity and specificity of discoveries.

\subsection{Related Literature}
There have been many proposals for estimating a single precision matrix in high-dimensional data settings.
A popular approach is to amend \eqref{eq:OLL} with an $\ell_1$-penalty \citep{YL2007,Banerjee2008,Friedman2008,Yuan2008b}.
The solution to this penalized problem is generally referred to as the \emph{graphical lasso} and it is popular as it performs automatic model selection, i.e., the resulting estimate is sparse.
It is heavily used in Gaussian graphical modeling (GGM) as the support of a Gaussian precision matrix represents a Markov random field \citep{Lauritz96}.

The $\ell_1$-approach has been extended to deal with more than a single sample-group.
\citet{DINGO} employed a two-class approach that first extracts a global precision matrix by the graphical lasso after which precision regressions are employed to find local differences.
\citet{Zhao} also regard the two-class setting but, in contrast to many other approaches, focus on direct estimation of the difference between two precision matrices.
Many works also move beyond the two-class setting.
\citet{GLMZ2011} have proposed a parametrization of class-specific precision matrices that expresses the individual elements as a product of shared and class-specific factors.
They include $\ell_1$-penalties on both the shared and class-specific factors in order to jointly estimate the sparse precision matrices (representing graphical models).
The penalty on the shared factors promotes a shared sparsity structure while the penalty on the class-specific factors promotes class-specific deviations from the shared sparsity structure.
\citet{Danaher2013} have generalized these efforts by proposing the \emph{joint graphical lasso} which allows for various penalty structures.
They study two particular choices: the \emph{group graphical lasso} that encourages a shared sparsity structure across the class-specific precision matrices, and the \emph{fused graphical lasso} that promotes a shared sparsity structure as well as shared precision element-values.

The methods that move beyond the two-class setting have in common that they (implicitly) assume the same degree of similarity between all possible pairs of precision matrices.
Two recent works provide an important generalization by allowing for varying degrees of similarity: \citet{PSV2015} and \citet{SS16}.
These works permit, respectively from a Bayesian and frequentist perspective, for the pair-specific similarities to be estimated from the data.
Our motivation is related to these works (see Section \ref{motivate.sec}).

A hypothesis testing literature on multiple high-dimensional precision matrices has developed concurrently with the estimation literature.
Generally, the testing approaches are supported by penalized estimation.
As in estimation, the approaches can be demarcated by either a global or a local thrust \citep{Cai}.
The former focuses on testing the overall difference between two precision matrices.
The latter focuses on the simultaneous testing of the non-redundant individual entries of the difference matrix between two precision matrices.
\citet{SM17} provide a two-sample global testing approach under a sparsity assumption.
\citet{XCC15} provide both a global test as well as local testing through a (sparse) regression approach.
See \citet{Cai} for a review of recent work in testing for high-dimensional covariance and precision structures.

\subsection{Motivation of Approach}\label{motivate.sec}
Testing of high-dimensional precision matrices is generally only powerful when the alternative is sparse.
However, sparsity need not necessarily be a tenable assumption.
Moreover, the testing approaches are confined to two-class settings.
Hence, we focus on estimation.
Our goal is to provide a multiple class joint-estimation method that does not depend on a sparsity assumption and that allows for the flexible incorporation of prior information.
We motivate our approach below.

While simultaneous estimation and model selection can be deemed elegant, automatic sparsity is not always an asset.
It may be that one is intrinsically interested in more accurate representations of class-specific precision matrices in the high-dimensional situation.
By `intrinsically' we mean a representation that does not assume a (specific) sparsity pattern or structure.
Such representations are useful in enabling in the high-dimensional setting (standard) statistical applications directly dependent on the precision matrix, such as covariance-regularized regression \citep{Wit09} or discriminant analysis \citep{Price2014}.
One is then not after sparse representations, but rather (relatively) low-variance representations of the precision(s) in high-dimension.
It is then natural to prefer usage of a regularization method that shrinks the estimated elements of the precision matrices proportionally.

In addition, when indeed considering network representations of data (such that some level of sparsity is ultimately desired), one need not necessarily prefer the encouragement of sparsity through an $\ell_1$-approach.
It is well-known that $\ell_1$-based support recovery and estimation is consistent only under the assumption that the true (differential) graphical model is (very) sparse.
The $\ell_1$-penalty is unable to retrieve the sparsity pattern when the number of truly non-null elements exceeds the available sample size \citep{VanWieringen2014}.
This can be termed undesirable as there is accumulating evidence that many networks traditionally represented by graphical models, such as biochemical pathways governing disease aetiology and progression, are dense \citep{Omnigenic}.
In such a situation one may wish to couple a non-sparsity-inducing penalty with a post-hoc selection step allowing for probabilistic control over element selection \citep{VanWieringen2014}.
We therefore consider $\ell_2$ or ridge-type penalization.

The $\ell_2$-approach we consider will be \emph{targeted} in the sense that it allows for the specification of (possibly class-specific) target matrices that may encode prior information.
The motivation for including targets in general is that well-informed choices of the target can greatly improve the estimation in terms of loss/risk (Section \ref{Sims.sec}).
In addition, our framework also allows for varying degrees of similarity between (all possible) pairs of class-specific precision matrices through the incorporation of a penalty matrix (Section \ref{GenFused.sec}).
The diagonal elements of this matrix determine the rates of shrinkage of the class-specific precision matrices towards their corresponding targets while the off-diagonal entries determine the rates of pair-specific fusion.
The proposed framework is thus flexible in the sense that it allows for the incorporation of prior information along two roads as well as their interplay: (i) via the target matrices, and (ii) via the penalty matrix.
At one end of the spectrum we can include weak prior information through uninformative shared target matrices while letting the similarities between all pairs of precision matrices be subsequently determined by the data \citep[analogously to][]{PSV2015,SS16}.
At the other end we can include strong prior knowledge through informative class-specific target matrices while imposing restrictions on class-specific similarities by imposing (exclusion) constraints on the penalty matrix.

\subsection{Overview}\label{Overview.sec}
Section \ref{GenFused.sec} presents the \emph{targeted fused ridge estimation} framework.
The proposed fused $\ell_2$-penalty allows for the simultaneous estimation of multiple precision matrices from high-dimensional data classes that chiefly share the same structure but that may differentiate in locations of interest.
The usage of the mentioned target and penalty matrices makes the framework flexible and general.
It contains the recent work of \citet{Price2014} and \citet{VanWieringen2014} as special cases.
It may also be viewed as an $\ell_2$-generalization of the work of \citet{Danaher2013}.
Moreover, the framework can be viewed as bridging the work of \citet{Danaher2013} and \citet{SS16}, by allowing varying degrees of class-specific similarities, ranging from completely fixed for all possible pairs to completely data-determined for all possible pairs.
In the same vein, it may be viewed as a computationally feasible alternative to the work of \citet{PSV2015}, as it allows for the incorporation of prior information without having to formally specify prior distributions.
As such it evades the computational burden of a full Bayes approach.

The method is contingent upon the selection of penalty values and target matrices, topics that are treated in Section \ref{PenTar.sec}.
This section shows how---through the penalty values and target matrices---varying levels of specificity may be incorporated.
Section \ref{sec:posthoc} then focuses on the graphical interpretation of precision matrices.
It shows how the fused ridge precision estimates may be coupled with post-hoc support determination in order to arrive at multiple graphical models.
We will refer to this coupling as the \emph{fused graphical ridge}.
This then serves as a basis for integrative or meta-analytic network modeling.
Section \ref{Sims.sec} then assesses the performance of the proposed estimator through extensive simulation experiments.
These simulations show that the inclusion of target matrices can improve estimation efficiency.
Section \ref{Illustrate.sec} illustrates the techniques by applying it in a large scale integrative study of gene expression data of diffuse large B-cell lymphoma.
The focus is then on finding common motifs and motif differences in network representations of (deregulated) molecular pathways.
The analysis shows the added value of the targeted fusion approach to integration by juxtaposing it with a nonintegrative approach.
Moreover, it shows how pilot data and database information can be combined to provide effective target matrices.
Section \ref{Discuss.sec} concludes with a discussion.

\subsection{Notation}
Some additional notation must be introduced. Throughout the text and supplementary material, we use the following notation for certain matrix properties and sets: We use $\vA \succ \vec{0}$ and $\vB \succeq \vec{0}$ to denote symmetric positive definite and positive semi-definite matrices $\vA$ and $\vB$, respectively.
By $\bbR$, $\bbR_+$, and $\bbR_{++}$ we denote the real numbers, the non-negative real numbers, and the strictly positive real numbers, respectively.
In notational analogue, $\calS^p$, $\calS^p_+$, and $\calS^p_{++}$ are used to denote the space of $p\times p$ real symmetric matrices, the real symmetric positive semi-definite matrices, and real symmetric positive definite matrices, respectively. That is, e.g., $\calS_{++}^p = \{\vX \in \bbR^{p \times p} : \vX = \vX^\top \wedge \vX \succ \vec{0}\}$.
Negative subscripts similarly denote negative reals and negative definiteness. By $\vA \geq \vB$ and similar we denote \emph{element-wise} relations, i.e., $(\vA)_{jq} \geq (\vB)_{jq}$ for all $(j,q)$.
Matrix subscripts will usually denote class membership, e.g., $\vA_g$ denotes (the realization of) matrix $\vA$ in class $g$.
For notational brevity we will often use the shorthand $\{\vA_g\}$ to denote the set $\{\vA_g\}_{g=1}^{G}$.

The following notation is used throughout for operations: We write $\diag(\vA)$ for the column vector composed of the diagonal of $\vA$ and
$\vect(\vA)$ for the vectorization operator which stacks the columns of $\vA$ on top of each other.
Moreover, $\circ$ will denote the Hadamard product while $\otimes$ refers to the Kronecker product.

We will also repeatedly make use of several special matrices and functions.
We let $\vI_p$ denote the ($p\times p$)-dimensional identity matrix.
Similarly, $\vJ_p$ will denote the ($p\times p$)-dimensional all-ones matrix.
In addition, $\vec{0}$ will denote the null-matrix, the dimensions of which should be clear from the context.
Lastly, $\sqfnorm{\missingarg}$ and $\bbOne[\missingarg]$ will stand for the squared Frobenius norm and the indicator function, respectively.

\section{Targeted Fused Ridge Estimation}\label{GenFused.sec}
In this section we first give a general formulation of the targeted fused ridge estimation problem (Section \ref{Prob.sec}).
Next, the maximizing class-specific argument is explored as well as its properties (Section \ref{sec:properties}).
Last, an algorithm is presented with which the general, multiple-class solution can be obtained (Section \ref{sec:Algo}).

\subsection{A General Penalized Log-Likelihood Problem}
\label{Prob.sec}
Suppose $G$ classes of $(n_g \times p)$-dimensional data exist and that the samples within each class are i.i.d.\ normally distributed.
The log-likelihood for the data takes the following form under the additional assumption that all $n_\summed$ observations are independent:
\begin{equation}
  \label{eq:loglik}
  \calL\left(\{\vOmega_g\}; \{\vS_g\}\right)
    \propto \sum_g n_g
      \bigl\{ \ln\deter{\vOmega_g} - \tr(\vS_g\vOmega_g) \bigr\}.
\end{equation}
We desire to obtain estimates $\{\hvOmega_g\} \in \calS^p_{++}$ of the precision matrices for each class.
Though not a requirement, we primarily consider situations in which $p > n_g$ for all $g$, necessitating the need for regularization.
To this end, amend \eqref{eq:loglik} with the \emph{fused ridge penalty} given by
\par\nobreak
{\small
 \setlength{\abovedisplayskip}{6pt}
 \setlength{\belowdisplayskip}{\abovedisplayskip}
 \setlength{\abovedisplayshortskip}{0pt}
 \setlength{\belowdisplayshortskip}{3pt}
  \begin{equation}
     f^\text{FR}\left(\{\vOmega_g\}; \{\lambda_{g_1 g_2}\}, \{\vT_g\}\right)
     =   \sum_g \frac{\lambda_{gg}}{2} \sqfnorm[\big]{\vOmega_g {-} \vT_g}
      +  \sum_{\mathclap{g_1, g_2}} \frac{\lambda_{g_1 g_2}}{4}
        \sqfnorm[\big]{ (\vOmega_{g_1} {-} \vT_{g_1}) {-}
                          (\vOmega_{g_2} {-} \vT_{g_2}) },
    \label{eq:FR}
  \end{equation}
}%
where the $\vT_g \in \calS_+^p$ indicate known class-specific \emph{target matrices} (see also Section \ref{sec:Tselec}), the $\lambda_{gg} \in \bbR_{++}$ denote class-specific \emph{ridge penalty parameters}, and the $\lambda_{g_1 g_2} \in \bbR_+$ are pair-specific \emph{fusion penalty parameters} subject to the requirement that $\lambda_{g_1 g_2} = \lambda_{g_2 g_1}$.
All penalties can then be conveniently summarized into a non-negative symmetric matrix $\vLambda = [\lambda_{g_1 g_2}]$ which we call the \emph{penalty matrix}.
The diagonal of $\vLambda$ corresponds to the class-specific ridge penalties whereas off-diagonal entries are the pair-specific fusion penalties.
The rationale and use of the penalty matrix is motivated further in Section \ref{sec:penaltygraph}.
Combining \eqref{eq:loglik} and \eqref{eq:FR} yields a general targeted fused ridge estimation problem:
\par\nobreak
{\small
 \setlength{\abovedisplayskip}{6pt}
 \setlength{\belowdisplayskip}{\abovedisplayskip}
 \setlength{\abovedisplayshortskip}{0pt}
 \setlength{\belowdisplayshortskip}{3pt}
  \begin{equation}
    \label{eq:argmax0}
    \argmax_{\{\vOmega_g\} \in \calS_{++}^p}
    \left\{
    \calL\left(\{\vOmega_g\}; \{\vS_g\}\right)
    - \sum_g \frac{\lambda_{gg}}{2} \sqfnorm[\big]{ \vOmega_g {-} \vT_g }
    - \sum_{\mathclap{g_1, g_2}} \frac{\lambda_{g_1 g_2}}{4}
    \sqfnorm[\big]{ (\vOmega_{g_1} {-} \vT_{g_1}) {-}
                    (\vOmega_{g_2} {-} \vT_{g_2}) }
    \right\}.
  \end{equation}
}%
The problem of \eqref{eq:argmax0} is strictly concave.
Furthermore, it is worth noting that non-zero fusion penalties, $\lambda_{g_1 g_2} > 0$ for all $g_1 \neq g_2$, alone will not guarantee uniqueness when $p > n_\summed$: In high dimensions, all ridge penalties $\lambda_{gg}$ should be strictly positive to ensure identifiability.
These and other properties of the estimation problem are reviewed in Section \ref{sec:properties}.

The problem stated in \eqref{eq:argmax0} is very general.
We shall sometimes consider a single common ridge penalty $\lambda_{gg} = \lambda$ for all $g$, as well as a common fusion penalty $\lambda_{g_1 g_2} = \lambda_f$ for all class pairs $g_1 \neq g_2$ (cf., however, Section \ref{sec:penaltygraph}) such that $\vLambda = \lambda\vI_G + \lambda_f(\vJ_G-\vI_G)$.
This simplification leads to the first special case:
\begin{equation*}
  \argmax_{\{\vOmega_g\} \in \calS_{++}^p}
  \left\{
    \calL\left(\{\vOmega_g\}; \{\vS_g\}\right)
    - \frac{\lambda}{2}\sum_g  \sqfnorm[\big]{ \vOmega_g \!- \vT_g }
    - \frac{\lambda_f}{4}\sum_{\mathclap{g_1, g_2}}
      \sqfnorm[\big]{ (\vOmega_{g_1} \!- \vT_{g_1})  -
                      (\vOmega_{g_2} \!- \vT_{g_2}) }
  \right\}.
\end{equation*}
Here and analogous to \eqref{eq:argmax0}, $\lambda$ controls the rate of shrinkage of each precision $\vOmega_g$ towards the corresponding target $\vT_g$ \citep{VanWieringen2014}, while $\lambda_f$ determines the retainment of entry-wise similarities between $(\vOmega_{g_1} \!- \vT_{g_1})$ and $(\vOmega_{g_2} \!- \vT_{g_2})$ for all class pairs $g_1 \neq g_2$.

When $\vT_g = \vT$ for all $g$, the problem further simplifies to
\begin{equation}
  \label{eq:argmax2}
  \argmax_{\{\vOmega_g\} \in \calS_{++}^p}
  \left\{
    \calL\left(\{\vOmega_g\}; \{\vS_g\}\right)
    - \frac{\lambda}{2}\sum_g
      \sqfnorm[\big]{ \vOmega_g \!- \vT }
    - \frac{\lambda_f}{4}\sum_{\mathclap{g_1, g_2}}
      \sqfnorm[\big]{ \vOmega_{g_1} \!- \vOmega_{g_2}  }
  \right\},
\end{equation}
where the targets are seen to disappear from the fusion term.
Lastly, when $\vT = \vec{0}$ the problem \eqref{eq:argmax2} reduces to its simplest form recently considered by \citet{Price2014}.
Appendix \ref{app:geometric} studies, in order to support an intuitive feel for the fused ridge estimation problem, its geometric interpretation in this latter context.

\subsection{Estimator and Properties}\label{sec:properties}
There is no explicit solution to \eqref{eq:argmax0} except for certain special cases and thus an iterative optimization procedure is needed for its general solution.
As described in Section \ref{sec:Algo}, we employ a coordinate ascent procedure which relies on the concavity of the penalized likelihood (see Lemma~\ref{lem:concavity} in Appendix \ref{sec:Support}) and repeated use of the following result, whose proof (as indeed all proofs) has been deferred to Appendix \ref{sec:Proofs}:

\begin{proposition}
\label{prop:fusedridge}%
Let $\{\vT_g\} \in \calS_+^p$ and let $\vLambda \in \calS^G$ be a fixed penalty matrix such that $\vLambda \geq \vec{0}$ and $\diag(\vLambda) > \vec{0}$.
Furthermore, assume that $\vOmega_g$ is positive definite and fixed for all $g\neq g_0$.
The maximizing argument for class $g_0$ of the optimization problem \eqref{eq:argmax0} is then given by
\begin{gather}
  \label{eq:update}
  \hvOmega_{g_0}\bigl(\vLambda, \{\vOmega_g\}_{g {\neq} g_0} \bigr)
  =
  \left\{
    \left[
      \bar{\lambda}_{g_0} \vI_p
      + \frac{1}{4}\big(\bar{\vS}_{g_0} - \bar{\lambda}_{g_0}\bar{\vT}_{g_0}\big)^2
    \right]^{1/2}
    + \frac{1}{2}\big(\bar{\vS}_{g_0} - \bar{\lambda}_{g_0} \bar{\vT}_{g_0}\big)
  \right\}^{-1},
  \intertext{where}
  \bar{\vS}_{g_0}
  = \vS_{g_0} - \sum_{g \neq g_0}\frac{\lambda_{g g_0}}{n_{g_0}} (\vOmega_g \!- \vT_g),
  \label{eq:barupdate}
  \quad
  \bar{\vT}_{g_0} = \vT_{g_0},
  \andwhere
  \bar{\lambda}_{g_0} = \frac{\lambda_{g_0\summed}}{n_{g_0}},
\end{gather}
with $\lambda_{g_0\summed} = \sum_{g} \lambda_{g g_0}$ denoting the sum of the \nth{g_0} column (or row) of $\vLambda$.
\end{proposition}

\begin{remark}
Defining $\bar{\vT}_{g_0} = \vT_{g_0}$ in Proposition \ref{prop:fusedridge} may be deemed redundant.
However, it allows us to state equivalent alternatives to \eqref{eq:barupdate} without confusing notation.
See Section \ref{sec:Algo} as well as Appendix \ref{sec:Proofs} and Section 1 of the Supplementary Material.
\end{remark}

\begin{remark}
The target matrices from Proposition \ref{prop:fusedridge} may be chosen nonnegative definite.
However, choosing n.d.\ targets may lead to ill-conditioned estimates in the limit.
From a shrinkage perspective we thus prefer to choose $\{\vT_g\} \in \calS_{++}^p$.
See Section \ref{sec:Tselec}.
\end{remark}

Proposition~\ref{prop:fusedridge} provides a function for updating the estimate of the \nth{g_0} class while fixing the remaining parameters. As a special case, consider the following. If all off-diagonal elements of $\vLambda$ are zero no `class fusion' of the estimates takes place and the maximization problem decouples into $G$ individual, disjoint ridge estimations: See Corollary \ref{prop:fusedridge2} in Appendix \ref{sec:Proofs}. The next result summarizes some properties of \eqref{eq:update}:

\begin{proposition}
\label{prop:fusedridge3}%
Consider the estimator of Proposition \ref{prop:fusedridge} and its accompanying assumptions.
Let $\hvOmega_{g} \equiv \hvOmega_{g}\bigl(\vLambda, \{\vOmega_{g'}\}_{g' {\neq} g}\bigr)$ be the precision matrix estimate of the \nth{g} class.
For this estimator, the following properties hold:
\begin{enumerate}[i.]
  \item $\hvOmega_g \succ \vec{0}$ for all $\lambda_{gg} \in \bbR_{++}$;\vspace{-.1cm}
  \label{prop:fusedridge3item1}
  \item $\lim\limits_{\lambda_{g g} \to 0^+} \hvOmega_g = \vS_g^{-1}$ if $\sum_{g' \neq g} \lambda_{gg'} = 0$ and $p \leq n_g$;\vspace{-.2cm}
  \label{prop:fusedridge3item2}
  \item $\lim\limits_{\lambda_{g g} \to \infty^-} \hvOmega_g  = \vT_g$ if $\lambda_{g g'} < \infty$ for all $g'\neq g$;\vspace{-.2cm}
  \label{prop:fusedridge3item3}
  \item $\lim\limits_{\lambda_{g_1 g_2} \to \infty^-}(\hvOmega_{g_1} - \vT_{g_1})
  = \lim\limits_{\lambda_{g_1 g_2} \to \infty^-} (\hvOmega_{g_2} - \vT_{g_2})$ if $\lambda_{g_1' g_2'} < \infty$ for all $\{g_1',g_2'\} \neq \{g_1,g_2\}$.
  \label{prop:fusedridge3item4}
\end{enumerate}
\end{proposition}

The first item of Proposition \ref{prop:fusedridge3} implies that strictly positive $\lambda_{gg}$ are sufficient to guarantee positive definite estimates from the ridge estimator.
The second item implies that if `class fusion' is absent, then one obtains the standard MLE $\vS_g^{-1}$ as the right-hand limit for group $g$, whose existence is only guaranteed when $p \leq n_g$.
The third item shows that the fused ridge precision estimator for class $g$ is shrunken exactly to its target matrix when the ridge penalty tends to infinity while the fusion penalties do not.
The last item shows that the precision estimators of any two classes tend to a common estimate when the fusion penalty between them tends to infinity while all remaining penalty parameters remain finite.

The attractiveness of the general estimator hinges upon the efficiency by which it can be obtained. We state a result useful in this respect before turning to our computational approach in Section \ref{sec:Algo}:
\begin{proposition}
\label{prop:InvLess}%
Let $\hvOmega_{g} \equiv \hvOmega_{g}\bigl(\vLambda, \{\vOmega_{g'}\}_{g' {\neq} g}\bigr)$ be the precision matrix estimate \eqref{eq:update} for the \nth{g} class and define $[\hvOmega_{g}]^{-1} \equiv \hvSigma_{g}$.
The estimate $\hvOmega_{g}$ can then be obtained without inversion through:
\begin{equation}\nonumber
  \hvOmega_{g}
  =
  \frac{1}{\bar{\lambda}_{g}}
    \left[ \hvSigma_{g} - (\bar{\vS}_{g} - \bar{\lambda}_{g}   \bar{\vT}_{g}) \right]
  =
  \frac{1}{\bar{\lambda}_{g}}
  \left\{
    \left[
      \bar{\lambda}_{g} \vI_p
      + \frac{1}{4}\big(\bar{\vS}_{g} - \bar{\lambda}_{g}\bar{\vT}_{g_0}\big)^2
    \right]^{1/2}
    - \frac{1}{2}\big(\bar{\vS}_{g} - \bar{\lambda}_{g} \bar{\vT}_{g}\big)
  \right\}.
\end{equation}
\end{proposition}

\begin{remark}
Note that Proposition \ref{prop:InvLess} implies that our framework also immediately provides for regularized class-specific estimates of covariance matrices as $\hvSigma_{g} = \bar{\lambda}_{g}\hvOmega_{g} + (\bar{\vS}_{g} - \bar{\lambda}_{g} \bar{\vT}_{g})$.
Its properties are analogous to those stated in Proposition \ref{prop:fusedridge3}.
\end{remark}

\subsection{Algorithm}\label{sec:Algo}
Equation \eqref{eq:update} allows for updating the precision estimate $\hvOmega_{g}$ of class $g$ by plugging in the remaining $\hvOmega_g'$, $g' \neq g$, and assuming them fixed.
Hence, from initial estimates, all precision estimates may be iteratively updated until some convergence criterion is reached.
We propose a block coordinate ascent procedure to solve \eqref{eq:argmax0} by repeated use of the results in Proposition \ref{prop:fusedridge}.
This procedure is outlined in Algorithm \ref{alg:fusedridge}.
By the strict concavity of the problem in \eqref{eq:argmax0}, the procedure guarantees that, contingent upon convergence, the unique maximizer is attained when considering all $\hvOmega_g$ jointly.
Moreover, we can state the following result:

\begin{proposition}
\label{prop:PosRealm}%
The gradient ascent procedure given in Algorithm \ref{alg:fusedridge} will always stay within the realm of positive definite matrices $\calS_{++}^p$.
\end{proposition}

The procedure is implemented in the \texttt{rags2ridges} package within the {\R} statistical language \citep{R}.
This implementation focuses on \emph{stability} and \emph{efficiency}. With regard to the former: Equivalent (in terms of the obtained estimator) alternatives to \eqref{eq:barupdate} can be derived that are numerically more stable for extreme values of $\vLambda$.
The most apparent such alternative is:
\begin{equation}
  \label{eq:barupdate1}
  \bar{\vS}_{g_0} = \vS_{g_0},
  \quad
  \bar{\vT}_{g_0} = \vT_{g_0} + \sum_{g \neq g_0}
          \frac{\lambda_{g g_0}}{\lambda_{g_0\summed}}
          (\vOmega_g \!- \vT_g),
  \andwhere
  \bar{\lambda}_{g_0}
  = \frac{\lambda_{g_0\summed}}{n_{g_0}}.
\end{equation}
It `updates' the target $\bar{\vT}_g$ instead of the sample covariance $\bar{\vS}_g$ and has the intuitive interpretation that the target matrix for a given class in the fused case is a combination of the actual class target matrix and the `target corrected' estimates of remaining classes.
The implementation makes use of this alternative where appropriate.
See Section 1 of the Supplementary Material for details on alternative updating schemes.

\begin{algorithm}[H]
\caption{Pseudocode for the fused ridge block coordinate ascent procedure.}
\label{alg:fusedridge}
\begin{algorithmic}[1]
\State \algorithmicrequire{
\State \emph{Sufficient data:} $(\vS_1, n_1), \ldots, (\vS_G, n_G)$
\State \emph{Penalty matrix:} $\vLambda$
\State \emph{Convergence criterion:} $\varepsilon > 0$
}
\State \algorithmicensure{
\State \emph{Estimates:} $\hvOmega_1, \ldots, \hvOmega_G$
}
\Procedure{ridgeP.fused}{$\vS_1, \ldots, \vS_G, n_1, \ldots, n_G, \vLambda, \varepsilon$}
\State \label{lst:Initial} \emph{Initialize}: $\hvOmega_g^{(0)}$ for all $g$.
  \For {$c = 1, 2, 3, \ldots$}
    \For {$g = 1, 2, \ldots, G$}
      \State \label{lst:UpdateStep} Update $\hvOmega_g^{(c)} :=
          \hvOmega_g
            \big(
              \vLambda,
              \hvOmega{}_1^{(c)}, \ldots, \hvOmega_{g-1}^{(c)},
              \hvOmega{}_{g+1}^{(c-1)}, \ldots, \hvOmega_G^{(c-1)}
            \big)$
            by \eqref{eq:update}.
    \EndFor
     \If {$\max_g\!\Big\{ \frac{\sqfnorm{\hvOmega{}_g^{(c)} - \hvOmega{}_g^{(c-1)}}}
                               {\sqfnorm{\hvOmega{}_g^{(c)}}}  \Big\} <
                                  \varepsilon$}
        \State \Return $\big(\hvOmega_1^{(c)}, \ldots, \hvOmega_G^{(c)}\big)$
    \EndIf
  \EndFor
\EndProcedure
\end{algorithmic}
\end{algorithm}

The worst-case asymptotic time complexity of the procedure is $\mathcal{O}(p^{3})$ due to the necessity of the matrix square root.
Efficiency is then secured through various roads.
First, in certain special cases closed-form solutions to \eqref{eq:argmax0} exist.
When appropriate, these explicit solutions are used.
Moreover, these solutions may provide warm-starts for the general problem.
See Section 2 of the Supplementary Material for details on estimation in these special cases.
Second, the result from Proposition \ref{prop:InvLess} is used, meaning that the relatively expensive operation of matrix inversion is avoided.
Third, additional computational speed was achieved by implementing core operations in \textsf{C++} via the {\R}-packages \texttt{Rcpp} and \texttt{RcppArmadillo} \citep{Sanderson2010, Eddelbuettel2011, RcppArmadillo, Rcpp2013}. These efforts make analyzes with large $p$ feasible.
Throughout, we will initialize the algorithm with $\hvOmega_g^{(0)} = p/\tr(\vS_\summed)\cdot\vI_p$ for all $g$.

\section{Penalty and Target Selection}\label{PenTar.sec}
In this section we discuss selection of the penalty parameters and the target matrices.
First, we discuss, by way of examples, how the penalty matrix connects to a penalty-graph and how its structure may encode prior information in the analysis of various study-designs (Section \ref{sec:penaltygraph}).
Next, we present several computational approaches to select optimal values for the parameters in the (possibly structured) penalty matrix (Section \ref{sec:PenSelec}).
Last, we give several considerations in choosing target matrices (Section \ref{sec:Tselec}).

\subsection{The Penalty Graph and Analysis of Factorial Designs}
\label{sec:penaltygraph}
Equality of all class-specific ridge penalties $\lambda_{gg}$ is deemed restrictive, as is equality of all pair-specific fusion penalties $\lambda_{g_1 g_2}$.
In many settings, such as the analysis of factorial designs, finer control over the individual values of $\lambda_{gg}$ and $\lambda_{g_1 g_2}$ befits the analysis.
This will be motivated by several examples of increasing complexity.
In order to do so, some additional notation is developed:
The penalties of $\vLambda$ can be summarized by a node- and edge-weighted graph $\calP = (W, H)$ where the vertex set $W$ corresponds to the possible classes and the edge set $H$ corresponds to the similarities to be retained.
The weight of node $g \in W$ is given by $\lambda_{g g}$ and
the weight of edge $(g_1, g_2)\in H$ is then given by $\lambda_{g_1 g_2}$.
We refer to $\calP$ as the \emph{penalty graph} associated with the penalty matrix $\vLambda$.
The penalty graph $\calP$ is simple and undirected as the penalty matrix is symmetric.
In the examples below we generally assume $p > n_{\bullet}$.

\begin{example}
Consider $G = 2$ classes or subtypes ($\mathrm{ST}$) of diffuse large B-cell lymphoma (DLBCL) patients with tumors resembling either so-called activated B-cells ($\ABC$) or germinal centre B-cells ($\GCB$).
Patients with the latter subtype have superior overall survival \citep{Alizadeh2000}.
As the $\GCB$ phenotype is more common than $\ABC$, one might imagine a scenario where the two class sample sizes are sufficiently different such that $n_\GCB \gg n_\ABC$.
Numeric procedures to obtain a common ridge penalty (see, e.g., Section \ref{sec:PenSelec}) would then be dominated by the smaller group.
Hence, choosing non-equal class ridge penalties for each group will allow for a better analysis.
In such a case, the following penalty graph and matrix would be suitable:

\begin{equation}
\label{eq:ex1}
\begin{tikzpicture}[node distance = 2mm, auto,
  baseline ={(0,-3.5pt)},
  main_node/.style={circle,draw,minimum size=1em,inner sep=1pt}]
    \node (P) at (-1, 0) {$\calP =$};
    \node [main_node, label={[yshift=0.1cm]$\ABC$} ] (n1) at (0,0) {$\lambda_{11}$};
    \node [main_node, label={[yshift=0.1cm]$\GCB$} ] (n2) at (2,0) {$\lambda_{22}$};

    \path
      (n1) edge node [above] {$\lambda_f$} (n2);
\end{tikzpicture}
\qquad
\vLambda =
\begin{bmatrix}
  \lambda_{11} & \lambda_f \\
  \lambda_f & \lambda_{22}
\end{bmatrix}.
\end{equation}
\end{example}

\begin{example}
\label{ex:2}
Consider data from a one-way factorial design where the factor is ordinal with classes $\mathrm{A}$, $\mathrm{B}$, and $\mathrm{C}$.
For simplicity, we choose the same ridge penalty $\lambda$ for each class.
Say we have prior information that $\mathrm{A}$ is closer to $\mathrm{B}$ and $\mathrm{B}$ is closer to $\mathrm{C}$ than $\mathrm{A}$ is to $\mathrm{C}$.
The fusion penalty on the pairs containing the intermediate level $\mathrm{B}$ might then be allowed to be stronger.
The following penalty graph and matrix are thus sensible:

\begin{equation}
\label{eq:ex2}
\begin{tikzpicture}[node distance = 2mm, auto,
  baseline={([yshift=-.5ex]current bounding box.center)},
  main_node/.style={circle,draw,minimum size=1em,inner sep=2pt}]
    \node (P) at (-1, 0) {$\calP =$};
    \node [main_node, label={[yshift=0.05cm]$\mathrm{A}$}] (n1) at (0,0) {$\lambda$};
    \node [main_node, label={[yshift=0.05cm]$\mathrm{C}$}] (n2) at (4,0) {$\lambda$};
    \node [main_node, label={[yshift=0.05cm]$\mathrm{B}$}] (n3) at (2,0) {$\lambda$};

    \path
      (n1) edge node [above, midway] {$\lambda_\mathrm{B}$} (n3)
      (n3) edge node [above, midway] {$\lambda_\mathrm{B}$} (n2)
      (n1) edge [bend right=15] node [below, midway] {$\lambda_\mathrm{AC}$} (n2);

\end{tikzpicture}
\qquad
\vLambda =
\begin{bmatrix}
\lambda              & \lambda_\mathrm{B} & \lambda_\mathrm{AC}\\
\lambda_\mathrm{B}   & \lambda            & \lambda_\mathrm{B}\\
\lambda_\mathrm{AC}  & \lambda_\mathrm{B} & \lambda
\end{bmatrix}.
\end{equation}

\noindent Depending on the application, one might even omit the direct shrinkage between $\mathrm{A}$ and $\mathrm{C}$ by fixing $\lambda_\mathrm{AC} = 0$.
A similar penalty scheme might also be relevant if one class of the factor is an unknown mix of the remaining classes and one wishes to borrow statistical power from such a class.
\end{example}

\begin{example}
\label{ex:3}
In two-way or $n$-way factorial designs one might wish to retain similarities in the `direction' of each factor along with a factor-specific penalty.
Consider, say, 3 oncogenomic data sets ($\mathrm{DS}_1$, $\mathrm{DS}_2$, $\mathrm{DS}_3$) regarding $\ABC$ and $\GCB$ DLBCL cancer patients.
This yields a total of $G = 6$ classes of data.
One choice of penalization of this $2$ by $3$ design is represented by the penalty graph and matrix below:

\begin{equation}
\begin{tikzpicture}[node distance = 2mm, auto,
  baseline ={(0,-3.5pt)},
  main_node/.style={circle,draw,minimum size=1em,inner sep=2pt}]

    \node (P) at (-1.25, 0) {$\calP =$};
    \node [main_node] (n1) at (0,0.625) {$\lambda$};
    \node [main_node] (n2) at (2,0.625) {$\lambda$};
    \node [main_node] (n3) at (4,0.625) {$\lambda$};
    \node [main_node] (n4) at (0,-0.625) {$\lambda$};
    \node [main_node] (n5) at (2,-0.625) {$\lambda$};
    \node [main_node] (n6) at (4,-0.625) {$\lambda$};

    \draw (n1) -- (n2) node [below, midway] {\scriptsize$\lamDS$};
    \draw (n2) -- (n3) node [below, midway] {\scriptsize$\lamDS$};
    \draw (n4) -- (n5) node [above, midway] {\scriptsize$\lamDS$};
    \draw (n5) -- (n6) node [above, midway] {\scriptsize$\lamDS$};
    \draw (n1) -- (n4) node [left, midway] {\scriptsize$\lamST$};
    \draw (n2) -- (n5) node [left, midway] {\scriptsize$\lamST$};
    \draw (n3) -- (n6) node [left, midway] {\scriptsize$\lamST$};

    \path
      (n1) edge [bend left=15] node [above, near start] {\scriptsize$\lamDS$} (n3)
      (n4) edge [bend right=15] node [below, near start] {\scriptsize$\lamDS$} (n6);

    \node [above=0.1cm of n1] (DS1) {$\mathrm{DS}_1$};
    \node [above=0.1cm of n2] (DS2) {$\mathrm{DS}_2$};
    \node [above=0.1cm of n3] (DS3) {$\mathrm{DS}_3$};
    \node [left=0.05cm of n1] (GCB) {$\GCB$};
    \node [left=0.05cm of n4] (ABC) {$\ABC$};

\end{tikzpicture}
\qquad
\vLambda =
\begin{bmatrix}
  \lambda   & \lamDS    & \lamDS    & \lamST    & 0         & 0\\
  \lamDS    & \lambda   & \lamDS    & 0         & \lamST    & 0\\
  \lamDS    & \lamDS    & \lambda   & 0         & 0         & \lamST\\
  \lamST    & 0         & 0         & \lambda   & \lamDS    & \lamDS\\
  0         & \lamST    & 0         & \lamDS    & \lambda   & \lamDS\\
  0         & 0         & \lamST    & \lamDS    & \lamDS    & \lambda
\end{bmatrix}.
\label{eq:ERpenaltygraph}
\end{equation}

\noindent This example would favor similarities (with the same force) only between pairs sharing a common level in each factor.
This finer control allows users, or the employed algorithm, to penalize differences between data sets more (or less) strongly than differences between the $\ABC$ and $\GCB$ sub-classes.
This corresponds to not applying direct shrinkage of interaction effects which is of interest in some situations.
\end{example}

While the penalty graph primarily serves as an intuitive overview, it does provide some aid in the construction of the penalty matrix for multifactorial designs.
For example, the construction of the penalty matrix \eqref{eq:ERpenaltygraph} in Example~\ref{ex:3} corresponds to a Cartesian graph product of two complete graphs similar to those given in \eqref{eq:ex1} and \eqref{eq:ex2}.
We state that $\calP$ and $\vLambda$ should be chosen carefully in conjunction with the choice of target matrices.
Ideally, only strictly necessary penalization parameters (from the perspective of the desired analysis) should be introduced.
Each additional penalty introduced will increase the difficulty of finding the optimal penalty values by increasing the dimension of the search-space.

\subsection{Selection of Penalty Parameters}\label{sec:PenSelec}
As the $\ell_2$-penalty does not automatically induce sparsity in the estimate, it is natural to seek loss efficiency.
We then use cross-validation (CV) for penalty parameter selection due to its relation to the minimization of the Kullback-Leibler divergence and its predictive accuracy stemming from its data-driven nature.
We randomly divide the data of each class into $k = 1, \ldots, K$ disjoint subsets of approximately the same size.
Previously, we have defined $\hvOmega_{g} \equiv \hvOmega_{g}\bigl(\vLambda, \{\vOmega_{g'}\}_{g' {\neq} g}\bigr)$ to be the precision matrix estimate of the \nth{g} class.
Let $\hvOmega{}_g^{\neg k}$ be the analogous estimate (with similar notational dependencies) for class $g$ based on all samples not in $k$.
Also, let $\vS_g^{k}$ denote the sample covariance matrix for class $g$ based on the data in subset $k$ and let $n_g^{k}$ denote the size of subset $k$ in class $g$.
The $K$-fold CV score for our fused regularized precision estimate based on the fixed penalty $\vLambda$ can then be given as:
\begin{equation*}
  \mathrm{KCV}(\vLambda) =
  \frac{1}{KG} \sum_{g = 1}^G \sum_{k = 1}^{K} n_g^{k} \left[-\ln|\hvOmega{}_g^{\neg k}| + \tr(\hvOmega{}_g^{\neg k}\vS_g^{k})\right]
  = -\frac{1}{KG} \sum_{g = 1}^G \sum_{k = 1}^{K}
    \calL_{g}^k\bigl(\hvOmega_g^{\neg k}; \vS_{g}^{k} \bigr).
\end{equation*}
One would then choose $\vLambda^\ast$ such that
\begin{equation}
  \vLambda^\ast
  = \argmin_{\vLambda}
     \mathrm{KCV}(\vLambda), \quad\text{subject to:} \quad \vLambda \geq \vec{0} \wedge \diag(\vLambda) > \vec{0}.
\end{equation}
The least biased predictive accuracy can be obtained by choosing $K = n_g$ such that $n_g^{k} = 1$.
This would give the fused version of leave-one-out CV (LOOCV).
Unfortunately, LOOCV is computationally demanding for large $p$ and/or large $n_g$.
We propose to select the penalties by the computationally expensive LOOCV only if adequate computational power is available.
In cases where it is not, we propose two alternatives.

Our first  alternative is a special version of the LOOCV scheme that significantly reduces the computational cost.
The \emph{special} LOOCV ($\SLOOCV$) is computed much like the LOOCV.
However, only the class estimate in the class of the omitted datum is updated.
More specifically, the $\SLOOCV$ problem is given by:
\begin{equation}
  \label{eq:argminSLOOCV}
  \vLambda^\diamond
  = \argmin_{\vLambda}
     \mathrm{SLOOCV}(\vLambda), \quad\text{subject to:} \quad \vLambda \geq \vec{0} \wedge \diag(\vLambda) > \vec{0},
\end{equation}
with
\begin{equation}\nonumber
  \SLOOCV(\vLambda)
  = -\frac{1}{n_\bullet} \sum_{g = 1}^G \sum_{i = 1}^{n_g}
    \calL_{g}^{i}\bigl(\widetilde{\vOmega}{}_g^{\neg i}; \vS_{g}^{i} \bigr).
\end{equation}
The estimate $\widetilde{\vOmega}{}_g^{\neg i}$ in \eqref{eq:argminSLOOCV} is obtained by updating only $\hvOmega_g$ using Proposition~\ref{prop:fusedridge}.
For all other $g' \neq g$, $\widetilde{\vOmega}{}_{g'}^{\neg i} = \hvOmega_g$.
The motivation for the SLOOCV is that a single observation in a given class $g$ does not exert heavy direct influence on the estimates in the other classes.
This way the number of fused ridge estimations for each given $\vLambda$ and each given leave-one-out sample is reduced from $n_\summed$ to $G$ estimations.
Our second and fastest alternative is an approximation of the fused LOOCV score.
This approximation can be used as an alternative to (S)LOOCV when the class sample sizes are relatively large (precisely the scenario where LOOCV is unfeasible).
See Section 3 of the Supplementary Material for detailed information on this approximation.

\subsection{Choice of Target Matrices}\label{sec:Tselec}
The target matrices $\{\vT_g\}$ can be used to encode prior information and their choice is highly dependent on the application at hand.
As they influence the efficacy as well as the amount of bias of the estimate, it is of some importance to make a well-informed choice.
Here, we describe several options of increasing level of informativeness, showcasing the flexibility of target specification.

The limited fused ridge problem in \citet{Price2014} corresponds to choosing the common target $\vT_g = \vT = \vec{0}$.
This can be considered the least informative target possible.
We generally argue against the use of the non positive definite target $\vT = \vec{0}$, as it implies shrinking the class precision matrices towards the null matrix and thus towards infinite variance.

In some situations one may wish to penalize the diagonal elements of the precision matrices at a different rate than the off-diagonal elements.
Specifying $\vT_g = (\mathbf{S}_{g} \circ \mathbf{I}_p)^{-1}$ would be equivalent to shrinking the precision estimate for class $g$ towards a diagonal matrix carrying the inverse variances of $\mathbf{S}_{g}$ and, hence, (from the precision-perspective) letting the diagonal elements of $\mathbf{S}_{g}$ go unpenalized.
Such a target can be scaled to give varying rates of shrinkage for the (off-)diagonal elements.
That is, one could specify $\gamma_g(\mathbf{S}_{g} \circ \mathbf{I}_p)^{-1}$ with $\gamma_g \in [0, \infty)$, although from an empirical perspective it would make sense to choose $\gamma_g \in [0, 1]$.
In the special case when $\vT_g = \vT$ for all $g$ one could choose $\vT = \gamma(\mathbf{S}_{\summed} \circ \mathbf{I}_p)^{-1}$.
When choosing $\gamma_g = 0$ for all $g$, the common target $\vT_g = \vT = \vec{0}$ ensues.

In the non-fused setting, the consideration of a scalar target matrix $\vT = \alpha\vI_p$ for some $\alpha \in [0, \infty)$ leads to a computational benefit stemming from the property of rotation equivariance \citep{VanWieringen2014}: Under such targets the ridge estimator only operates on the eigenvalues of the sample covariance matrix.
This benefit transfers to the fused setting for the estimator described in Proposition~\ref{prop:fusedridge}.
To see this let $\mathbf{V}_{g}\mathbf{D}(\bar{\vS}_{g})\mathbf{V}_{g}^{\mathrm{T}}$ be the spectral decomposition of $\bar{\vS}_{g}$ with $\mathbf{D}(\bar{\vS}_{g})$ denoting a diagonal matrix with the eigenvalues of $\bar{\vS}_{g}$ on the diagonal and where $\mathbf{V}_{g}$ denotes the matrix that contains the corresponding eigenvectors as columns.
Naturally, the orthogonality of $\mathbf{V}_{g}$ implies $\mathbf{V}_{g}\mathbf{V}_{g}^{\mathrm{T}} = \mathbf{V}_{g}^{\mathrm{T}}\mathbf{V}_{g} = \mathbf{I}_p$.
Now, note that, if $\mathbf{T}_g = \alpha_g \mathbf{I}_p$, we can write $\hvOmega_{g}\bigl(\vLambda, \{\vOmega_{g'}\}_{g' {\neq} g} \bigr)$ as:
\begin{equation}\nonumber
  \label{eq:updateEqui}
  \mathbf{V}_{g}\left\{
    \left[
      \bar{\lambda}_{g} \vI_p
      + \frac{1}{4}\big(\mathbf{D}(\bar{\vS}_{g}) - \bar{\lambda}_{g}\alpha_g \mathbf{I}_p\big)^2
    \right]^{1/2}
    + \frac{1}{2}\big(\mathbf{D}(\bar{\vS}_{g}) - \bar{\lambda}_{g}\alpha_g \mathbf{I}_p\big)
  \right\}^{-1}\mathbf{V}_{g}^{\mathrm{T}}.
\end{equation}
Letting $d(\cdot)_{jj}$ denote the $j$th eigenvalue of the matrix terms in brackets we thus have that:
\begin{equation}\nonumber
  \label{eq:updateEquiEV}
  d\left[\hvOmega_{g}\bigl(\vLambda, \{\vOmega_{g'}\}_{g' {\neq} g} \bigr) \right]_{jj} =
  \left\{ \sqrt{\bar{\lambda}_{g} + \frac{1}{4}\left[ d(\bar{\vS}_{g})_{jj} - \bar{\lambda}_{g}\alpha_g \right]^{2}} +  \frac{1}{2}\left[d(\bar{\vS}_{g})_{jj} - \bar{\lambda}_{g}\alpha_g \right]  \right\}^{-1}.
\end{equation}
Proposition \ref{prop:fusedridge3}.\emph{\ref{prop:fusedridge3item3}} then implies that if $\lambda_{gg'} < \infty$ for all $g'\neq g$, $d\left[\hvOmega_{g}\bigl(\vLambda, \{\vOmega_{g'}\}_{g' {\neq} g} \bigr) \right]_{jj} \rightarrow \alpha_g$ as $\lambda_{gg} \rightarrow \infty^{-}$, for all $j$.
Hence, using scalar target matrices implies shrinking the eigenvalues of the class-specific estimated precision matrix to the central value $\alpha_g$.
One may consider $\vT_g = \alpha_g\vI_p$ with $\alpha_g \in [0, \infty)$ for each $g$.
The rotation equivariance property dictates that it is sensible to choose $\alpha_g$ based on empirical information regarding the eigenvalues of $\vS_g$.
One such choice could be the average of the reciprocals of the non-zero eigenvalues of $\vS_g$.
A straightforward alternative would be to choose $\alpha_g = [\tr(\vS_g)/p]^{-1}$.
In the special case of \eqref{eq:argmax2} where all $\alpha_g = \alpha$ the analogous choice would be $\alpha = [\tr(\vS_\bullet)/p]^{-1}$.
The limited fused ridge problem in \citet{Price2014} corresponds to choosing $\alpha_g = 0$ for all $g$, such that (again) a common target $\vT_g = \vT = \vec{0}$ is employed.

More informative targets would move beyond diagonal targets such as the scalar matrix.
An example would be the consideration of factor-specific targets for factorial designs.
Recalling Example \ref{ex:3}, one might deem the data set factor to be a `nuisance factor'.
Hence, one might choose different targets $\vT_{\GCB}$ and $\vT_{\ABC}$ based on training data or the pooled estimates of the $\GCB$ and $\ABC$ samples, respectively.
In general, the usage of pilot training data or (pathway) database information (or both) allows for the construction of target matrices with higher specificity.
We illustrate how to construct (topology-specific) targets from database information in the DLBCL application of Section \ref{Illustrate.sec}.

\section{Fused Graphical Modeling}
\label{sec:posthoc}
In this section we focus on the graphical interpretation of precision matrices.
First, a simple score test to assess the necessity of fusing is introduced (Section \ref{sec:FuseYN}).
Afterwards, the well-known basics of graphical modeling are given, linking the support of a precision matrix to a conditional independence graph (Section \ref{sec:GMM}).
Next, a simple empirical Bayes procedure for support determination is explained (Section \ref{sec:SelectEdge}).
Last, we introduce several simple metrics for the identification of commonalities and differences between two or more conditional independence graphs (Section \ref{sec:commonAndDifferentialNetworks}).

\subsection{To Fuse or Not to Fuse}
\label{sec:FuseYN}
As a preliminary step to downstream modeling one might consider testing the hypothesis of no class heterogeneity---and therefore the necessity of fusing---amongst the class-specific precision matrices.
Effectively, one then wishes to test the null-hypothesis $H_0 : \vOmega_1 = \ldots =  \vOmega_G$.
Under $H_0 $ an explicit estimator is available in which the fused penalty parameters play no role, cf.\ Section 2.2 of the Supplementary Material.
Here we suggest a score test \citep{BeraBil01} for the evaluation of $H_0$ in conjunction with a way to generate its null distribution in order to assess its observational extremity.

A score test is convenient as it only requires estimation under the null hypothesis, allowing us to exploit the availability of an explicit estimator.
The score statistic equals:
\begin{equation*}
  U = \left.- \sum_{g=1}^G
  \left(\frac{\partial \calL(\{\vOmega_g\}; \{\vS_g\})}{\partial \vOmega_g}\right)^\top
  \left(\frac{\partial^2 \calL(\{\vOmega_g\}; \{\vS_g\})}{\partial \vOmega_g \partial \vOmega_g^\top}\right)^{-1}
  \frac{\partial \calL(\{\vOmega_g\}; \{\vS_g\})}{\partial \vOmega_g}
  \right|_{\vOmega_g = \hvOmega^{H_0}},
\end{equation*}
where $\hvOmega^{H_0}$ denotes the precision estimate under $H_0$ given in equation (S4) of the Supplementary Material, which holds for all classes $g$.
The gradient can be considered in vectorized form and is readily available from \eqref{eq:loglikderiv}.
The Hessian of the log-likelihood equals
$\partial^2 \calL/(\partial \vOmega_g \partial \vOmega_g^\top) = - \vOmega_g^{-1} \otimes \vOmega_g^{-1}$.
For practical purposes of evaluating the score statistic, we employ the identity
  $(\mathbf{A}^\top \otimes \mathbf{B}) \vect(\mathbf{C}) = \vect(\mathbf{B} \mathbf{C} \mathbf{A})$
which avoids the manipulation of $(p^2 \times p^2)$-dimensional matrices.
Hence, the test statistic $U$ is computed by
\begin{equation*}
  \hat{U}
  = \sum_{g=1}^G \vect(\hvX_g)^\top\vect(\hvOmega^{H_0} \hvX_g \hvOmega^{H_0})
  = \sum_{g=1}^G \tr\bigl[ \hvX_g (\hvOmega^{H_0} \hvX_g \hvOmega^{H_0})\bigr],
\end{equation*}
where $\hvX_g = n_g \{2[(\hvOmega^{H_0})^{-1} - \vS_g] - [(\hvOmega^{H_0})^{-1} - \vS_g] \circ \vI_p \}$.

The null distribution of $U$ can be generated by permutation of the class labels: one permutes the class labels, followed by re-estimation of $\vOmega$ under $H_0$ and the re-calculation of the test statistic.
The observed test statistic (under $H_0$) $\hat{U}$ is obtained from the non-permuted class labels and the regular fused estimator.
The $p$-value is readily obtained by comparing the observed test statistic $\hat{U}$ to the null distribution obtained from the test statistic under permuted class labels.
We note that the test is conditional on the choice of $\lambda_{gg}$.

\subsection{Graphical Modeling}
\label{sec:GMM}
A contemporary use for precision matrices is found in the reconstruction and analysis of networks through graphical modeling.
Graphical models merge probability distributions of random vectors with graphs that express the conditional (in)dependencies between the constituent random variables.
In the fusion setting one might think that the class precisions share a (partly) common origin (conditional independence graph) to which fusion appeals.
We focus on class-specific graphs $\calG_g = (V, E_g)$ with a finite set of vertices (or nodes) $V$ and set of edges $E_g$.
The vertices correspond to a collection of random variables and we consider the same set $V = \{Y_{1},\ldots,Y_{p}\}$ of cardinality $p$ for all classes $g$.
That is, we consider the same $p$ variables in all $G$ classes.
The edge set $E_g$ is a collection of pairs of distinct vertices $( Y_j, Y_{j'} )$ that are connected by an undirected edge and this collection may differ between classes.
In case we assume $\{Y_{1},\ldots,Y_{p}\} \sim \calN_p(\vec{0}, \vSigma_g)$ for all classes $g$ we are considering multiple Gaussian graphical models.

Conditional independence between a pair of variables in the Gaussian graphical model corresponds to zero entries in the (class-specific) precision matrix.
Let $\hvOmega_g$ denote a generic estimate of the precision matrix in class $g$.
Then the following relations hold for all pairs $\{Y_{j}, Y_{j'}\} \in \mathcal{V}$ with $j \neq j'$:
\begin{equation*}
  (\hvOmega_g)_{jj'} = \omega_{jj'}^{(g)} = 0
  \quad \Longleftrightarrow \quad
  Y_j \independent Y_{j'} \bigm| V \setminus \bigl\{ Y_j, Y_{j'} \bigr\} ~\mbox{in class}~ g
  \quad \Longleftrightarrow \quad
  ( Y_j, Y_{j'} ) \not\in E_g.
\end{equation*}
Hence, determining the (in)dependence structure of the variables for class $g$---or equivalently the edge set $E_g$ of $\calG_g$---amounts to determining the support of $\hvOmega_g$.

\subsection{Edge Selection}\label{sec:SelectEdge}

We stress that support determination may be skipped entirely as the estimated precision matrices can be interpreted as complete (weighted) graphs.
For more sparse graphical representations we resort to support determination by a local false discovery rate (lFDR) procedure \citep{EfronLocFDR} proposed by \citet{SS05}.
This procedure assumes that the nonredundant off-diagonal entries of the partial correlation matrix
\begin{equation*}
  (\hvP_g)_{jj'}
  = -\hat{\omega}{}_{jj'}^{(g)}
    \left(
      \hat{\omega}{}_{jj}^{(g)}\hat{\omega}{}_{j'j'}^{(g)}
    \right)^{-\frac{1}{2}}
\end{equation*}
follow a mixture distribution representing null and present edges.
The null-distribution is known to be a scaled beta-distribution \citep[cf.][]{SS05a} which allows for estimating the lFDR:
\begin{align*}
  \widehat{\mathrm{lFDR}}{}_{jj'}^{(g)}
  = P\!\Big(
      ( Y_j, Y_{j'}) \not\in E_g
      \;\Big|\;
      (\hvP_g)_{jj'}
    \Big),
\end{align*}
which gives the empirical posterior probability that the edge between $Y_j$ and $Y_{j'}$ is null in class $g$ conditional on the observed corresponding partial correlation.
The analogous probability that an edge is present can be obtained by considering $1 - \widehat{\mathrm{lFDR}}{}_{jj'}^{(g)}$.
See \citet{EfronLocFDR,SS05,VanWieringen2014} for further details on the lFDR procedure.
Our strategy will be to select for each class only those edges for which $1 - \widehat{\mathrm{lFDR}}{}_{jj'}^{(g)}$ surpasses a certain threshold.
\citet{SS05} recommend, on the basis of the observation that the ``majority of the non-null cases lie well within the 0.2 FDR cutoff limits" \citep{Efron2005}, to select an edge to be present when $1 - \widehat{\mathrm{lFDR}}{}_{jj'}^{(g)} > .8$.
We will choose the cut-off for edge-presence somewhat more conservative in our simulations and applications (see Sections \ref{Sims.sec} and \ref{Illustrate.sec}).
The two-step procedure of regularization followed by subsequent support determination has the advantage that it enables probabilistic statements about the inclusion (or exclusion) of edges.

\subsection{Common and Differential (Sub-)Networks}
\label{sec:commonAndDifferentialNetworks}
After estimation and sparsification of the class precision matrices the identification of commonalities and differences between the graphical estimates are of natural interest.
Here we consider some (summary) measures to aid such identifications.
Assume in the following that multiple graphical models have been identified by the sparsified estimates $\hvOmega_1^{0}, \ldots, \hvOmega_G^{0}$ and that the corresponding graphs are denoted by $\calG_1, \ldots, \calG_G$.

An obvious method of comparison is by pairwise graph differences or intersections.
We use the \emph{differential network} $\calG_{g_1\setminus g_2} = (V, E_{g_1} \setminus E_{g_2})$ between class $g_1$ and $g_2$ to provide an overview of edges present in one class but not the other.
The \emph{common network} $\calG_{1\cap 2} = (V, E_1 \cap E_2)$ is composed of the edges present  in both graphs.
We also define the \emph{edge-weighted total network} of $m \leq G$ graphs $\calG_1, \ldots, \calG_m$ as the graph formed by the union $\calG_{1 \cup \cdots \cup m} = (V, E_1 \cup \cdots \cup E_m)$ where the weight $w_{jj'}$ of the edge $e_{jj'}$ is given by the cardinality of the set
$\{g \in \{1, \ldots, m\} : e_{jj'} \in E_g\}$.
More simply, $\calG_{1 \cup \cdots  \cup m}$ is determined by summing the adjacency matrices of $\calG_1$ to $\calG_m$.
Analogously, the \emph{signed edge-weighted total network} takes into account the stability of the sign of an edge over the classes by summing signed adjacency matrices.
Naturally, the classes can also be compared by one or more summary statistics at node-, edge-, and network-level per class \citep[cf.][]{Newman10}.

We also propose the idea of `network rewiring'.
Suppose an investigator is interested in the specific interaction between genes $A$ and $B$ for classes $g_1$ and $g_2$.
The desire is to characterize the dependency between genes $A$ and $B$ and determine the differences between the two classes.
To do so, we suggest using the decomposition of the covariance of $A$ and $B$ into the individual contributions of all paths between $A$ and $B$.
A path $z$ between $A$ and $B$ of length $t_z$ in a graph for class $g$ is, following \citet{Lauritz96}, defined to be a sequence $A = v_0, \ldots, v_{t_{z}} = B$ of distinct vertices such that $(v_{d-1}, v_d) \in E_g$ for all $d = 1, \ldots, t_z$.
The possibility of the mentioned decomposition was shown by \citet{Jones2005} and, in terms of $\hvOmega_g^{0} = [\omega_{jj'}]$, can be stated as:
\begin{equation}
  \label{eq:covdecomp}
  \Cov(A, B)
  = \sum_{z \in \mathcal{Z}_{AB}}
  (-1)^{t_{z}+1}
  \omega_{Av_1}\omega_{v_1v_2}\omega_{v_2v_3} \cdots
    \omega_{v_{t_{z}-2}v_{t_{z}-1}}\omega_{v_{t_{z}-1}B}
    \frac{\deter{(\hvOmega_g^{0})_{\neg P}}}{\deter{\hvOmega_g^{0}}},
\end{equation}
where $\mathcal{Z}_{AB}$ is the set of all paths between $A$ and $B$ and $(\hvOmega_g^{0})_{\neg P}$ denotes the matrix $\hvOmega_g^{0}$ with rows and columns corresponding to the vertices of the path $z$ removed.
Each \emph{term} of the covariance decomposition in \eqref{eq:covdecomp} can be interpreted as the flow of information through a given path $z$ between $A$ and $B$ in $\calG_g$.
Imagine performing this decomposition for $A$ and $B$ in both $\hvOmega_{g_1}^{0}$ and $\hvOmega_{g_2}^{0}$.
For each path, we can then identify whether it runs through the common network $\calG_{g_1\cap g_2}$, or uses the differential networks $\calG_{g_2\setminus g_1}, \calG_{g_1\setminus g_2}$ unique to the classes.
The paths that pass through the differential networks can be thought of as a `rewiring' between the groups (in particular compared to the common network).
In summary, the covariance between a node pair can be separated into a component that is common and a component that is differential (or rewired).

\begin{example}
Suppose we have the following two graphs for classes $g_1 = 1$ and $g_2 = 2$:
\begin{equation*}
\begin{tikzpicture}[node distance = 2mm, auto,
  main_node/.style={circle,draw,minimum size=1.2em,inner sep=0pt]
  }]

    \node (G1) at (-1.6, 0.5) {$\calG_1 =$};
    \node[main_node] (n1) at (-0.866,0.5) {$A$};
    \node[main_node] (n2) at (0,1) {$B$};
    \node[main_node] (n3) at (1,1) {$3$};
    \node[main_node] (n4) at (1,0) {$4$};
    \node[main_node] (n5) at (0,0) {$5$};

    \draw (n1) -- (n2);
    \draw (n1) -- (n5) -- (n2);
    \draw (n1) -- (n5) -- (n4) -- (n2);
    \draw (n5) -- (n3);
\end{tikzpicture}
\qquad
\begin{tikzpicture}[node distance = 2mm, auto,
  main_node/.style={circle,draw,minimum size=1.2em,inner sep=0pt]
  }]

    \node (G2) at (-1.6, 0.5) {$\calG_2 =$};
    \node[main_node] (n1) at (-0.866,0.5) {$A$};
    \node[main_node] (n2) at (0,1) {$B$};
    \node[main_node] (n3) at (1,1) {$3$};
    \node[main_node] (n4) at (1,0) {$4$};
    \node[main_node] (n5) at (0,0) {$5$};

    \draw (n1) -- (n5) -- (n2);
    \draw (n1) -- (n5) -- (n4) -- (n3) -- (n2);
\end{tikzpicture}
\end{equation*}
\noindent and consider the covariance between node $A$ and $B$.
In $\calG_1$ the covariance $\Cov(Y_A, Y_B)$ is decomposed into contributions by the paths $(A,B)$, $(A,5,B)$, and $(A,5,4,B)$.
Similarly for $\calG_2$, the contributions are from paths $(A,5,B)$ and $(A,5,4,3,B)$.
Thus $(A,5,B)$ is the only shared path.
Depending on the size of the contributions we might conclude that network 1 has some `rewired pathways' compared to the other.
This method gives a concise overview of the estimated interactions between two given genes, which genes mediate or moderate these interactions, as well as how the interaction patterns differ across the classes.
In turn this might suggest candidate genes for perturbation or knock-down experiments.
\end{example}


\section{Simulation Study}\label{Sims.sec}
In this section we explore and measure the performance of the fused estimator and its behavior in four different scenarios.
Performance is measured primarily by the squared Frobenius loss,
\begin{equation*}
  L_F^{(g)}\bigl(\hvOmega_g(\vLambda), \vOmega_g\bigr)
  = \sqfnorm[\big]{ \hvOmega_g(\vLambda) - \vOmega_g },
\end{equation*}
between the class precision estimate and the true population class precision matrix.
However, the performance is also assessed in terms of the quadratic loss,
\begin{equation*}
  L_Q^{(g)}\bigl(\hvOmega_g(\vLambda), \vOmega_g\bigr)
  = \sqfnorm[\big]{ \hvOmega_g(\vLambda)\vOmega_g^{-1} - \vI_p }.
\end{equation*}
The risk defined as the expected loss associated with an estimator, say,
\begin{equation*}
  \mathcal{R}_F\bigl\{ \hvOmega_g(\vLambda) \bigr\}
   = \bbE\Bigl[ L_F^{(g)}\bigl(\hvOmega_g(\vLambda), \vOmega_g\bigr) \Bigr],
\end{equation*}
is robustly approximated by the median loss over a repeated number of simulations and corresponding estimations.

We designed six simulation scenarios to explore the properties and performance of the fused ridge estimator and alternatives.
  Scenario 1 evaluates the fused ridge estimator under two choices of the penalty matrix,
  the non-fused ridge estimate applied individually to the classes, and
  the non-fused ridge estimate using the pooled covariance matrix when
    (1a)~$\vOmega_1 = \vOmega_2$ and
    (1b)~$\vOmega_1 \neq \vOmega_2$.
  Scenario 2 evaluates the fused ridge estimator under different choices of targets:
    $\vT_1 = \vT_2 = \vec{0}$,
    $\vT_1 = \vT_2 = \alpha\vI_p$ with different choices of $\alpha$, and
    $\vT_1 = \vT_2 = \vOmega$.
  Scenario 3 evaluates the fused ridge estimator for varying network topologies and degrees of class homogeneity.
  Specifically, for
    (3a)~scale-free topology and
    (3b)~small-world topology,
  each with
    (3i)~low class homogeneity and
    (3ii)~high class homogeneity.
  Scenario 4 investigates the fused estimator under non-equal class sample sizes.
  Scenario 5 compares the fused ridge estimator to the fused graphical lasso \citep{Danaher2013} estimator.
  Scenario 6 compares the fused ridge estimator to the Laplacian Shrinkage for Inverse Covariance matrices from Heterogenous populations \citep[LASICH;][]{SS16} estimator and a Bayesian Multiple Gaussian Graphical Modeling \citep[BMGGM;][]{PSV2015} approach.
Except for scenario 4, we make no distinction between the loss in different classes.
Except for scenario 1, we use penalty matrices of the form $\vLambda = \lambda\vI_G + \lambda_f(\vJ_G-\vI_G)$.

\subsection{Scenario 1: Fusion Versus no Fusion}
Scenario 1 explores the loss-efficiency of the fused estimate versus non-fused estimates as a function of the class sample size $n_g$ for fixed $p$ and hence for different $p/n_\summed$ ratios.
Banded population precision matrices are simulated from $G = 2$ classes.
We set $p = 100$ and
\begin{equation}
  \label{eq:banded}
  (\vOmega_g)_{jj'}
  =  \frac{k + 1}{\abs{j - j'} + 1} \bbOne\bigl[\abs{j - j'} \leq k\bigr]
\end{equation}
with $k$ non-zero off-diagonal bands.
The sub-scenario
(1a) $\vOmega_1 = \vOmega_2$ uses $k = 15$ bands whereas
(1b) $\vOmega_1 \neq \vOmega_2$ uses $k = 15$ bands for $\vOmega_1$ and $k = 2$ bands for $\vOmega_2$.
Hence, identical and very different population precision matrices are considered, respectively.

\begin{figure}[p]
  \centering
  \includegraphics[width=\textwidth]{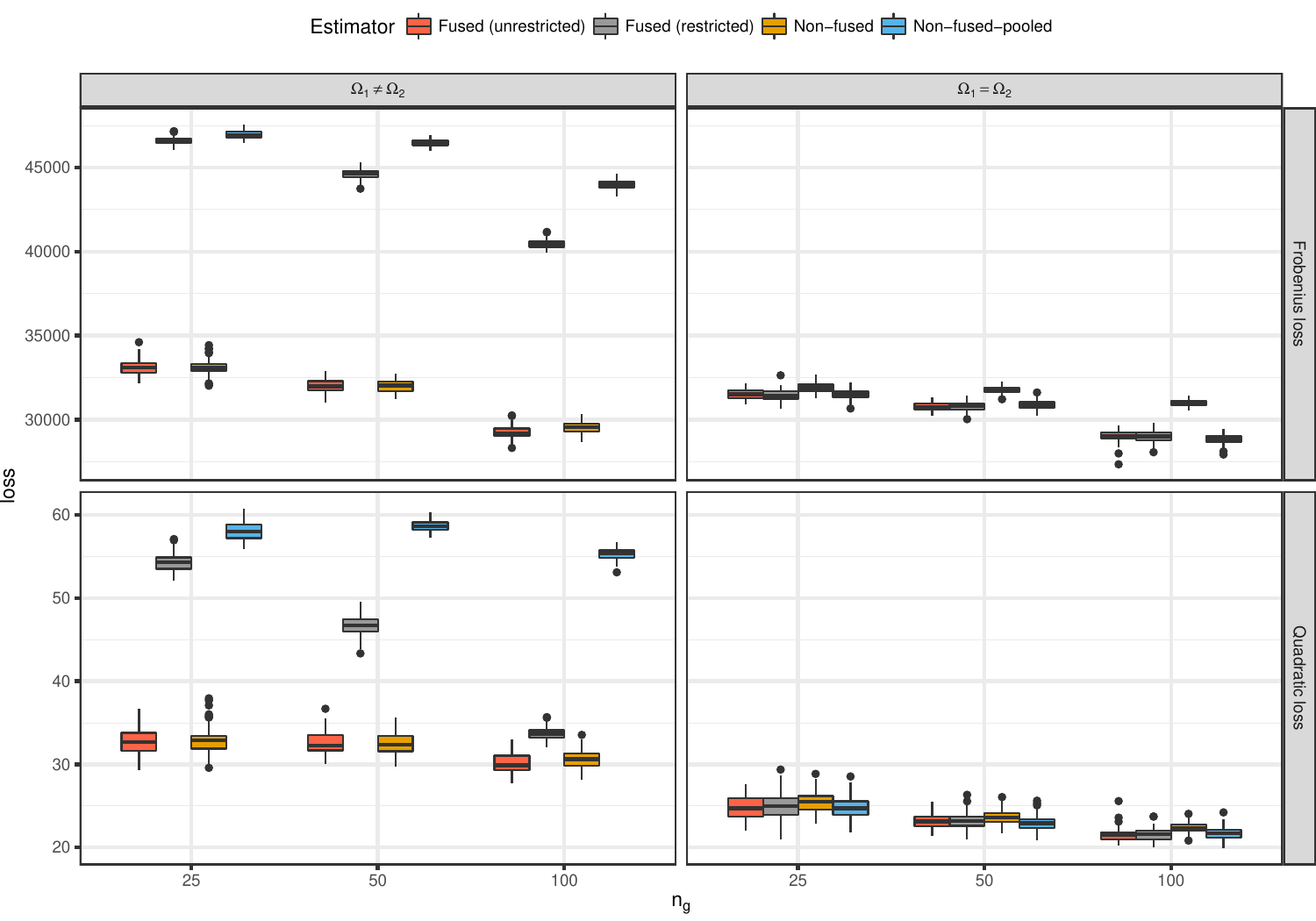}
  \caption{
  Results for simulation Scenario 1, depicting the losses against the class samples size for different ridge estimators under unequal and equal class population matrices.
  $G = 2$ classes are considered with banded population precision matrices of variable-dimension $p = 100$.
  The left-hand panels represent the $\vOmega_1 \neq \vOmega_2$ scenario.
  The right-hand panels represent the $\vOmega_1 = \vOmega_2$ scenario.
  The upper panels depict the results under the Frobenius loss.
  The lower panels depict the results under the quadratic loss.
  The considered class sample sizes are $n_g \in \{25,50,100\}$ and the losses were computed for the fused ridge estimator with an unrestricted penalty matrix, the fused ridge estimator with a restricted penalty matrix such that the ridge penalty is shared across classes, the regular non-fused ridge estimator applied separately to each class, and the regular non-fused ridge estimator using the pooled estimate $\vS_\summed$.
  In all cases $\vT_1 = \vT_2 = \alpha_{\summed 2}\vI_p$ with $\alpha_{\summed 2} = p/\tr(\vS_\summed)$, i.e., $\alpha_{\summed2}$ represents the inverse of the averaged eigenvalues of $\mathbf{S}_{\summed}$.
  Note that the boxplots in the figure (for each class sample size $n_g$) are ordered according to the legend (given at the top of the image).
  }
  \label{fig:plot_fig1}
\end{figure}

For $n_g = 25, 50, 100$ the loss over $100$ repeated runs was computed.
In each run, the optimal \emph{unrestricted} penalty matrix $\vLambda$ was determined by LOOCV.
The losses were computed for
(1i) the fused ridge estimator with an unrestricted penalty matrix,
(1ii) the fused ridge estimator with a restricted penalty matrix such that $\lambda_{11} = \lambda_{22}$,
(1iii) the regular non-fused ridge estimator applied separately to each class, and
(1iv) the regular non-fused ridge estimator using the pooled estimate $\vS_\summed$.
In all cases the targets $\vT_1 = \vT_2 = \alpha_{\summed 2}\vI_p$ were used with $\alpha_{\summed 2} = p/\tr(\vS_\summed)$.
The risk and quartile losses for scenario 1 are seen in the boxplots of Figure \ref{fig:plot_fig1}.

Generally, the \emph{unrestricted} fused estimates are found to perform at least as well as the (superior of the) \emph{non-fused} estimates.
This can be expected as the fused ridge estimate might be regarded as an interpolation between using the non-fused ridge estimator on the pooled data and within each class separately.
Hence, the LOOCV procedure is thus able to capture and select the appropriate penalties both when the underlying population matrices are very similar and when they are very dissimilar.
In the case of differing class population precision matrices, the \emph{restricted} fused ridge estimator (that uses the single ridge penalty $\lambda_{11} = \lambda_{22}$) performs somewhat intermediately, indicating again the added value of the flexible penalty setup.
It is unsurprising that the non-fused estimate using the pooled covariance matrix is superior in scenario (1b), where $\vOmega_1 = \vOmega_2$, as it is the explicit estimator in this scenario, cf.\ Section~2.2 of the Supplementary Material.


\subsection{Scenario 2: Target Versus no Target}
Scenario 2 investigates the added value of the targeted approach to fused precision matrix estimation compared to that of setting $\vT_g = \vec{0}$ which reduces to the special-case considered by \citet{Price2014}.
We simulated data sets with $G = 2$ classes and $p = 50$ variables from three topologies:
(2i) banded precision matrices (as given in Equation \ref{eq:banded}) with $k = 25$ bands;
(2ii) precision matrices representing star-graphs, and
(2iii) precision matrices based on Erd\"{o}s-R\'{e}nyi random graph games \citep{RandomGraph}.
For topology (2ii) the first variable represents the internal (hub) node and the values of the off-diagonal entries $(1, j)$ and $(j, 1)$ taper-off by $1/(j + 1)$.
For (2iii) each edge is present with probability $1/p$ and non-zero off-diagonal values are taken to be $.25$.
Performance was evaluated using
(2a)~$\vT_1 = \vT_2 = \vec{0}$,
(2b)~$\vT_g = \alpha_\summed\vI_p$,
(2c)~$\vT_g = \alpha_{\summed 2}\vI_p$, and
(2d)~the spot-on target $\vT_1 = \vT_2 = \vOmega$.
We set $\alpha_{\summed} = [\sum_j(\mathbf{S}_{\summed})_{jj}^{-1}]/p$ and $\alpha_{\summed 2}$ is defined as above.
Risks were estimated by the losses for each class for each of $n_g = 25, 50, 100$ class sample sizes over $100$ simulation repetitions.
The optimal penalties where determined by LOOCV with penalty matrices of the form $\vLambda = \lambda\vI_G + \lambda_f(\vJ_G-\vI_G)$.

\begin{figure}[h!]
  \centering
  \includegraphics[width=\textwidth]{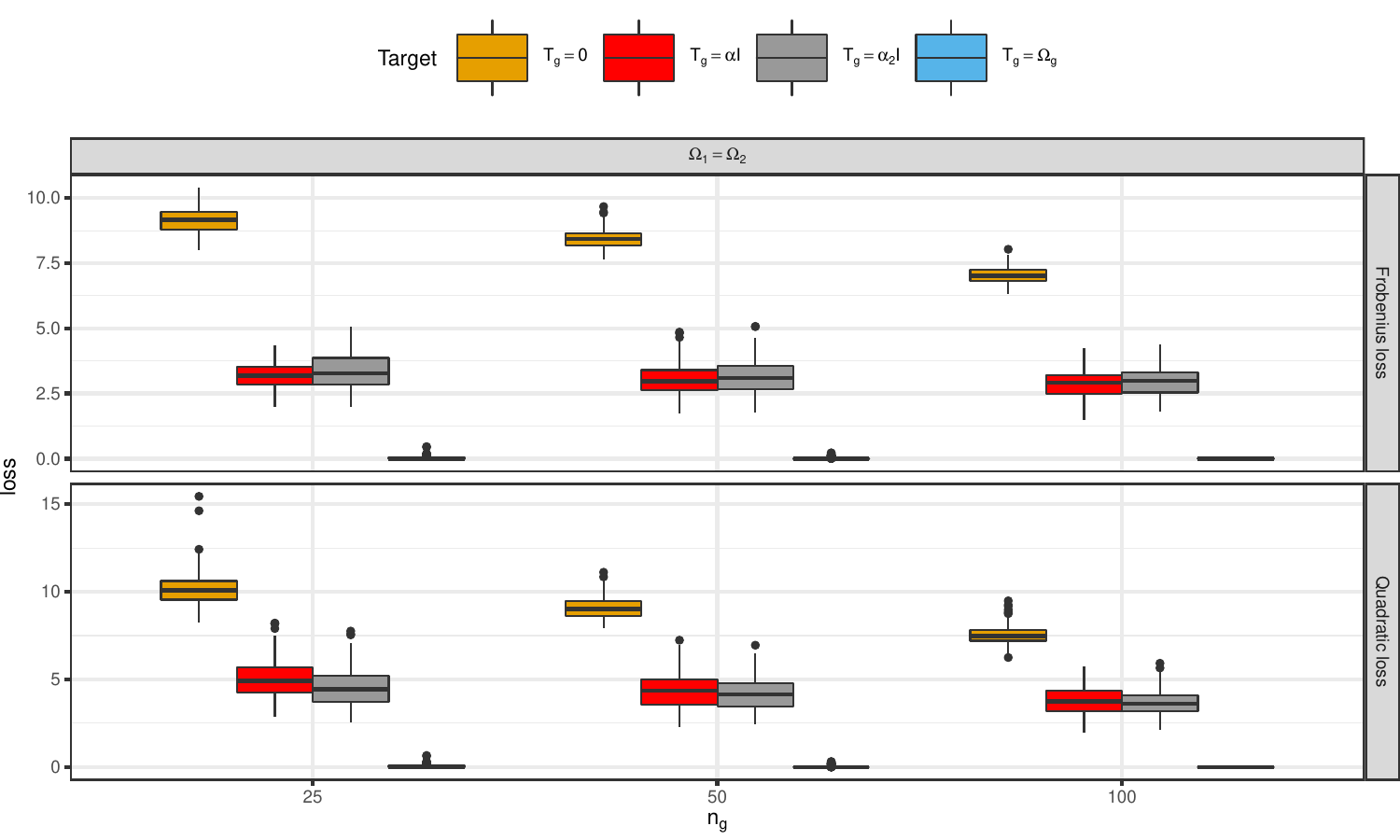}
  \caption{
  Results for simulation Scenario 2iii, depicting the comparison of the targeted versus the un-targeted approach in the random-graph population setting.
    We consider $G = 2$ classes with the population precision matrix $\vOmega$ for each class being a Erd\"{o}s-R\'{e}nyi random graph matrix with $p = 50$.
    Each edge is present with probability $1/p$.
    Non-zero off-diagonal values are taken to be $.25$.
    The upper panel depicts the results under the Frobenius loss while the lower panel depicts the results under the quadratic loss.
    The considered class sample sizes are $n_g \in \{25,50,100\}$.
    The target matrix is taken to be equal over classes, i.e., $\mathbf{T}_1 = \mathbf{T}_2$.
    The un-targeted situation is represented by $\mathbf{T}_g = \boldsymbol{0}$.
    The most informative target is the spot-on target $\vT_g = \vOmega$.
    Two diagonal targets are also considered: $\vT_g = \alpha_{\summed}\vI_p$, with $\alpha_{\summed} = [\sum_j(\mathbf{S}_{\summed})_{jj}^{-1}]/p$; and $\vT_g = \alpha_{\summed2}\vI_p$, with $\alpha_{\summed2} = p/\mathrm{tr}(\mathbf{S}_{\summed})$.
    Hence, $\alpha_{\summed}$ represents the average of the inverse marginal variances of $\mathbf{S}_{\summed}$ and $\alpha_{\summed2}$ represents the inverse of the averaged eigenvalues of $\mathbf{S}_{\summed}$.
    Note that the boxplots in the figure (for each class sample size $n_g$) are ordered according to the legend (given at the top of the image).
  }
  \label{fig:plot_fig2}
\end{figure}

The results for the random-graph topology are shown in the boxplots in Figure~\ref{fig:plot_fig2}.
The results for the star-graph and banded matrix topologies can be found in Section~4 of the Supplementary Material.
As expected, the spot-on target shows superior performance in terms of loss in all cases.
Diagonal targets also improve estimation efficiency relative to the null target.
This latter observation holds for all considered topologies and both types of diagonal target, across the considered sample sizes and loss types.
Only in scenario (2i) under the Frobenius loss is the null target preferred over the diagonal targets.
Perhaps this is not surprising:
For the Frobenius norm the slowest rate of convergence of the estimator comes from the diagonal entries \citep{Rothman12,Maurya}.
From the losses as defined above we get that, in a sense, the Frobenius norm emphasizes proportionality, while the quadratic norm emphasizes the diagonal.
The situation in scenario (2i) is actually quite dense: A banded matrix with 25 bands.
As the Frobenius loss emphasizes proportionality and is slow to converge in terms of diagonal entries it will then favor $\mathbf{T} = \boldsymbol{0}$.
Because when emphasizing proportionality, the $\mathbf{T} = \boldsymbol{0}$ target will keep the estimate longer in a state that resembles a matrix with many bands.

Hence, we conclude that, in general, informative targets are preferred over null targets, even when the informative target is as simple as a scalar matrix (given that the scalar is, in a sense, well-chosen).
Overall, the results suggests that well-informed choices of the target can greatly improve the estimation and that the algorithm will put emphasis on the target if it reflects the truth.
Such behavior is also seen analytically in the ridge estimator of \citet{SS05} inferred from their closed expression of the optimal penalty.
Such behavior also corresponds to the observation that positive definite target matrices will tend to preserve data signal \citep{VanWieringen2014}.

As the null-target scenario corresponds to the case of \citet{Price2014}, we performed a secondary timing benchmark of their accompanying \texttt{RidgeFusion} package compared to \texttt{rags2ridges}.
We evaluated estimation time of each package on a single simulated data set with
$p = 50$,
$G = 2$, and
$n_1 = n_2 = 10$ using a banded matrix as before.
The average estimation times over 100 model fits where
9.3 and 25.4
milliseconds
for packages
\texttt{rags2ridges} and
\texttt{RidgeFusion}, respectively.
This approximates a factor
2.74
speed-up for a single model fit.
The timing was done using the package \texttt{microbenchmark} \citep{Mersmann2014} and the estimates from each package were in agreement within expected numerical precision.

\subsection{Scenario 3: Varying Topology and Class (Dis)Similarity}
Scenario 3 investigates the fused estimator with $G = 3$ classes for (3i) high and (3ii) low class homogeneity and two different latent random graph topologies on $p = 100$ variables.
The topologies are the (3a) `small-world' and the (3b) `scale-free' topology generated by Watts-Strogatz and Barab\'{a}si graph games, respectively \citep{Watts1998,Barabasi1999}.
The former generates topologies where all node degrees are similar while the latter game generates networks with (few) highly connected hubs.
From the generated topology, we construct a latent precision matrix $\vPsi$ with diagonal elements set to 1 and the non-zero off-diagonal entries dictated by the network topology set to $0.1$.

The two topologies are motivated as they imitate many real phenomena and processes.
Small-world topologies approximate systems such as power grids, the neural network of the worm C.~elegans, and the social networks of film actors \citep{Watts1998, Mei2011}.
Conversely, scale-free topologies approximate many social networks, protein-protein interaction networks, airline networks, the world wide web, and the internet \citep{Barabasi1999, Barabasi2009}.

We control the inter-class homogeneity using a latent inverse Wishart distribution for each class covariance matrix as considered by \citet{Bilgrau2015b}.
That is, we let
\begin{align}
  \label{eq:invwishart}
  \vSigma_g = \vOmega_g^{-1} \sim \mathcal{W}_p^{-1}\Big((\nu - p - 1)\vPhi^{-1}, \nu\Big),
  \quad \nu > p + 1
\end{align}
where $\mathcal{W}_p^{-1}(\vTheta, \nu)$ denotes an inverse Wishart distribution with scale matrix $\vTheta$ and $\nu$ degrees of freedom.
The parametrization implies the expected value $\bbE[\vSigma_g] = \bbE[\vOmega_g^{-1}] = \vPhi^{-1}$ and thus $\vPhi$ defines the latent expected topology.
We simulate from a multivariate normal distribution as before conditional on the realized covariance $\vSigma_g$.
In \eqref{eq:invwishart}, the parameter $\nu$ controls the inter-class homogeneity.
Large $\nu$ imply that $\vOmega_1 \approx \vOmega_2 \approx \vOmega_3$ and thus a large class homogeneity.
Small values of $\nu \to (p + 1)^+$ imply large heterogeneity.

\begin{figure}[h]
  \centering
  \includegraphics[width=\textwidth]{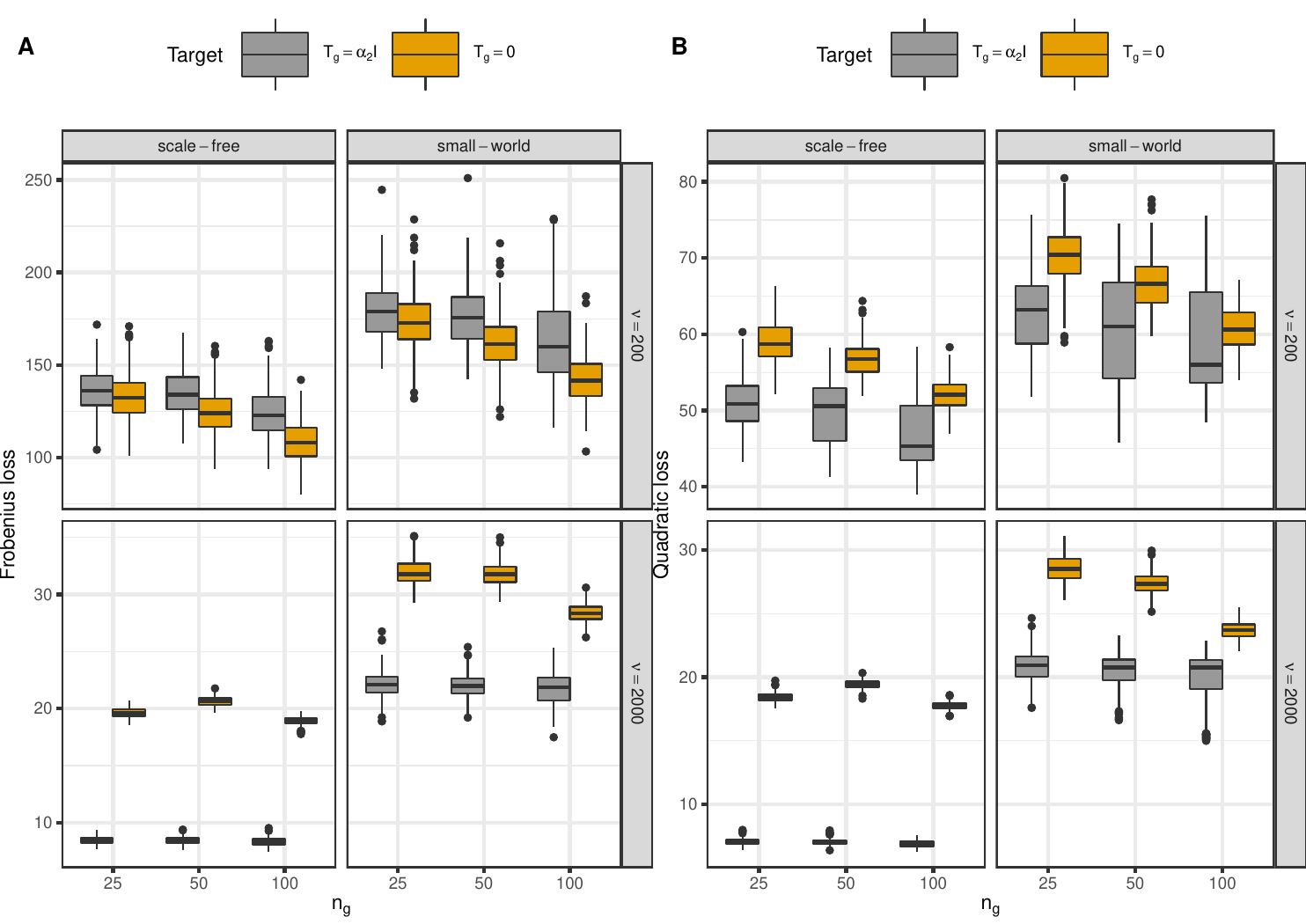}
  \caption{Results for simulation Scenario 3.
  Panel A depicts the boxplots of \emph{Frobenius} losses for each combination of network topology, degree of class similarity, choice of target, and class sample-size.
    Panel B depicts the boxplots of \emph{quadratic} losses for each combination of network topology, degree of class similarity, choice of target, and class sample-size.
  Note that the boxplots in the figure (for each class sample size $n_g$) are ordered according to the legend (given at the top of the image).}
  \label{fig:plot_sim3}
\end{figure}

For the simulations, we chose (i) $\nu = 200$ and (ii) $\nu = 2000$.
Again we fitted the model using both the zero target as well as the scalar matrix target described above using the reciprocal value of the mean eigenvalue, i.e., $\vT_1 = \vT_2 = \vT_3 = \alpha \vI_p$ for both $\alpha = 0$ and $\alpha = \alpha_{\summed 2} = p/\tr(\vS_\summed)$.
The estimation was repeated 100 times for each combination of high/low class similarity, network topology, choice of target, and class sample-size $n_1 = n_2 = n_3 = 25, 50, 100$.
Panels A and B of Figure \ref{fig:plot_sim3} show box-plots of the results.

First, the loss is seen to be dependent on the network topology, irrespective of the loss function.
Second, as expected, the loss is strongly influenced by the degree of class (dis)similarity where a higher homogeneity yields a lower loss.
Intuitively, this makes sense as the estimator can borrow strength across the classes and effectively increase the degrees of freedom in each class.
Third, the targeted approach has a superior loss in all cases with a high class homogeneity and thus the gain in loss-efficiency is greater for the targeted approach.
For low class homogeneity, the targeted approach performs comparatively to the zero target with respect to the Frobenius loss while it is seemingly better in terms of quadratic loss.
Measured by quadratic loss, the targeted approach nearly always outperforms the zero target.

\subsection{Scenario 4: Unequal Class Sizes}
\begin{figure}[h]
  \centering
  \includegraphics[width=\textwidth]{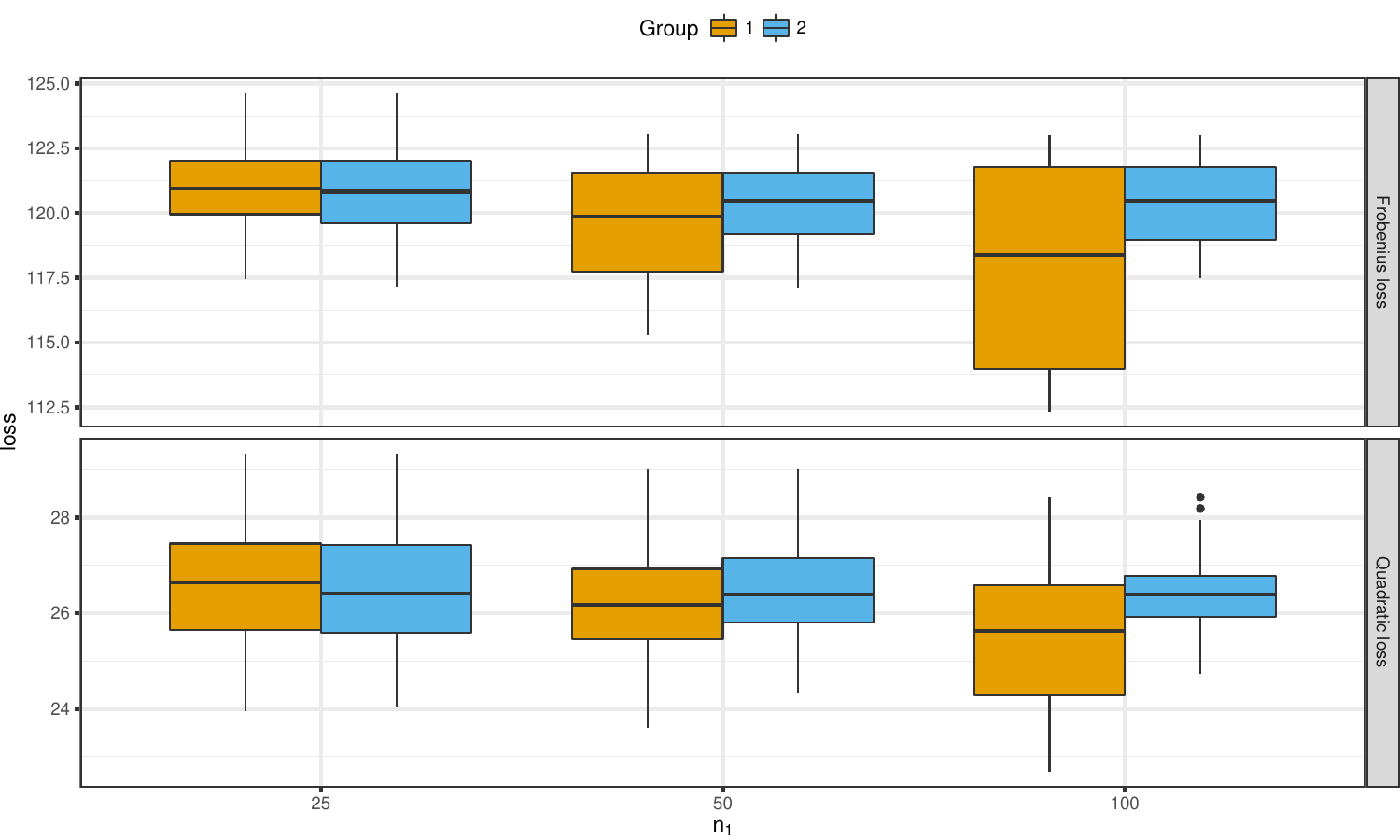}
  \caption{
  Results for simulation Scenario 4: Depicting the loss as a function of sample size of class 1 with fixed sample size for class 2.
  The upper panel depicts the results under the Frobenius loss while the lower panel depicts the results under the quadratic loss.}
  \label{fig:plot_sim4}
\end{figure}

Scenario 4 explores the fused estimator under unequal class sample sizes.
We simulated data from banded precision matrices with $k = 8$ non-zero off-diagonal bands, $G = 2$, and $p = 100$.
The number of samples in class 2 was fixed at $n_2 = 30$ while the number of samples in class 1 were varied: $n_1 = 25, 50, 100$.
The target matrices are specified such that $\vT_1 = \vT_2 = \alpha_{\summed 2}\vI_p$.
The results of the simulation are shown in Figure \ref{fig:plot_sim4}.
Note that we consider the Frobenius and quadratic loss within each class separately here.

Not surprisingly, the fused estimator performs better (for both classes) when $n_\summed$ increases.
Perhaps more surprising: there seems to be no substantial difference in loss for groups $n_1$ and $n_2$, suggesting that the fusion indeed borrows strength from the larger class.
A loss difference is only visible in the most extreme case where $n_1 = 100$ and $n_2 = 30$.
The relative difference however is not considered large.

\subsection{Scenario 5: Comparison to the Fused Graphical Lasso}
Scenario 5 compares the targeted fused ridge estimator with the fused graphical lasso estimator \citep{Danaher2013}.
We consider $G = 2$ classes with (initially) $\vOmega_1 = \vOmega_2$.
We then simulated data sets with $p = 50$ variables from two topologies: (i) random topology generated by the Erd\"{o}s-R\'{e}nyi random graph game \citep{RandomGraph}, and (ii) scale-free topology generated by the Barab\'{a}si graph game \citep{Barabasi1999}.
In this simulation the dimension $p$ is chosen to be $50$ in order to keep computation times appreciable (the lasso can be slow in dense situations).
For each topology, the density (parameter) is varied.
For the Erd\"{o}s-R\'{e}nyi random graph game we consider edge presence with probability $P \in \{1/p, .25, .35\}$, indicating increasingly dense topologies.
For the Barab\'{a}si graph game we consider linear preferential attachment and the number of edges to add in each time step $\# E \in \{ 1,3,5 \}$.
In each time-step of the Barab\'{a}si graph game algorithm (Barab\'{a}si \& Albert, 1999), $\# E$ edges are added.
Hence, higher values of $\# E$ result in more dense topologies.
Under both considered topologies the off-diagonal nonzero elements are chosen to be of value $.15$.
The fused graphical lasso is initiated such that the diagonal elements (for each class) are preserved.
For the fuse ridge we choose $\vT_g = \alpha_{g}\vI_p$, with $\alpha_g = p/\tr(\vS_g)$.
Hence, the target employed by the fused lasso is most likely advantageous with respect to loss.
For each setting we consider a 2-dimensional grid of ridge and fusion penalties.

For the fused ridge we consider the ridge-penalty $\lambda \in [.01, 1000]$ and the fusion-penalty $\lambda_f \in [1, 10,000]$.
For the fused graphical lasso we consider (abusing notation somewhat for notational brevity) the lasso-penalty $\lambda \in [.01, 100]$ and the fusion-penalty $\lambda_f \in [.1, 100]$.
The penalty-grids are probed by taking 30 $\log_{10}$-equidistant steps in each direction.
Risks are then estimated---for each $(\lambda,\lambda_f)$-combination nested within each combination of topology and corresponding density-parameter---by the median losses aggregated over the classes for each of $n_g = 25, 50$ class sample sizes over 100 simulation repetitions.
Hence, we obtain risk surfaces over the penalty-grid.

Figure \ref{FIG:riskLassoRidgeBGn25-1}, and Figures S3, and S4 (Section 5 of the Supplementary Material) visualize the results for the Barab\'{a}si graph game with $n_g = 25$ and with $\# E = 1$, $\# E = 3$, and $\# E = 5$, respectively.
These figures then give the Risk per $(\lambda,\lambda_f)$-combination.
The blue box in each figure indicates the $(\lambda,\lambda_f)$-combination that achieves the lowest Risk.
We make several observations on the basis of these figures.
The first is that the risk surface of the fused ridge estimator is smoother than the analogous surface of the fused graphical lasso.
This is to be expected as the ridge estimator provides proportional shrinkage.
Second, as the density of the topology increases, the ridge-penalty for which the lowest Risk is achieved expectedly decreases.
For very sparse situations, the ridge-penalty is large as it will tend to suppress signal to express sparsity.
Third, the fused-ridge-penalty (for which the lowest Risk is achieved) indeed expresses that the class-precision matrices stem from the same population.
Last, irrespective of the sparsity of the setting, we are able to find combinations of penalty-values that lead the fused ridge estimator to achieve lower Risk than the fused graphical lasso estimator.
This last observation is especially of note since we move through the penalty-space of the fused ridge in a more coarse-grained manner, which is advantageous to the fused graphical lasso.
Moreover, this last observation also holds irrespective of the chosen loss-type (Frobenius or quadratic).
Similar behavior is seen under $n_g = 50$ (Supplementary Figures S5--S7) and in the Erd\"{o}s-R\'{e}nyi random graph game setting (Supplementary Figures S8--S13).
These results are in line with observations made by \citet{VanWieringen2014} in the non-fused situation.

\begin{figure}[h!]
\centering
  \includegraphics[width=.85\textwidth]{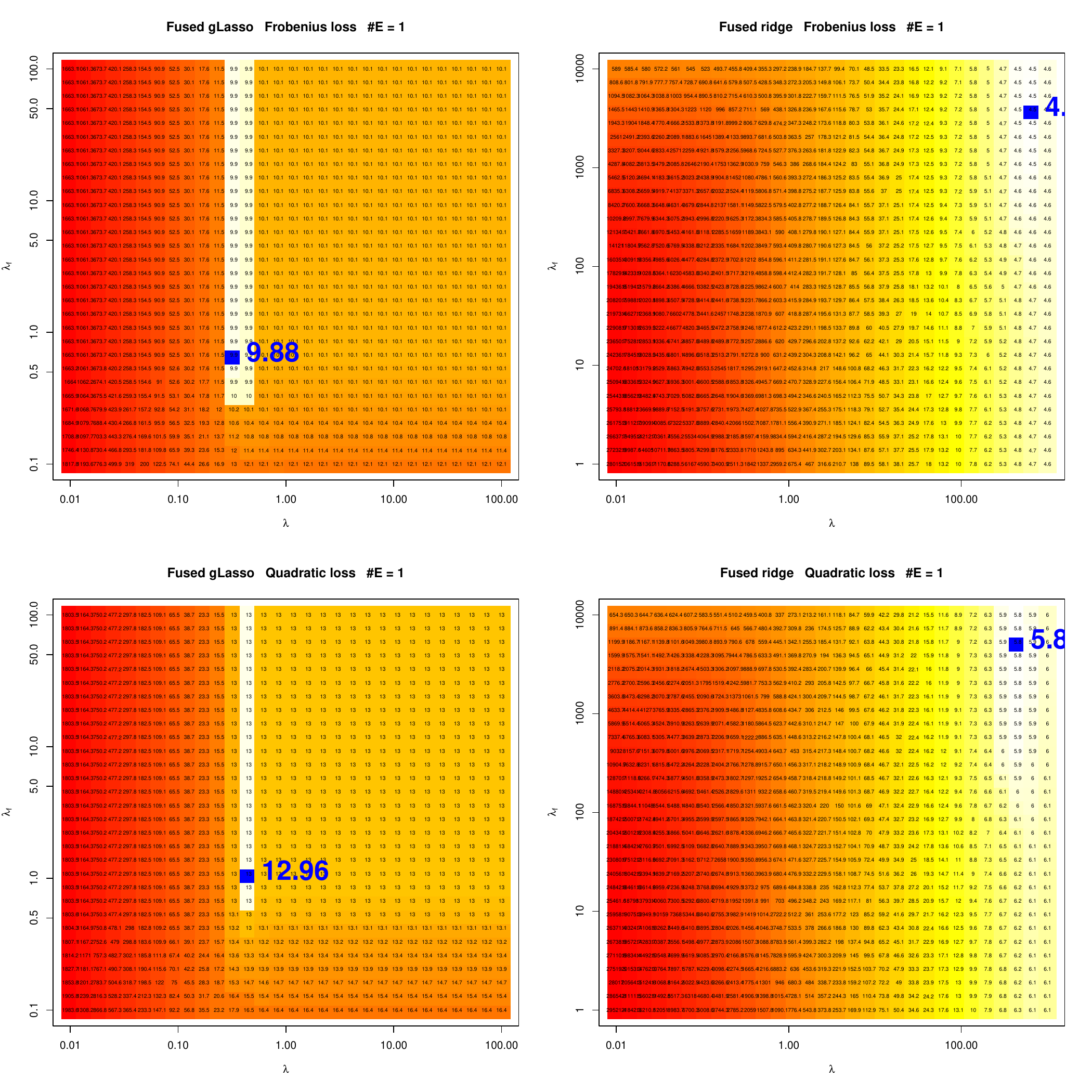}
    \caption{
    Comparison of the fused graphical lasso and the fused ridge estimator in the Barab\'{a}si graph game population setting with $n_g = 25$ and where the number of edges to add in each time step was taken to be $1$.
    Each square on the two-dimensional grid represents a $(\lambda,\lambda_f)$-combination.
    The number in each square represents the estimated Risk for the corresponding combination.
    The blue square (and corresponding number) indicate the lowest Risk achieved on the grid.
    Left-hand panels give the results for the fused graphical lasso.
    Right-hand panels give the results for the fused ridge estimator.
    Upper panels express the Risk surface under Frobenius loss.
    Lower panels express the Risk surface under quadratic loss.}
  \label{FIG:riskLassoRidgeBGn25-1}
\end{figure}

\begin{figure}[h!]
\centering
  \includegraphics[width=.85\textwidth]{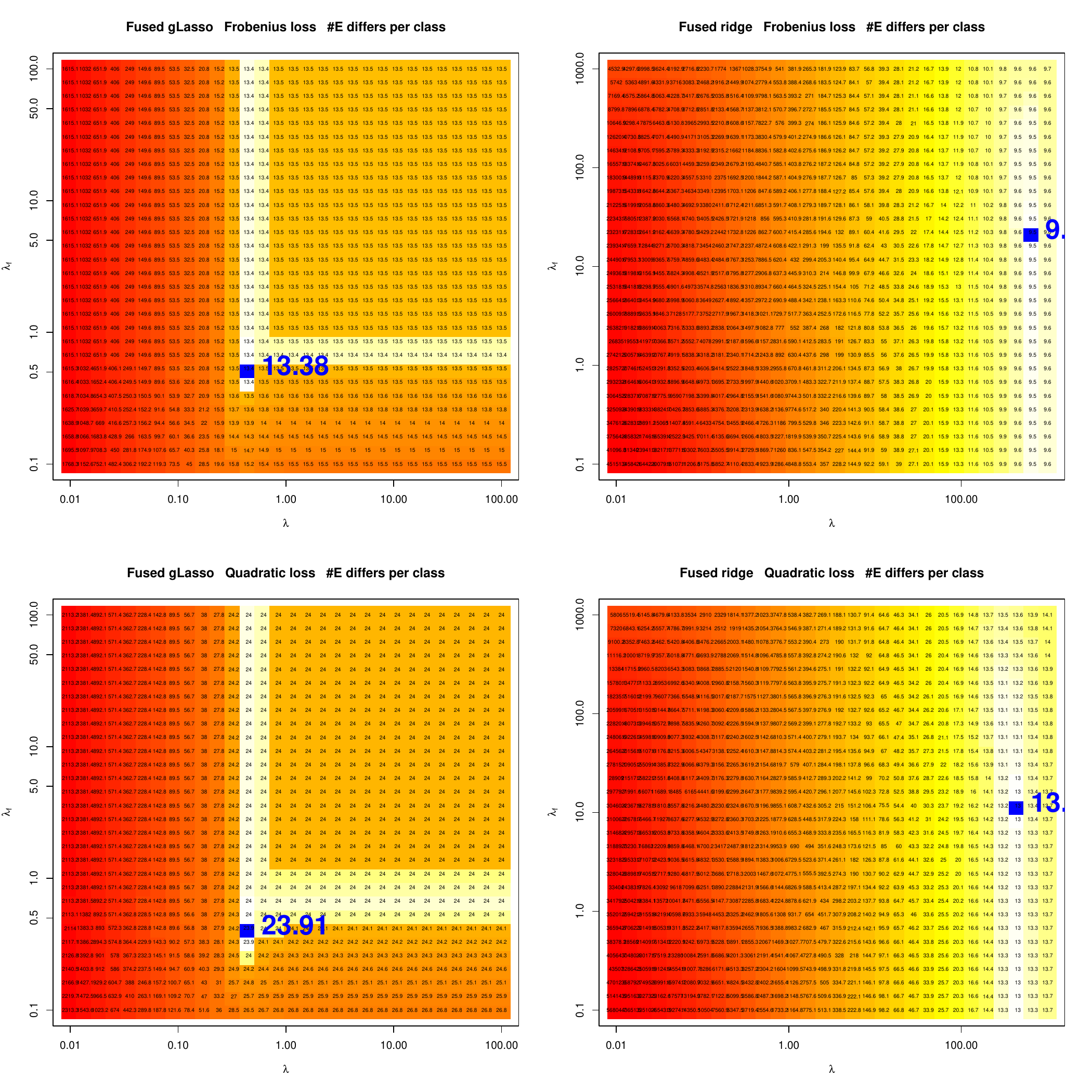}
    \caption{
    Comparison of the fused graphical lasso and the fused ridge estimator in the Barab\'{a}si graph game population setting with $n_g = 25$ under class dissimilarity.
    The the number of edges to add in each time step was taken to be $1$ for class 1 and $3$ for class 2.
    Each square on the two-dimensional grid represents a $(\lambda,\lambda_f)$-combination.
    The number in each square represents the estimated Risk for the corresponding combination.
    The blue square (and corresponding number) indicate the lowest Risk achieved on the grid.
    Left-hand panels give the results for the fused graphical lasso.
    Right-hand panels give the results for the fused ridge estimator.
    Upper panels express the Risk surface under Frobenius loss.
    Lower panels express the Risk surface under quadratic loss.}
  \label{FIG:glassoVsRidge-BGCDn25-1-3}
\end{figure}

We also consider an analogous simulation setting under class differences.
Again Erd\"{o}s-R\'{e}nyi and Barab\'{a}si random graph games were considered of the same variable-dimension.
But now the class 1 and class 2 data are not drawn from the same population.
In the Erd\"{o}s-R\'{e}nyi game the probability of edge presence was taken to be $1/p$ for class 1 and $.25$ for class 2.
In the Barab\'{a}si game the number of edges to add in each time step was taken to be $1$ for class 1 and $3$ for class 2.
Hence, in both settings the topology for class 1 was relatively sparse while the topology for class 2 was more dense.
For the fused ridge we consider the ridge-penalty $\lambda \in [.01, 1000]$ and the fusion-penalty $\lambda_f \in [.1, 1000]$.
For the fused graphical lasso we consider the lasso-penalty $\lambda \in [.01, 100]$ and the fusion-penalty $\lambda_f \in [.1, 100]$.
The class sample size $n_g$ was set to $25$.
Risks are then estimated---for each $(\lambda,\lambda_f)$-combination nested within setting---by the median losses aggregated over the classes over 100 simulation repetitions.
Figure \ref{FIG:glassoVsRidge-BGCDn25-1-3} contains the results of this exercise for the Barab\'{a}si game.
As expected, the fused-ridge penalty is relatively low, indicating that the class-precision matrices are indeed considered to stem from different populations.
Moreover, we are again able to find combinations of penalty-values that lead the fused ridge estimator to achieve lower Risk than the fused graphical lasso estimator.
Again, this observation holds irrespective of the chosen loss-type (Frobenius or quadratic).
And, again, similar behavior is seen in the Erd\"{o}s-R\'{e}nyi graph game setting (Supplementary Figure S14).

\subsection{Scenario 6: Comparison to LASICH and BMGGM}
The LASICH approach of \citet{SS16} and the BMGGM approach of \citet{PSV2015} can be seen as flexible generalizations of the fused graphical lasso.
These approaches allow for pair-specific similarities (between precision matrices) to be estimated from the data.
LASICH uses a Laplacian shrinkage approach while BMGGM uses a hierarchical Bayesian formulation that combines a Markov Random Field prior with a spike-and-slab prior.
Hence, these approaches thus also imply edge selection.
Scenario 6 then compares the targeted fused ridge estimator, as well as its coupling with post-hoc support determination, to the LASICH and BMGGM approaches.

We consider $G = 3$ classes.
We then simulated data sets with $p = 20$ variables from random topologies generated by the Erd\"{o}s-R\'{e}nyi random graph game \citep{RandomGraph}.
In this simulation the dimension $p$ is chosen to be $20$ in order to keep computation times appreciable.
The computation times of the full Bayesian BMGGM approach can become prohibitive for larger $p$.
Note that $p = 20$ concurs with the node-dimension in simulations performed by \citet{PSV2015}.
The density (parameter) is again varied.
For the Erd\"{o}s-R\'{e}nyi random graph game we consider edge presence with probability $P \in \{1/p, .35\}$, indicating relatively sparse and relatively dense topologies, respectively.
Moreover, for each setting of edge presence, we consider (i) $\vOmega_1 = \vOmega_2 = \vOmega_3$ and (ii) $\vOmega_1 \neq \vOmega_2 \neq \vOmega_3$.
For the setting in which the class precisions are equal the Erd\"{o}s-R\'{e}nyi game is run once and the resulting random graph is taken to be the population precision for all classes.
For the setting in which the class precisions are unequal the Erd\"{o}s-R\'{e}nyi game is run thrice and each resulting random graph is taken to be the population precision for one of the classes.
The edge presence and class similarity settings then define four sub-scenarios: (a) sparse equal class precisions, (b) dense equal class precisions, (c) sparse unequal class precisions, and (d) dense unequal class precisions.
The sample size for each class was taken to be $n_g = 15$.
In all sub-scenarios the off-diagonal nonzero elements are chosen to be of value $.15$.
For each estimation approach the estimation was repeated $50$ times for each combination of edge presence probability and class similarity.
We detail estimation specifics and assessment criteria below.

For the fused ridge approach we choose $\vT_g = \alpha_{g}\vI_p$, with $\alpha_g = p/\tr(\vS_g)$.
Moreover, the optimal penalties were determined by LOOCV.
Edge selection was performed using the lFDR procedure of Section \ref{sec:SelectEdge}.
More specifically, an edge in class $g$ was selected if $1 - \widehat{\mathrm{lFDR}}{}_{jj'}^{(g)} \geq .9$.
For the LASICH approach the $\rho_1$ and $\rho_2$ parameters are probed, analogous to the simulation in \citet{SS16}, over a 2-dimensional grid ranging, for both dimensions, from $1$ to $15$.
This takes note of the fact that LASICH performs well under relatively large values of the $\rho$ parameters \citep{SS16}.
The performance of LASICH was then assessed for that combination of $\rho$ parameters for which the performance was optimal (in terms of accuracy).
The BMGMM approach was used as in \citet{PSV2015}.
The joint estimation option was taken with $30,000$ MCMC iterations of which the first $10,000$ were discarded as burn-in.
For each class those edges were selected whose marginal posterior probability of inclusion $> .5$.

\begin{sloppypar}
The approaches are assessed with respect to Frobenius and quadratic loss, accuracy, as well as runtimes.
Accuracy, in terms of graph retrieval, is determined as $\mathrm{(TP+TN)/(TP+TN+FP+FN)}$, where $\mathrm{TP}$ represents the true positives, $\mathrm{TN}$ represents the true negatives, $\mathrm{FP}$ represents the false positives, and $\mathrm{FN}$ represents the false negatives (all in terms of edges).
Runtimes for the methods were recorded in seconds for each simulation.
\end{sloppypar}

Figure \ref{FIG:CompareLASICH_BMGGM} and Figure S15 (Section 6 of the Supplementary Material) visualize the results.
We make several observations on the basis of these figures.
The loss (upper panels Figure \ref{FIG:CompareLASICH_BMGGM}) for all methods is higher for dense compared to sparse settings.
The fused ridge and the LASICH approaches are competitive in terms of loss.
In terms of loss ranking: fused ridge slightly outperforms LASICH whom both outperform BMGGM.
As the class sample sizes are quite low the model likelihood is unlikely to dominate the prior information, resulting in higher loss for the BMGGM approach.
These observations on loss hold for both the Frobenius and the quadratic loss.
In addition we see, with regard to accuracy of graph retrieval (lower-left panel Figure \ref{FIG:CompareLASICH_BMGGM}), that the fused graphical ridge and LASICH approaches are on a par, both outperforming the BMGGM approach in all sub-scenarios.
The accuracy performance of all approaches is lower for the dense situations compared to the sparse situations.
For the fused graphical ridge approach this can (at least in part) be attributed to the stringency of the lFDR threshold used for edge-retention.
A stringent threshold might be very suited for sparse graphs, but as the density of the true graph rises it might become too stringent.
In all, post-hoc edge selection seems a viable option for graph inferral.
However, in balancing graph density and stringency of thresholding it would be beneficial if one has some a priori information on the density of the system that is under study.
The lower-right panel of Figure \ref{FIG:CompareLASICH_BMGGM} visualizes the runtimes over all sub-scenarios.
We see that the runtimes of the BMGGM approach become prohibitive when $p$ would get larger.
The LASICH approach is much faster and the fused ridge approach is the fastest.
These observations on runtimes also hold for the separate sub-scenarios (see Supplementary Figure S15).

\begin{figure}[p]
\centering
\begin{tabular}{cc}
  \includegraphics[width=.47\textwidth]{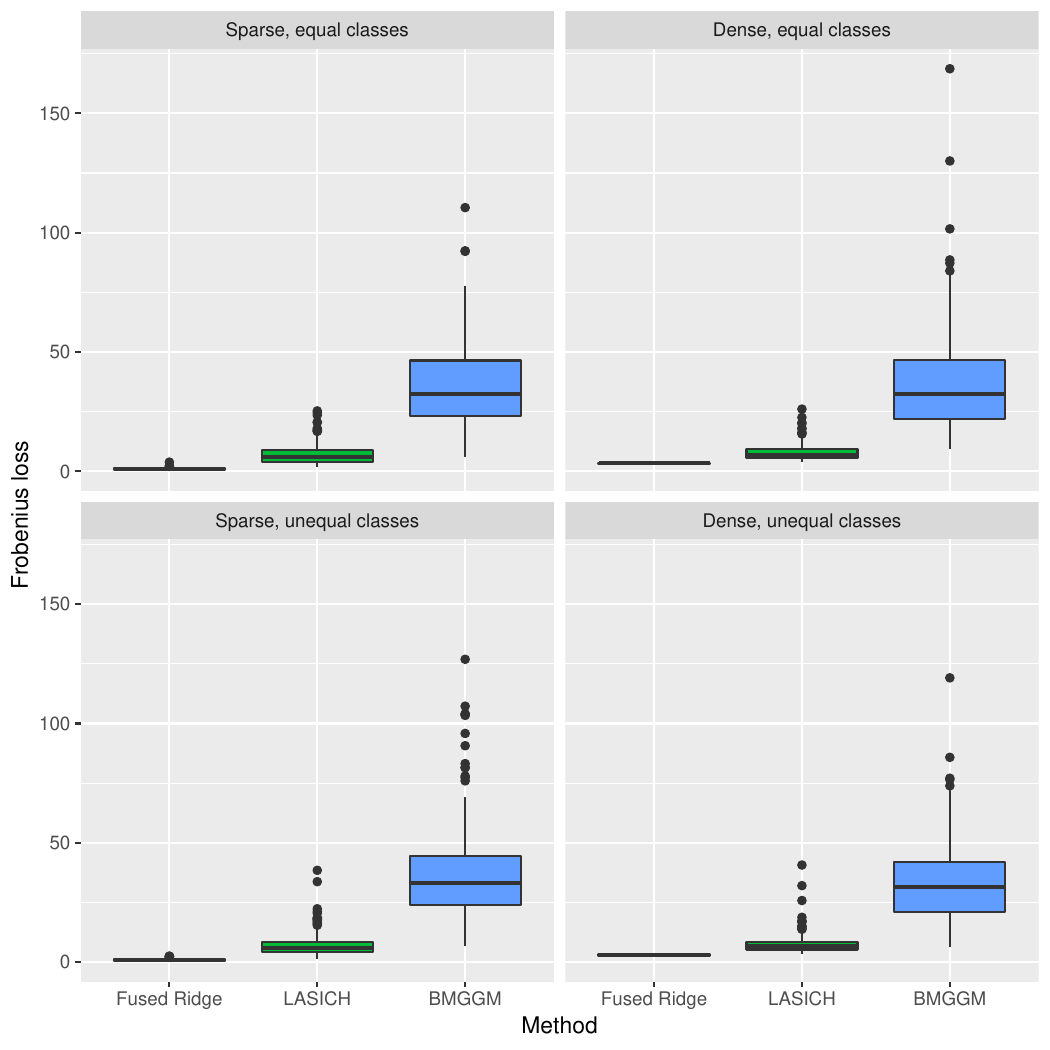}&
  \includegraphics[width=.47\textwidth]{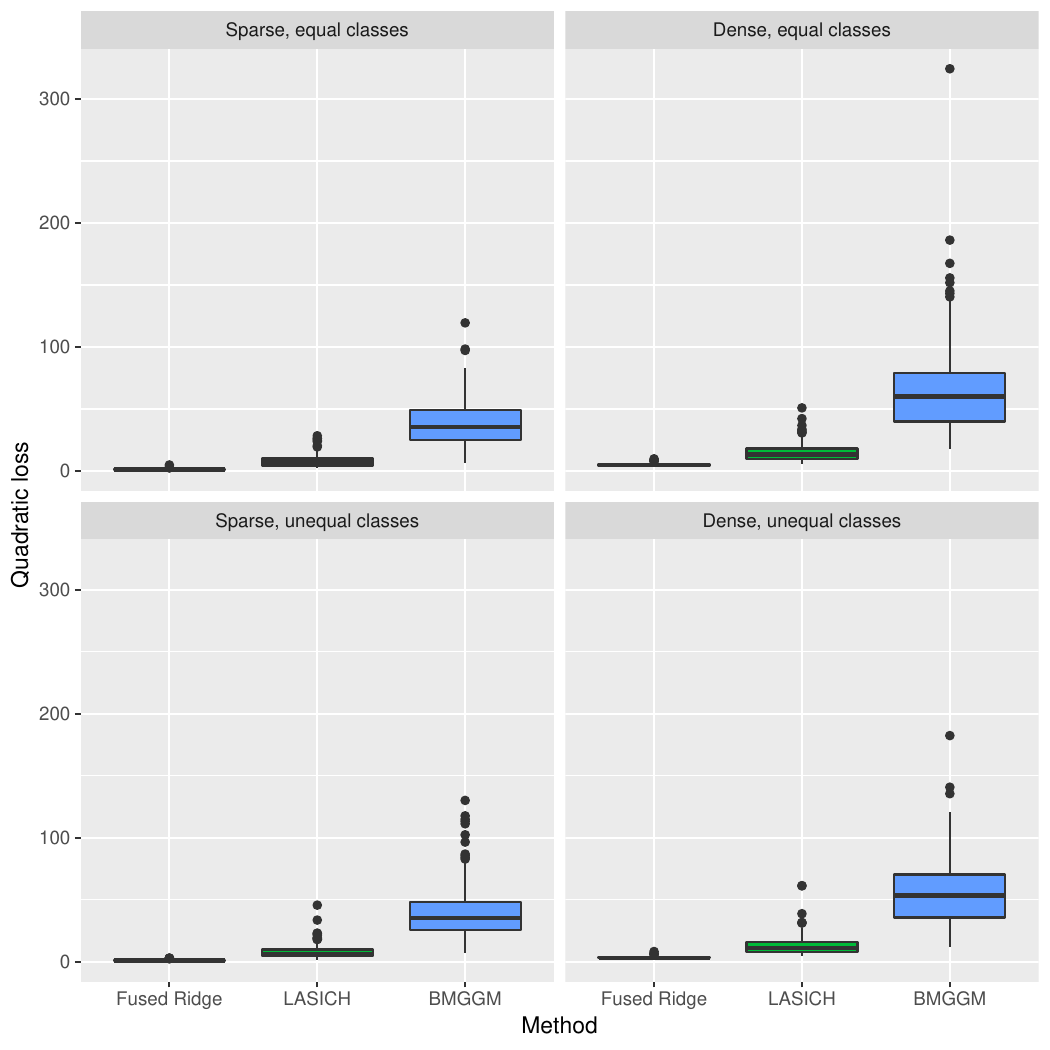}\\
  \includegraphics[width=.47\textwidth]{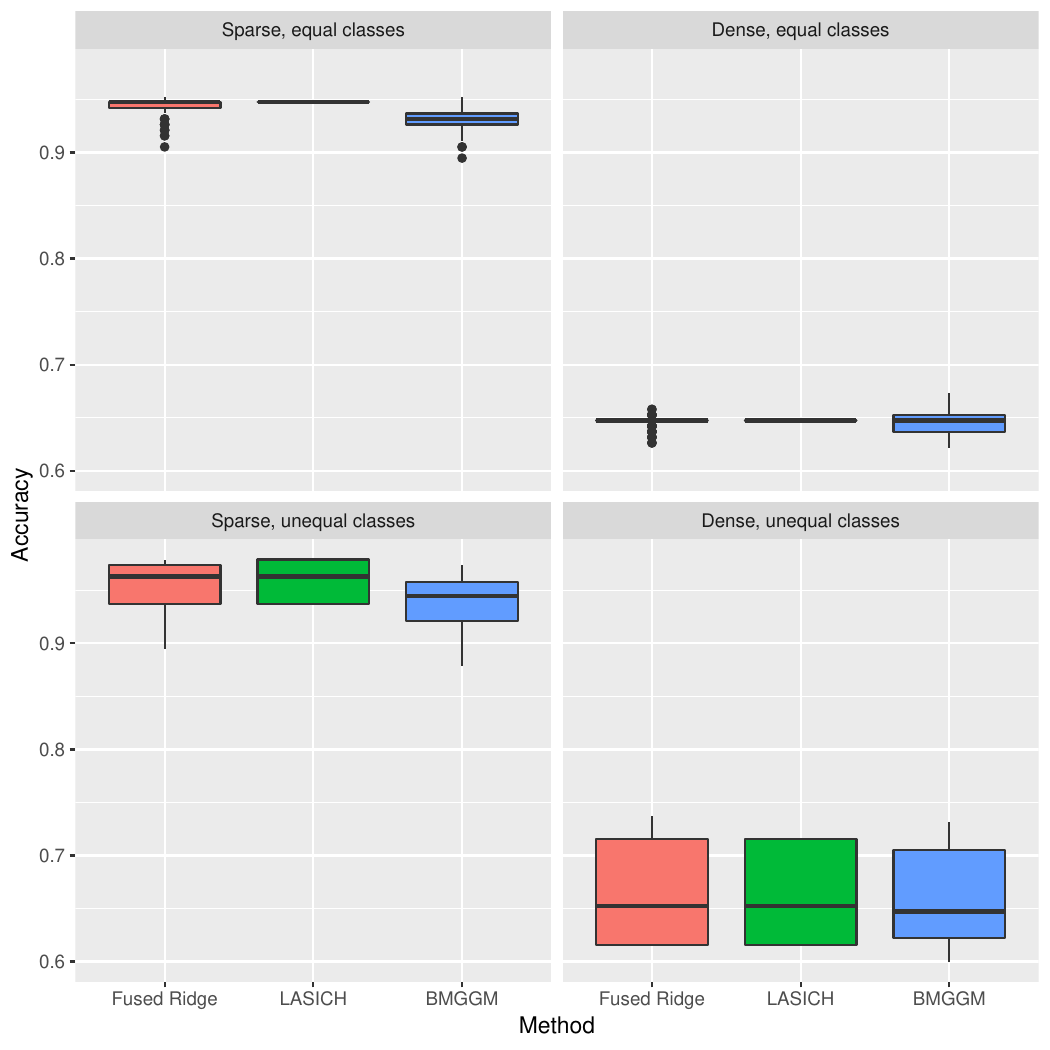}&
  \includegraphics[width=.47\textwidth]{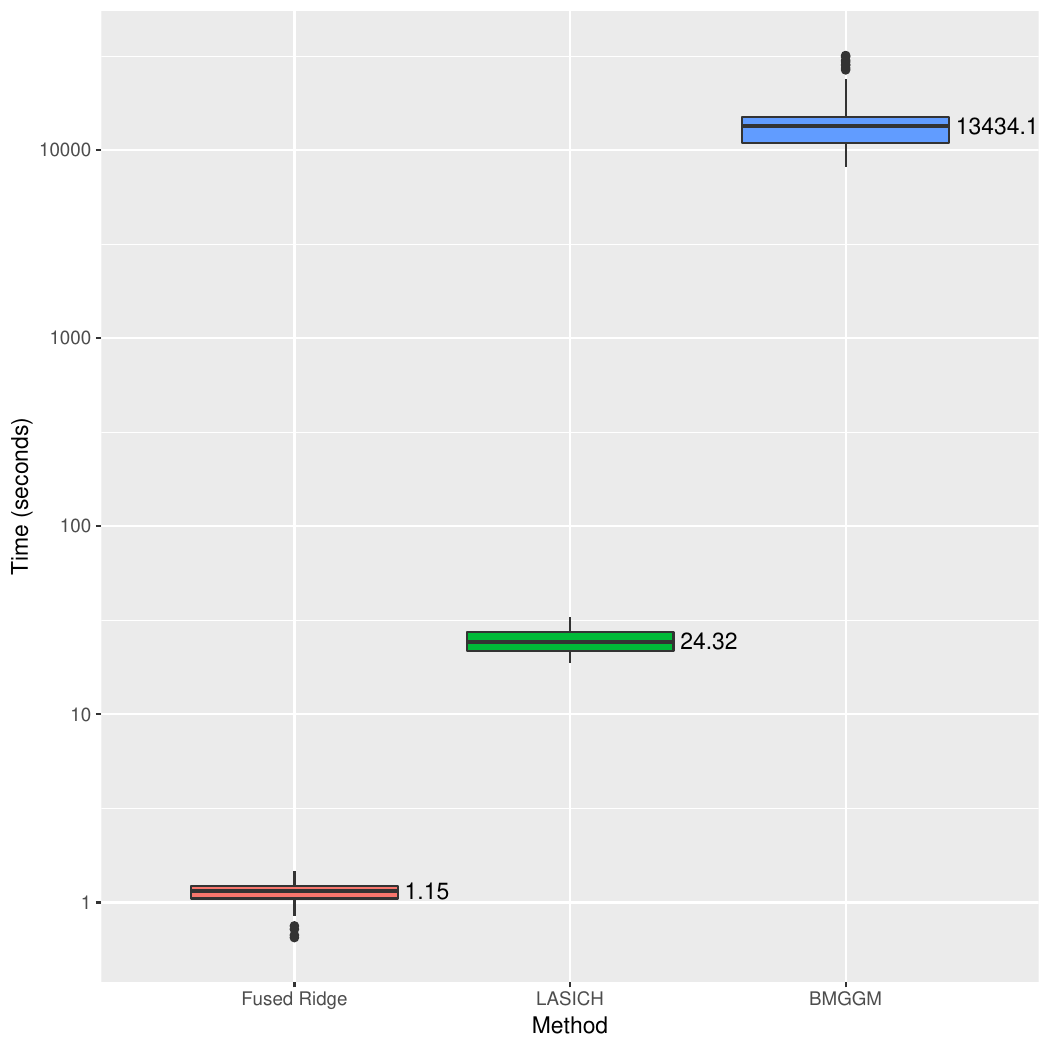}\\
\end{tabular}
    \caption{
    Results for simulation Scenario 6, depicting the comparison of the fused ridge estimator with the LASICH and BMGGM approaches.
    The upper panels depict the Frobenius loss (left-hand panel) and the quadratic loss (right-hand panel) for each of the four sub-scenarios.
    The lower-left panel depicts the accuracy results for each of the four sub-scenarios.
    The lower-right panel visualizes the runtimes over all sub-scenarios.
    Note that the $y$-axis for the lower-right panel has a logarithmic scale.
    The printed numbers above each boxplot then represent the median runtime for the respective method over all sub-scenarios.}
  \label{FIG:CompareLASICH_BMGGM}
\end{figure}

Based on the observations, we make the following recommendations.
There seems to be some merit in having probabilistic control over edge selection, given the adequate performance of both the fused ridge and BMGGM approaches in terms of accuracy.
BMGGM might then be the method of choice when one emphasizes posterior inference in a situation where $p$ is of moderate dimension.
However, BMGGM does not seem suited for fast exploration and large feature-dimensions.
For larger feature-dimensions LASICH and the fused ridge have the computational upper hand over BMGGM.
LASICH should then be preferred when class-membership is unknown.
LASICH can, when this is the case, infer class-membership based on hierarchical clustering.
However, when one has a good idea of class-membership and when one emphasizes both loss and accuracy, we recommend usage of the (computationally efficient) proposed fused (graphical) ridge approach.

\section{Applications}\label{Illustrate.sec}
Lymphoma refers to a group of cancers that originate in specific cells of the immune system such as white blood T- or B-cells.
Approximately $90\%$ of all lymphoma cases are non-Hodgkin's lymphomas---a diverse group of blood cancers excluding  Hodgkin's disease---of which the aggressive diffuse large B-cell lymphomas (DLBCL) constitutes the largest subgroup~\citep{Project1997}.
We showcase the usage of the fused ridge estimator through two analyzes of DLBCL data.

In DLBCL, there exists at least two major genetic subtypes of tumors named after their similarities in genetic expression with activated B-cells (ABC) and germinal centre B-cells (GCB).
A third \emph{umbrella} class, usually designated as Type III, contains tumors that cannot be classified as being either of the ABC or GCB subtype.
Patients with tumors of GCB class show a favorable clinical prognosis compared to that of ABC.
Even though the genetic subtypes have been known for more than a decade \citep{Alizadeh2000} and despite the appearance of refinements to the DLBCL classification system \citep{DykaerBoegsted2015}, DLBCL is still treated as a singular disease in daily clinical practice and the first differentiated treatment regimens have only recently started to appear in clinical trials \citep{Ruan2011,Nowakowski2015}.
Many known phenotypic differences between ABC and GCB are associative, which might underline the translational inertia.
Hence, the biological underpinnings and \emph{functional differences} between ABC and GCB are of central interest and the motivation for the analyzes below.

Incorrect regulation of the NF-$\kappa$B signaling pathway, among other things, is responsible for control of cell survival, and has been linked to cancer.
This pathway has certain known drivers of deregulation.
Aberrant interferon $\beta$ production due to recurrent oncogenic mutations in the central MYD88 gene interferes with cell cycle arrest and apoptosis \citep{Yang2012}.
It also well-known that BCL2, another member of the NF-$\kappa$B pathway, is deregulated in DLBCL \citep{Schuetz2012}.
Moreover, a deregulated NF-$\kappa$B pathway is a key hallmark distinguishing the poor prognostic ABC subclass from the good prognostic GCB subclass of DLBCL \citep{Roschewski2014}.
Our illustrative analyzes thus focus on the \emph{functional differences} between ABC and GCB in relation to the NF-$\kappa$B pathway.
Section \ref{sec:analysis1} investigates the DLBCL classes in the context of a single data set on the NF-$\kappa$B signalling pathway.
Section \ref{sec:dlbcl2} analyzes multiple  DLBCL NF-$\kappa$B data sets with a focus on finding common motifs and motif differences in network representations of pathway-deregulation.
These analyzes show the value of a fusion approach to integration.
In all analyzes we take the NF-$\kappa$B pathway and its constituent genes to be defined by the Kyoto Encyclopedia of Genes and Genomes (KEGG) database \citep{Kanehisa2000}.

\subsection{Nonintegrative Analysis of DLBCL Subclasses}
\label{sec:analysis1}
We first analyze the data from \citet{DykaerBoegsted2015}, consisting of $89$ DLBCL tumor samples.
These samples were RMA-normalized using custom brainarray chip definition files (CDF) \citep{Dai2005} and the {\R}-package \texttt{affy} \citep{Gautier2004}.
This preprocessing used Entrez gene identifiers (EID) by the National Center for Biotechnology Information (NCBI), which are also used by KEGG.
The usage of custom CDFs avoids the mapping problems between Affymetrix probeset IDs and KEGG.
Moreover, the custom CDFs can increase the robustness and precision of the expression estimates \citep{Lu2006, Sandberg2007}.
The RMA-preprocessing yielded 19,764 EIDs.
Subsequently, the features were reduced to the available 84 out of the 95 EIDs present in the KEGG NF-$\kappa$B pathway.
The samples were then partitioned, using the DLBCL automatic classifier (DAC) by \citet{Care2013}, into the three classes ABC $(n_1=31)$, III $(n_2=13)$, and GCB $(n_3=45)$, and gene-wise centered to have zero mean  within each class.

The analysis was performed with the following settings.
Target matrices for the groups were chosen to be scalar matrices with the scalar determined by the inverse of the average eigenvalue of the corresponding sample class covariance matrix, i.e.:
\begin{equation*}
  \vT_\text{ABC} = \alpha_1\vI_p, \quad
  \vT_\text{III} = \alpha_2\vI_p, \quad
  \vT_\text{GCB} = \alpha_3\vI_p,
  \where
  \alpha_g = \frac{p}{\tr(\vS_g)}.
\end{equation*}
These targets translate to a class-scaled `prior' of conditional independence for all genes in NF-$\kappa$B.
The optimal penalties were determined by
LOOCV
using the penalty matrix and graph given in \eqref{eq:DLBCLpenatlygraph1}.
Note that the penalty setup bears resemblance to Example~\ref{ex:2}.
Differing class-specific ridge penalties were allowed because of considerable differences in class sample size.
Direct shrinkage between ABC and GCB was disabled by fixing the corresponding pair-fusion penalty to zero.
The remaining fusion penalties were free to be estimated.
Usage of the Nelder-Mead optimization procedure then resulted in the optimal values given on the right-hand side of \eqref{eq:DLBCLpenatlygraph1} below:
\begin{equation}
\begin{footnotesize}
\begin{tikzpicture}[node distance = 2mm, auto,
  baseline={([yshift=-.75ex]current bounding box.center)},
  main_node/.style={circle,draw,minimum size=1em,inner sep=1pt}]
    \node [main_node, label={[yshift=0.05cm]$\text{ABC}$}] (n1) at (0,0) {$\lambda_{11}$};
    \node [main_node, label={[yshift=0.045cm]$\text{Type III}$}] (n2) at (1.5,0) {$\lambda_{22}$};
    \node [main_node, label={[yshift=0.05cm]$\text{GCB}$}] (n3) at (3,0) {$\lambda_{33}$};

    \path
      (n1) edge node [below, midway] {$\lambda_{12}$} (n2)
      (n2) edge node [below, midway] {$\lambda_{23}$} (n3);


\end{tikzpicture}
\quad
\label{eq:DLBCLpenatlygraph1}
\vLambda^\ast
=
\begin{bmatrix}
\lambda_{11}    & \lambda_{12} & 0\\
\lambda_{12}   & \lambda_{22}  & \lambda_{23}\\
0              & \lambda_{23} & \lambda_{33}
\end{bmatrix}
=
\begin{bmatrix} 2 & 1.5\times 10^{-3} & 0 \\ 1.5\times 10^{-3} & 2.7 & 2\times 10^{-3} \\ 0 & 2\times 10^{-3} & 2.3 \end{bmatrix}\begin{array}{l}\text{ABC} \\ \text{III} \\ \text{GCB}\end{array}.
\end{footnotesize}
\end{equation}

\begin{figure}[t!]
  \centering
  \includegraphics[width=\textwidth]{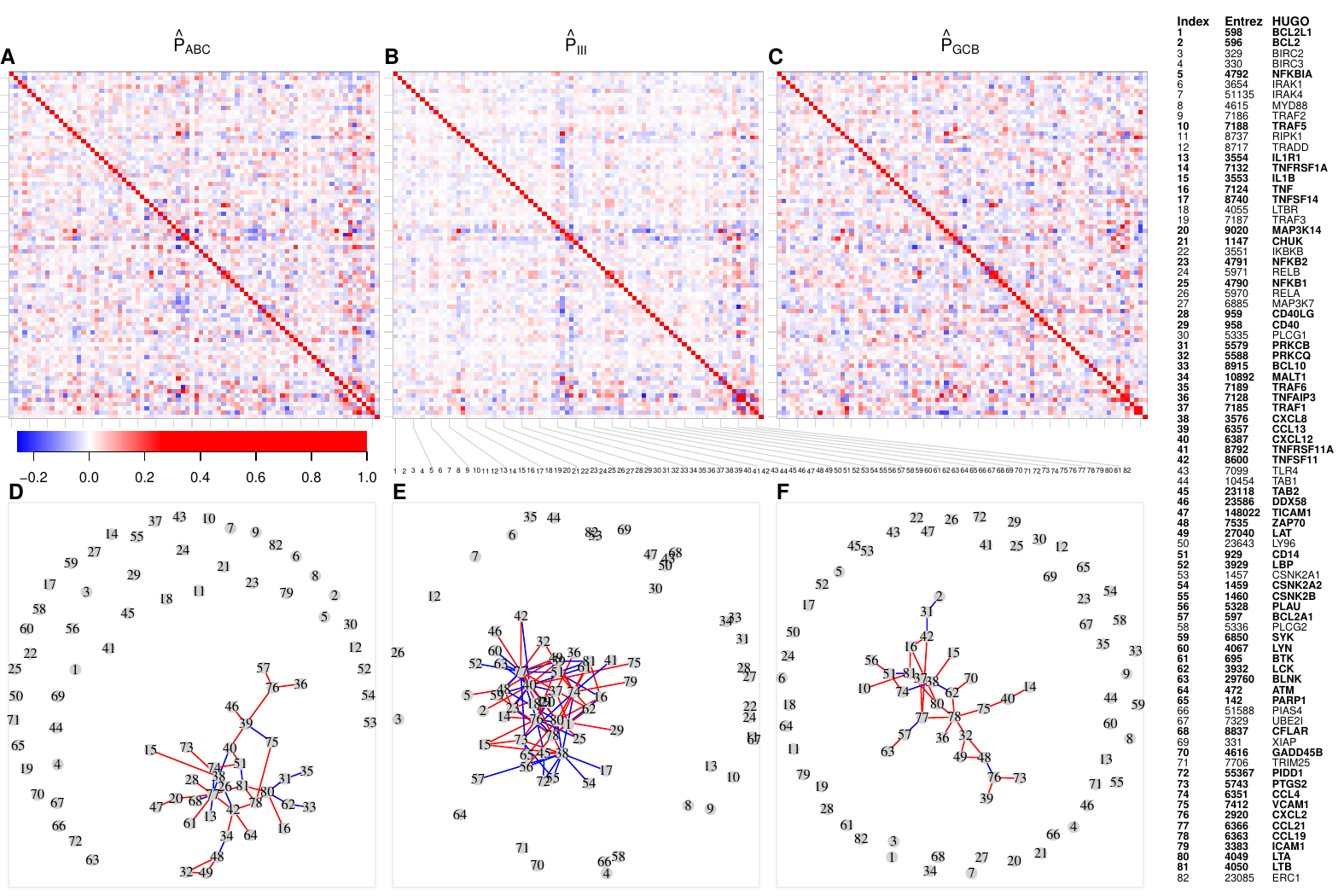}
  \caption{
    \emph{Top}: Heat maps and color key of the partial correlation matrices for the
    ABC (panel A), III (panel B), and GCB (panel C) classes in the NF-$\kappa$B signaling pathway on the \citet{DykaerBoegsted2015} data.
    \emph{Bottom}: Graphs corresponding to the sparsified precision matrices for the classes above.
    Red and blue edges correspond to positive and negative partial correlations, respectively.
    \emph{Far right-panel}: EID key and corresponding Human Genome Organization (HUGO) Gene Nomenclature Committee (HGNC) curated gene names of the NF-$\kappa$B signaling pathway genes.
    Genes that are connected in panels D--F are shown bold.
  }
  \label{fig:pw_analysis1}
\end{figure}

The ridge penalties of classes ABC and GCB are seen to be comparable in size.
The small size of the Type III class leads to a relatively larger penalty to ensure a well-conditioned and stable estimate.
The estimated fusion penalties are all relatively small, implying that heavy fusion is undesirable due to class-differences.
The three class-specific precision matrices were estimated under $\vLambda^\ast$ and subsequently scaled to partial correlation matrices.
Panels A--C of Figure~\ref{fig:pw_analysis1} visualize these partial correlation matrices.
In general, the ABC and GCB classes seem to carry more signal in both the negative and positive range vis-\`{a}-vis the Type III class.

Post-hoc support determination was carried out on the partial correlation matrices using the class-wise $\mathrm{lFDR}$ approach of Section \ref{sec:SelectEdge}.
The $1 - \mathrm{lFDR}$ threshold was chosen conservatively to
$0.99$, selecting
39, 85, 34 edges for classes
ABC, III, GCB, respectively.
The relatively high number of edges selected for the Type III class is (at least partly) due to the difficulty of determining the mixture distribution mentioned in Section \ref{sec:SelectEdge} when the overall partial correlation signal is relatively flat.
Panels D--E of Figure~\ref{fig:pw_analysis1} then show the conditional independence graphs corresponding to the sparsified partial correlation matrices.
We note that a single connected component is identified in each class, suggesting, at least for the ABC and GCB classes, a genuine biological signal.
A secondary supporting overview is provided in Table~\ref{tab:netstats}.

Table~\ref{tab:netstats} gives the most central genes in the graphs of Panels D--E by two measures of node centrality: degree and betweenness.
The node degree indicates the number of edges incident upon a particular node.
The betweenness centrality indicates in how many shortest paths between vertex pairs a particular node acts as an intermediate vertex.
Both measures are proxies for the importance of a feature.
See, e.g., \cite{Newman10} for an overview of these and other centrality measures.
It is seen that the CCL, CXCL, and TNF gene families are well-represented as central and connected nodes across all classes.
The gene CCL21 is very central in classes ABC and III, but less so in the GCB class.
From Panels D--E of Figure~\ref{fig:pw_analysis1} it is seen that BCL2 and BCL2A1 are only connected in the non-ABC classes.
Contrary to expectation, MYD88 is disconnected in all graphs.
The genes ZAP70, LAT, and LCK found in Figure \ref{fig:pw_analysis1} and Table~\ref{tab:netstats} are well-known T-cell specific genes involved in the initial T-cell receptor-mediated activation of NF-$\kappa$B in T-cells \citep{Bidere2009}.
From the differences in connectivity of these genes, different abundances of activated T-cells or different NF-$\kappa$B activation programs for ABC/GCB might be hypothesized.

\begin{table}[!tbp]
\begin{center}
\begin{footnotesize}
\begin{tabular}{llrclrclrclr}
\hline\hline
\multicolumn{1}{l}{\bfseries }&\multicolumn{2}{c}{\bfseries }&\multicolumn{1}{c}{\bfseries }&\multicolumn{2}{c}{\bfseries ABC}&\multicolumn{1}{c}{\bfseries }&\multicolumn{2}{c}{\bfseries III}&\multicolumn{1}{c}{\bfseries }&\multicolumn{2}{c}{\bfseries GCB}\tabularnewline
\cline{5-6} \cline{8-9} \cline{11-12}
\multicolumn{1}{l}{}&\multicolumn{1}{c}{EID}&\multicolumn{1}{c}{Index}&\multicolumn{1}{c}{}&\multicolumn{1}{c}{Degree}&\multicolumn{1}{c}{Betw.}&\multicolumn{1}{c}{}&\multicolumn{1}{c}{Degree}&\multicolumn{1}{c}{Betw.}&\multicolumn{1}{c}{}&\multicolumn{1}{c}{Degree}&\multicolumn{1}{c}{Betw.}\tabularnewline
\hline
CCL21&6366&$77$&&$9\:(5^+, 4^-)$&$202.0$&&$17\:(9^+, 8^-)$&$297.00$&&$4\:(3^+, 1^-)$&$106$\tabularnewline
CXCL8&3576&$38$&&$5\:(2^+, 3^-)$&$126.0$&&$12\:(4^+, 8^-)$&$234.00$&&$4\:(1^+, 3^-)$&$ 56$\tabularnewline
CCL19&6363&$78$&&$4\:(4^+, 0^-)$&$120.0$&&$10\:(6^+, 4^-)$&$ 91.70$&&$6\:(6^+, 0^-)$&$230$\tabularnewline
LTA&4049&$80$&&$5\:(3^+, 2^-)$&$143.0$&&$10\:(6^+, 4^-)$&$195.00$&&$3\:(3^+, 0^-)$&$ 56$\tabularnewline
CXCL12&6387&$40$&&$3\:(2^+, 1^-)$&$ 84.2$&&$12\:(5^+, 7^-)$&$187.00$&&$2\:(2^+, 0^-)$&$ 27$\tabularnewline
CXCL2&2920&$76$&&$3\:(3^+, 0^-)$&$ 61.0$&&$11\:(5^+, 6^-)$&$196.00$&&$3\:(2^+, 1^-)$&$ 53$\tabularnewline
LTB&4050&$81$&&$4\:(3^+, 1^-)$&$ 85.5$&&$5\:(3^+, 2^-)$&$  4.24$&&$6\:(3^+, 3^-)$&$ 98$\tabularnewline
CD14&929&$51$&&$3\:(2^+, 1^-)$&$ 20.2$&&$6\:(3^+, 3^-)$&$ 25.90$&&$3\:(2^+, 1^-)$&$ 32$\tabularnewline
CCL4&6351&$74$&&$2\:(1^+, 1^-)$&$  5.0$&&$8\:(5^+, 3^-)$&$118.00$&&$2\:(1^+, 1^-)$&$  4$\tabularnewline
ZAP70&7535&$48$&&$3\:(2^+, 1^-)$&$ 60.0$&&$5\:(4^+, 1^-)$&$ 50.70$&&$3\:(2^+, 1^-)$&$ 75$\tabularnewline
CCL13&6357&$39$&&$4\:(3^+, 1^-)$&$119.0$&&$5\:(3^+, 2^-)$&$ 19.70$&&$1\:(1^+, 0^-)$&$  0$\tabularnewline
TNFSF11&8600&$42$&&$5\:(4^+, 1^-)$&$160.0$&&$2\:(1^+, 1^-)$&$  0.00$&&$3\:(2^+, 1^-)$&$ 55$\tabularnewline
TNF&7124&$16$&&$1\:(1^+, 0^-)$&$  0.0$&&$4\:(2^+, 2^-)$&$  1.68$&&$3\:(3^+, 0^-)$&$ 24$\tabularnewline
LAT&27040&$49$&&$2\:(2^+, 0^-)$&$  0.0$&&$4\:(4^+, 0^-)$&$ 15.80$&&$2\:(2^+, 0^-)$&$  0$\tabularnewline
LCK&3932&$62$&&$2\:(0^+, 2^-)$&$ 31.0$&&$3\:(3^+, 0^-)$&$ 10.00$&&$3\:(2^+, 1^-)$&$ 64$\tabularnewline
\hline
\end{tabular}
\end{footnotesize}
\end{center}
\caption{The most central genes, their EID, and their plot index. For each class and node, the degree (with the number of positive and negative edges connected to that node in parentheses) and the betweenness centrality is shown. Only the 15 genes with the highest degrees summed over each class are shown.\label{tab:netstats}}
\end{table}

\subsection{Integrative DLBCL Analysis}
\label{sec:dlbcl2}
We now expand the analysis of the previous section to show the advantages of integration by fusion.
A large number of DLBCL gene expression profile (GEP) data sets is freely available at the NCBI Gene Expression Omnibus (GEO) website \citep{Barrett2013}.
We obtained 11 large-scale DLBCL data sets whose GEO-accession numbers (based on various Affymetrix microarray platforms) can be found in the first column of Table~\ref{tab:dlbclinfo}.
One of the sets, with GEO-accession number GSE11318, is treated as a pilot/training data set for the construction of target matrices (see below).
The GSE10846 set is composed of two distinct data sets corresponding to two treatment regimens (R-CHOP and CHOP) as well as different time-periods of study.
Likewise, GSE34171 is composed of three data sets corresponding to the respective microarray platforms used: HG-U133A, HG-U133B, and HG-U133 plus 2.0.
As the samples on HG-U133A and HG-U133B were paired and run on \emph{both} platforms, the (overlapping) features were averaged to form a single virtual microarray comparable to that of HG-U133 plus 2.0.
Note that the \citet{DykaerBoegsted2015} data used in Section \ref{sec:analysis1} is part of the total batch under GEO-accession number GSE56315.
The sample sizes for the individual data sets vary in the range 78--495 and can also be found in Table~\ref{tab:dlbclinfo}.
The data yield a total of 2,276 samples making this, to our knowledge, the hitherto largest integrative DLBCL study.

\begin{table}[!tbp]
\begin{center}
\begin{tabular}{lrrcrrcrrcr}
\hline\hline
\multicolumn{1}{l}{\bfseries }&\multicolumn{2}{c}{\bfseries ABC}&\multicolumn{1}{c}{\bfseries }&\multicolumn{2}{c}{\bfseries Type III}&\multicolumn{1}{c}{\bfseries }&\multicolumn{2}{c}{\bfseries GBC}&\multicolumn{1}{c}{\bfseries }&\multicolumn{1}{c}{\bfseries }\tabularnewline
\cline{2-3} \cline{5-6} \cline{8-9}
\multicolumn{1}{l}{}&\multicolumn{1}{c}{$g$}&\multicolumn{1}{c}{$n_g$}&\multicolumn{1}{c}{}&\multicolumn{1}{c}{$g$}&\multicolumn{1}{c}{$n_g$}&\multicolumn{1}{c}{}&\multicolumn{1}{c}{$g$}&\multicolumn{1}{c}{$n_g$}&\multicolumn{1}{c}{}&\multicolumn{1}{c}{$\sum n_g$}\tabularnewline
\hline
{\bfseries Pilot data}&&&&&&&&&&\tabularnewline
   ~~GSE11318&\gray   $$&   $ 74$&   &\gray   $$&   $ 71$&   &\gray   $$&   $  27$&   &   $ 172$\tabularnewline
\hline
{\bfseries Data set}&&&&&&&&&&\tabularnewline
   ~~GSE56315&\gray   $ 1$&   $ 31$&   &\gray   $ 2$&   $ 13$&   &\gray   $ 3$&   $  45$&   &   $  89$\tabularnewline
   ~~GSE19246&\gray   $ 4$&   $ 51$&   &\gray   $ 5$&   $ 30$&   &\gray   $ 6$&   $  96$&   &   $ 177$\tabularnewline
   ~~GSE12195&\gray   $ 7$&   $ 40$&   &\gray   $ 8$&   $ 18$&   &\gray   $ 9$&   $  78$&   &   $ 136$\tabularnewline
   ~~GSE22895&\gray   $10$&   $ 31$&   &\gray   $11$&   $ 21$&   &\gray   $12$&   $  49$&   &   $ 101$\tabularnewline
   ~~GSE31312&\gray   $13$&   $146$&   &\gray   $14$&   $ 97$&   &\gray   $15$&   $ 224$&   &   $ 467$\tabularnewline
   ~~GSE10846.CHOP&\gray   $16$&   $ 64$&   &\gray   $17$&   $ 28$&   &\gray   $18$&   $  89$&   &   $ 181$\tabularnewline
   ~~GSE10846.RCHOP&\gray   $19$&   $ 75$&   &\gray   $20$&   $ 42$&   &\gray   $21$&   $ 116$&   &   $ 233$\tabularnewline
   ~~GSE34171.hgu133plus2&\gray   $22$&   $ 23$&   &\gray   $23$&   $ 15$&   &\gray   $24$&   $  52$&   &   $  90$\tabularnewline
   ~~GSE34171.hgu133AplusB&\gray   $25$&   $ 18$&   &\gray   $26$&   $ 17$&   &\gray   $27$&   $  43$&   &   $  78$\tabularnewline
   ~~GSE22470&\gray   $28$&   $ 86$&   &\gray   $29$&   $ 43$&   &\gray   $30$&   $ 142$&   &   $ 271$\tabularnewline
   ~~GSE4475&\gray   $31$&   $ 73$&   &\gray   $32$&   $ 20$&   &\gray   $33$&   $ 128$&   &   $ 221$\tabularnewline
   ~~$\sum n_g$&\gray   $$&   $638$&   &\gray   $$&   $344$&   &\gray   $$&   $1062$&   &   $2044$\tabularnewline
\hline
\end{tabular}
\end{center}
\caption{Overview of data sets, the defined classes, and the number of samples. In GSE31312, 28 samples were not classified with the DAC due to technical issues and hence do not appear in this table. In the pilot study GSE11318, 31 samples were primary mediastinal B-cell lymphoma and left out. Note also that the pilot data set GSE11318 was not classified by the DAC.\label{tab:dlbclinfo}}
\end{table}

Similar to above, all data sets were RMA-normalized using custom brainarray CDFs and the {\R}-package \texttt{affy}.
Again, NCBI EIDs were used to avoid non-bijective gene-ID translations between the array-platforms and the KEGG database.
The freely available {\R}-package \texttt{DLBCLdata} was created to automate the download and preprocessing of the data sets in a reproducible and convenient manner.
See the \texttt{DLBCLdata} documentation \citep{DLBCLdata} for more information.
Subsequently, the data sets were reduced to the intersecting 11,908 EIDs present on all platforms.
All samples in all data sets, except for the pilot study GSE11318, were classified as either ABC, GCB, or Type III using the DAC mentioned above.
The same classifier was used in all data sets to obtain a uniform classification scheme and thus maximize the comparability of the classes across data sets.
Subsequently, the features were reduced to the EIDs present in the NF-$\kappa$B pathway and gene-wise centered to have zero mean within each combination of DLBCL subtype and data set.
We thus have a two-way study design---DLBCL subtypes and multiple data sets---analogous to Example~\ref{ex:3}.
A concise overview of each of the $11 \times 3 = 33$ classes for the non-pilot data is provided in Table~\ref{tab:dlbclinfo}.

The target matrices were constructed from the pilot data in an attempt to use information in the directed representation $\calG_\mathrm{pw}$ of the NF-$\kappa$B pathway obtained from KEGG.
The directed graph represents direct and indirect causal interactions between the constituent genes.
It was obtained from the KEGG database via the {\R}-package \texttt{KEGGgraph} \citep{Zhang2015}.
A target matrix was constructed for each DLCBL subtype using the pilot data and the information from the directed topology by computing node contributions using multiple linear regression models.
That is, from an initial $\vT = \vec{0}$, we update $\vT$ for each node $\alpha \in V(\calG_\mathrm{pw})$ through the following sequence:
\begin{align*}
  T_{\alpha, \alpha}        &:= T_{\alpha, \alpha} + \tfrac{1}{\sigma^2} \\
  \vT_{\pa(\alpha), \alpha} &:= \vT_{\pa(\alpha), \alpha} + \tfrac{1}{\sigma^2} \vec{\beta}_{\pa(\alpha)} \\
  \vT_{\alpha, \pa(\alpha)} &:= \vT_{\alpha, \pa(\alpha)} + \tfrac{1}{\sigma^2} \vec{\beta}_{\pa(\alpha)} \\
  \vT_{\pa(\alpha), \pa(\alpha)}
    &:=  \vT_{\pa(\alpha), \pa(\alpha)} + \tfrac{1}{\sigma^2} \vec{\beta}_{\pa(\alpha)}\vec{\beta}_{\pa(\alpha)}^\top,
\end{align*}
where $\pa(\alpha)$ denotes the parents of node $\alpha$ in $\calG_\mathrm{pw}$, and where $\sigma$ and $\vec{\beta}$ are the residual standard error and regression coefficients obtained from the linear regression of $\alpha$ on $\pa(\alpha)$.
By this scheme the target matrix represents the conditional independence structure that would result from moralizing the directed graph.
If $\calG_\mathrm{pw}$ is acyclic then $\vT \succ 0$ is guaranteed.

The penalty setup bears resemblance to Example \ref{ex:3}.
The Type III class is considered closer to the ABC and GCB subtypes than ABC is to GCB.
Thus, the direct shrinkage between the ABC and GCB subtypes was fixed to zero.
Likewise, direct shrinkage between subtype and data set combinations was also disabled.
Hence, a common ridge penalty $\lambda$, a data set--data set shrinkage parameter $\lamDS$ and a subtype--subtype shrinkage parameter $\lamST$ were estimated.
The optimal penalties were determined by SLOOCV using the penalty matrix and graph given in \eqref{eq:DLBCLpenaltygraph} below:
\begin{equation}
\begin{footnotesize}
\begin{tikzpicture}[node distance = 2mm, auto,
  baseline={([yshift=-.5ex]current bounding box.center)},
  main_node/.style={circle,draw,minimum size=1em,inner sep=2pt}
  ]

    \node [main_node] (n1) at (0,0)  {$\lambda$};
    \node [main_node] (n2) at (2,0)  {$\lambda$};
    \node [main_node] (n3) at (4,0)  {$\lambda$};

    \node [main_node] (n4) at (0,-1) {$\lambda$};
    \node [main_node] (n5) at (2,-1) {$\lambda$};
    \node [main_node] (n6) at (4,-1) {$\lambda$};

    \node [minimum size=1em,inner sep=3pt] (n7) at (0,-2) {$\svdot$};
    \node [minimum size=1em,inner sep=3pt] (n8) at (2,-2) {$\svdot$};
    \node [minimum size=1em,inner sep=3pt] (n9) at (4,-2) {$\svdot$};

    \node [main_node] (n10) at (0,-3) {$\lambda$};
    \node [main_node] (n11) at (2,-3) {$\lambda$};
    \node [main_node] (n12) at (4,-3) {$\lambda$};

    \draw (n1) -- (n2) node [above, midway] {\scriptsize$\lamST$};
    \draw (n2) -- (n3) node [above, midway] {\scriptsize$\lamST$};
    \draw (n4) -- (n5) node [below, midway] {\scriptsize$\lamST$};
    \draw (n5) -- (n6) node [below, midway] {\scriptsize$\lamST$};
    \draw (n10) -- (n11) node [below, midway] {\scriptsize$\lamST$};
    \draw (n11) -- (n12) node [below, midway] {\scriptsize$\lamST$};

    \draw (n1) -- (n4) node [left, midway]  {\scriptsize$\lamDS$};
    \draw (n2) -- (n5) node [left, midway]  {\scriptsize$\lamDS$};
    \draw (n3) -- (n6) node [left, midway]  {\scriptsize$\lamDS$};
    \draw (n4) -- (n7) node [left, midway]  {\scriptsize$\lamDS$};
    \draw (n5) -- (n8) node [left, midway]  {\scriptsize$\lamDS$};
    \draw (n6) -- (n9) node [left, midway]  {\scriptsize$\lamDS$};
    \draw (n7) -- (n10) node [left, midway]  {\scriptsize$\lamDS$};
    \draw (n8) -- (n11) node [left, midway]  {\scriptsize$\lamDS$};
    \draw (n9) -- (n12) node [left, midway]  {\scriptsize$\lamDS$};

    \path
      (n1) edge [bend left=30] node [right, near start] {\scriptsize$\lamDS$} (n7)
      (n1) edge [bend left=30] node [right, midway] {\scriptsize$\lamDS$} (n10)
      (n4) edge [bend left=30] node [right, near end] {\scriptsize$\lamDS$} (n10)

      (n2) edge [bend left=30] node [right, near start] {\scriptsize$\lamDS$} (n8)
      (n2) edge [bend left=30] node [right, midway] {\scriptsize$\lamDS$} (n11)
      (n5) edge [bend left=30] node [right, near end] {\scriptsize$\lamDS$} (n11)

      (n3) edge [bend left=30] node [right, near start] {\scriptsize$\lamDS$} (n9)
      (n3) edge [bend left=30] node [right, midway] {\scriptsize$\lamDS$} (n12)
      (n6) edge [bend left=30] node [right, near end] {\scriptsize$\lamDS$} (n12);

    \node [above=0.1cm of n1] (ABC) {ABC};
    \node [above=0.1cm of n2] (T3) {Type III};
    \node [above=0.1cm of n3] (GCB) {GCB};
    \node [left=0.05cm of n1] (DS1) {$\mathrm{DS}_1$};
    \node [left=0.05cm of n4] (DS2) {$\mathrm{DS}_2$};
    \node [left=0.05cm of n10] (DS11) {$\mathrm{DS}_{11}$};


\end{tikzpicture}
\qquad
\vLambda =
\begin{bsmallmatrix}
  \lambda & \lamST & 0      & \lamDS & 0      & 0      & \cdots & \lamDS & 0      & 0      \\
  \lamST  &\lambda & \lamST & 0      & \lamDS & 0      & \cdots & 0      & \lamDS & 0      \\
  0       & \lamST &\lambda & 0      & 0      & \lamDS & \cdots & 0      & 0      & \lamDS \\
  \lamDS  & 0      & 0      &\lambda & \lamST & 0      & \cdots & \lamDS & 0      & 0      \\
  0       & \lamDS & 0      & \lamST &\lambda & \lamST & \cdots & 0      & \lamDS & 0      \\
  0       & 0      & \lamDS & 0      & \lamST &\lambda & \cdots & 0      & 0      & \lamDS \\
  \svdot  & \svdot & \svdot & \svdot & \svdot & \svdot & \sddot & \svdot & \svdot & \svdot \\
  \lamDS  & 0      & 0      & \lamDS & 0      & 0      & \cdots &\lambda & \lamST & 0      \\
  0       & \lamDS & 0      & 0      & \lamDS & 0      & \cdots & \lamST &\lambda & \lamST \\
  0       & 0      & \lamDS & 0      & 0      & \lamDS & \cdots & 0      & \lamST &\lambda
\end{bsmallmatrix}.
\label{eq:DLBCLpenaltygraph}
\end{footnotesize}
\end{equation}

\noindent The optimal penalties were found to be
$\lambda^\diamond = 2.2$
for the ridge penalty,
$\lambda^\diamond_\mathrm{DS} = 0.0022$
for the data set fusion penalty, and
$\lambda^\diamond_\mathrm{ST} = 0.00068$
for the subtype fusion penalty, respectively.

\begin{figure}[h!]
  \centering
  \includegraphics[width=\textwidth]{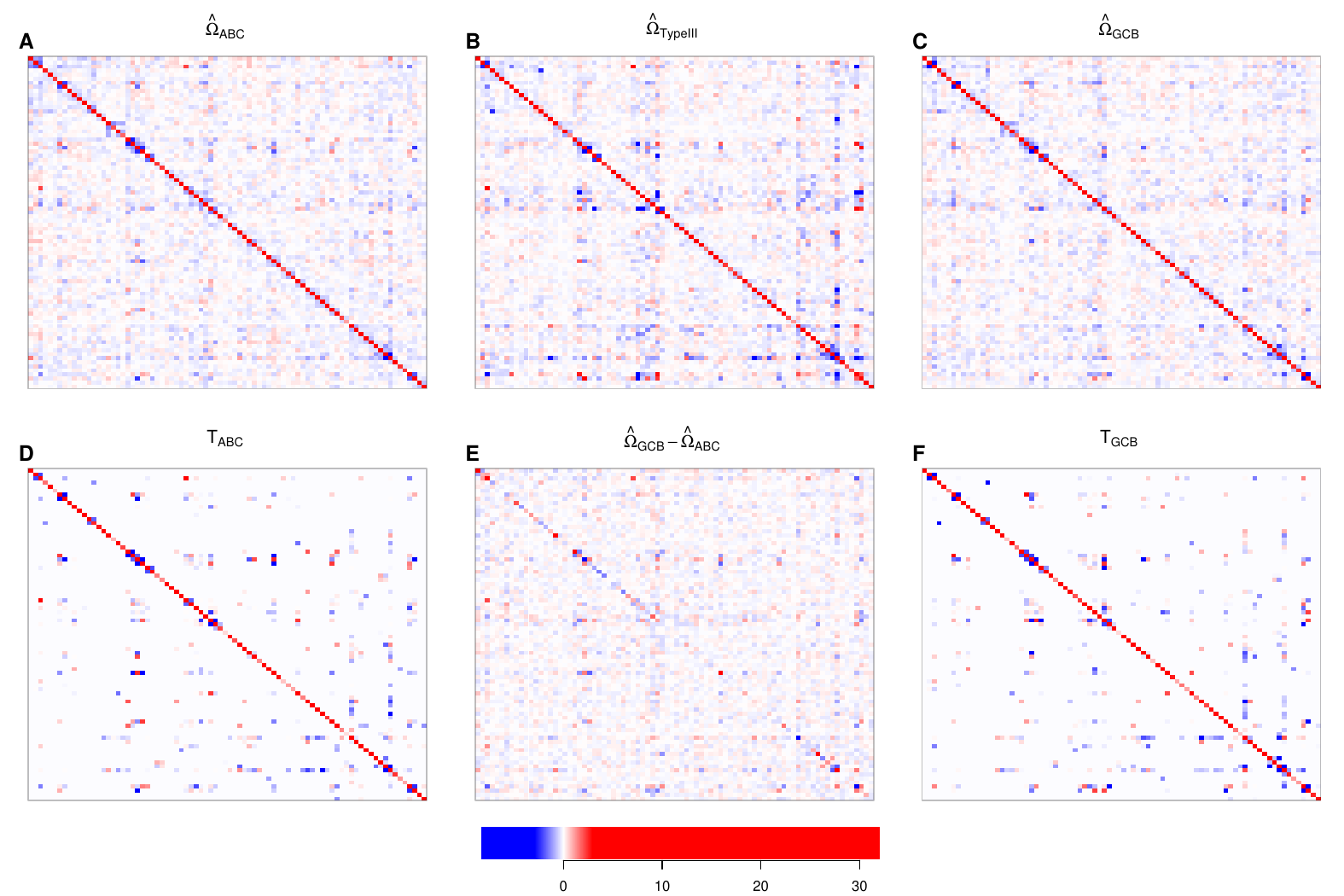}
  \caption{
    Summary of the estimated precision matrices for the NF-$\kappa$B pathway.
    \emph{Top row}: Heat maps of the estimated precision matrices pooled across data sets for each genetic subtype.
    \emph{Middle row from left to right:} The pooled target matrix for ABC, the difference between the pooled ABC and GCB estimates, and the pooled target matrix for GCB.
    \emph{Bottom:} The color key for the heat maps.
  }
  \label{fig:pw_prec_one}
\end{figure}

To summarize and visualize the 33 class precision estimates they were pooled within DLBCL subtype.
Panels A--C of Figure~\ref{fig:pw_prec_one} visualizes the 3 pooled estimates as heat maps.
Panels D and F visualize the constructed target matrices for the ABC and GCB subtypes, respectively.
Panel E then gives the difference between the pooled ABC and GCB estimates, indicating that they harbor differential signals to some degree.
We would like to capture the commonalities and differences with a differential network representation.

\begin{figure}[h!]
  \centering
  \includegraphics[width=\textwidth]{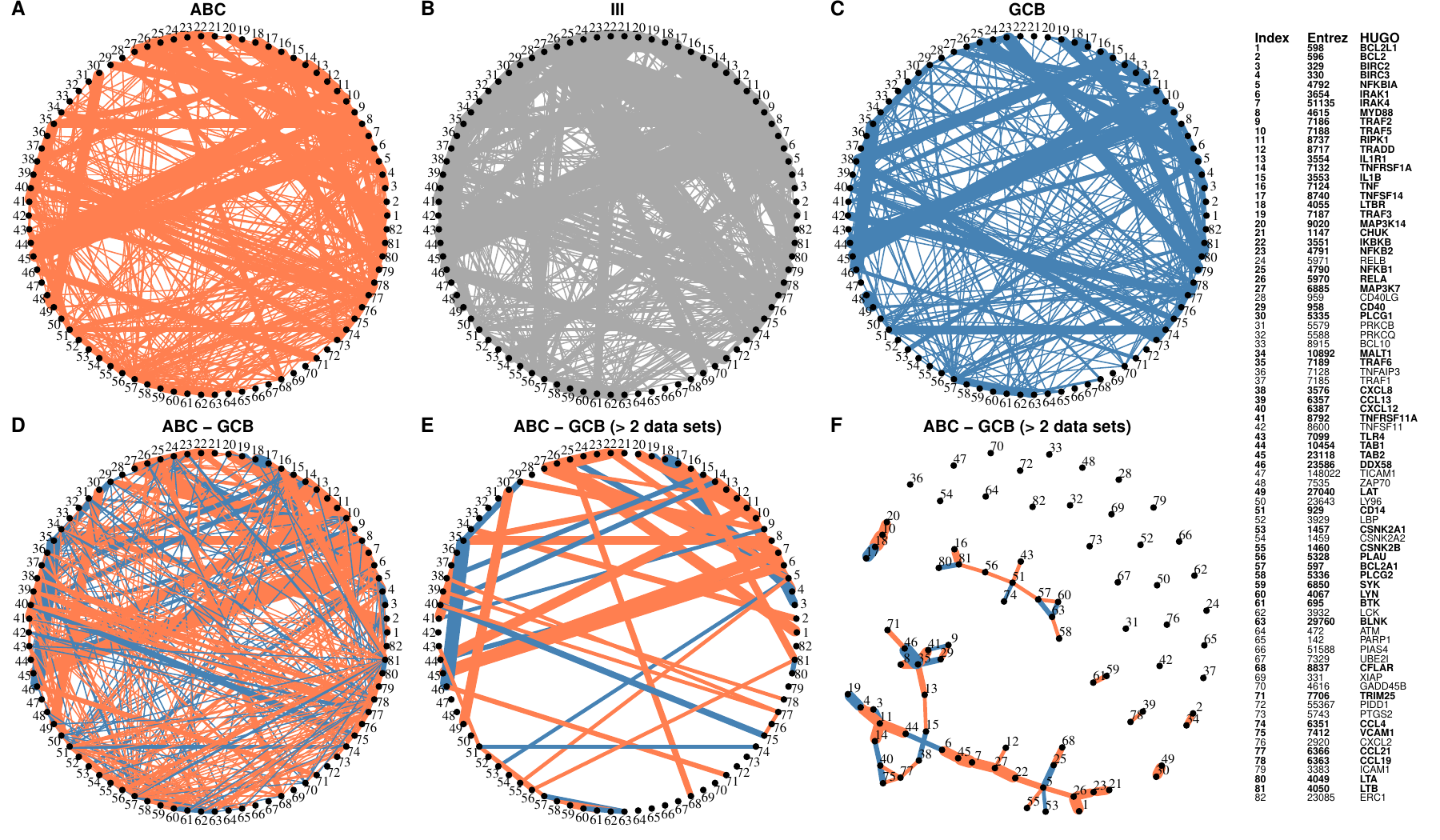}
  \caption{
    Summary of estimated GGMs for the NF-$\kappa$B pathway.
    \emph{Panels A--C}: Graphs obtained by adding the signed adjacency matrices for each subtype across the data sets.
    The edge widths are drawn proportional to the absolute edge weight.
    \emph{Panel D}: Graph obtained by subtracting the summarized signed adjacency matrix of GCB (panel A) from that of           ABC (panel C).
    Edge widths are drawn proportional to the absolute weight and colored according to the sign.
    Orange implies edges more present in ABC and blue implies edges more present in GCB.
    \emph{Panel E}: As the graph in panel D, however only edges with absolute weight $> 2$ are drawn.
    \emph{Panel F}: As the graph in panel E, but with an alternative layout.
    \emph{Far right-panel:} EID key and corresponding HGNC curated gene names of the NF-$\kappa$B pathway genes.
    Genes that are connected in panel F are shown bold.
  }
  \label{fig:pw_file_one}
\end{figure}

The estimated class-specific precision matrices were subsequently scaled to partial correlation matrices.
Each precision matrix was then sparsified using the lFDR procedure of Section \ref{sec:SelectEdge}.
Given the class an edge was selected whenever $1 - \widehat{\mathrm{lFDR}} \geq 0.999$.
To compactly visualize the the multiple GGMs we obtained \emph{signed edge-weighted total networks} mentioned in Section~\ref{sec:commonAndDifferentialNetworks}.
Clearly, for inconsistent connections the weight would vary around zero, while edges that are consistently selected as positive (negative) will have a large positive (negative) weight.
These meta-graphs are plotted in Figure~\ref{fig:pw_file_one}.
Panels A--C give the signed edge-weighted total networks for each subtype across the data sets.
They show that (within DLBCL subtypes) there are a number of edges that are highly concordant across all data sets.
To evaluate the greatest differences between the ABC and GCB subtypes, the signed edge-weighted total network of the latter was subtracted from the former.
The resulting graph $\calG_{\mathrm{ABC}-\mathrm{GCB}}$ can be found in Panel D.
Edges that are more stably present in the ABC subtype are represented in orange and the edges more stably present in the GCB subtype are represented in blue.
Panel F represents the graph from panel D with only those edges retained whose absolute weight exceeds $2$.
In a sense, the graph of panel F then represents the stable differential network.
The strongest connections here should suggest places of regulatory deregulation gained or lost across the two subtypes.
Interestingly, this differential network summary shows relatively large connected subgraphs suggesting differing regulatory mechanisms.


The graph in panel F of Figure~\ref{fig:pw_file_one} then conveys the added value of the integrative fusion approach.
Certain members of the CCL, CXCL, and TNF gene families who were highly central in the analysis of Section \ref{sec:analysis1} are still considered to be central here.
However, it is also seen that certain genes that garnered high centrality measures in the single data set analyzed in Section \ref{sec:analysis1} do not behave stably \emph{across} data sets, such as CXCL2.
In addition, the integrative analysis appoints the BCL2 gene family a central role, especially in relation to the ABC subtype.
This contrasts with Section \ref{sec:analysis1}, where the BCL2 gene family was not considered central and appeared to be connected mostly in the non-ABC classes.
Moreover, whereas the analysis of the single data set could not identify a signal for MYD88, the integrative analysis identifies MYD88 to be stably connected across data sets.
Especially the latter two observations are in line with current knowledge on deregulation in the NF-$\kappa$B pathway in DLBCL patients.
Also in accordance with the literature is the known interaction of LTA with LTB seen in panel F of Figure~\ref{fig:pw_file_one} \citep{WilliamsAbbott1997,Browning1997} which here appear to be differential between ABC/GCB.
Thus, borrowing information across classes enables a meta-analytic approach that can uncover information otherwise unobtainable through the analysis of single data sets.

\section{Discussion and Conclusion}\label{Discuss.sec}
We considered the problem of jointly estimating multiple inverse covariance matrices from high-dimensional data consisting of distinct classes.
A fused ridge estimator was proposed that generalizes previous contributions in two principal directions.
First, we introduced the use of targets in fused ridge precision estimation.
The targeted approach helps to stabilize the estimation procedure and allows for the incorporation of prior knowledge.
It also juxtaposes itself with various alternative penalized precision matrix estimators that pull the estimates towards the edge of the parameter space, i.e., who shrink towards the non-interpretable null matrix.
Second, instead of using a single ridge penalty and a single fusion penalty parameter for all classes, the approach grants the use of \emph{class-specific} ridge penalties and \emph{class-pair-specific} fusion penalties.
This results in a flexible shrinkage framework that (i) allows for class-specific tuning, that (ii) supports analyzes when a factorial design underlies the available classes, and that (iii) supports the appropriate handling of situations where some classes are high-dimensional whilst others are low-dimensional.
Targeted shrinkage and usage of a flexible penalty matrix might also benefit other procedures for precision matrix estimation such as the fused graphical lasso \citep{Danaher2013}.

The targeted fused ridge estimator was combined with post-hoc support determination, which serves as a basis for integrative or meta-analytic Gaussian graphical modeling.
This combination thus has applications in meta-, integrative-, and differential network analysis of multiple data sets or classes of data.
This meta-approach to network analysis has multiple motivations.
First, by combining data it can effectively increase the sample size in settings where samples are relatively scarce or expensive to produce.
In a sense it refocuses the otherwise declining attention to obtaining a sufficient amount of data---a tendency we perceive to be untenable.
Second, aggregation across multiple data sets decreases the likelihood of capturing idiosyncratic features (of individual data sets), thereby preventing over-fitting of the data.

Insightful summarization of the results is important for the feasibility of our approach to fused graphical modeling.
To this end we have proposed various basic tools to summarize commonalities and differences over multiple graphs.
These tools were subsequently used in a differential network analysis of the NF-$\kappa$B signaling pathway in DLBCL subtypes over multiple GEP data sets.
This application is not without critique, as it experiences a problem present in many GEP studies:
The classification of the DLBCL subtypes (ABC and GBC) is performed on the basis of the same GEP data on which the network analysis is executed.
This may be deemed methodologically undesirable.
However, we justify this double use of data as (a) the pathway of interest involves a selection of genes whereas the classification  uses all genes, and (b) the analysis investigates partial correlations and differential networks whereas the classification, in a sense, considers only differential expression.
Furthermore, as in all large-scale genetic screenings, the analyzes should be considered `tentative' and findings need to be validated in independent experiments.
Notwithstanding, the analyzes show that the fusion approach to network integration has merit in uncovering class-specific information on pathway deregulation.
Moreover, they exemplify the exploratory \emph{hypothesis generating} thrust of the framework we offer.

We see various inroad for further research.
With regard to estimation one could think of extending the framework to incorporate a fused version of the elastic net.
Mixed fusion, in the sense that one could do graphical lasso estimation with ridge fusion or ridge estimation with lasso fusion, might also be of interest.
From an applied perspective the desire is to expand the toolbox for insightful (visual) summarization of commonalities and differences over multiple graphs.
Moreover, it is of interest to explore improved ways for support determination.
The lFDR procedure, for example, could be expanded by considering all classes jointly.
Instead of applying the lFDR procedure to each class-specific precision matrix, one would then be interested in determining the proper mixture of a grand common null-distribution and multiple class-specific non-null distributions.
These inroads were out of the scope of current work, but we hope to explore them elsewhere.

\subsection{Software Implementation}
The fused ridge estimator and its accompanying estimation procedure is implemented in the \texttt{rags2ridges}-package \citep{rags} for the statistical language {\R}.
This package has many supporting functions for penalty parameter selection, graphical modeling, as well as network analysis.
We will report on its full functionality elsewhere.
The package is freely available from the Comprehensive {\R} Archive Network: \url{http://cran.r-project.org/}.

\acks{
Anders E. Bilgrau was supported by a grant from the Karen Elise Jensen Fonden, a travel grant from the Danish Cancer Society, and a visitor grant by the Dept.\ of Mathematics of the VU University Amsterdam.
Carel F.W. Peeters received funding from the European Community's Seventh Framework Programme (FP7, 2007-2013), Research Infrastructures action, under grant agreement No. FP7-269553 (EpiRadBio project).
The authors thank Karen Dybk{\ae}r of the Dept.\ of Haematology at Aalborg University Hospital, for her help on the biological interpretations in the DLBCL application.
The authors would also like to thank Ali Shojaie of the Dept.\ of Biostatistics, University of Washington, for making the LASICH code available.
Lastly, the Authors thank the Associate Editor and three anonymous reviewers, whose constructive comments have led to a considerable improvement in presentation.
}

\appendix

\section{Geometric Interpretation of the Fused Ridge Penalty}\label{app:geometric}
\begin{figure}[t!]
  \centering
  \includegraphics[width=\textwidth]{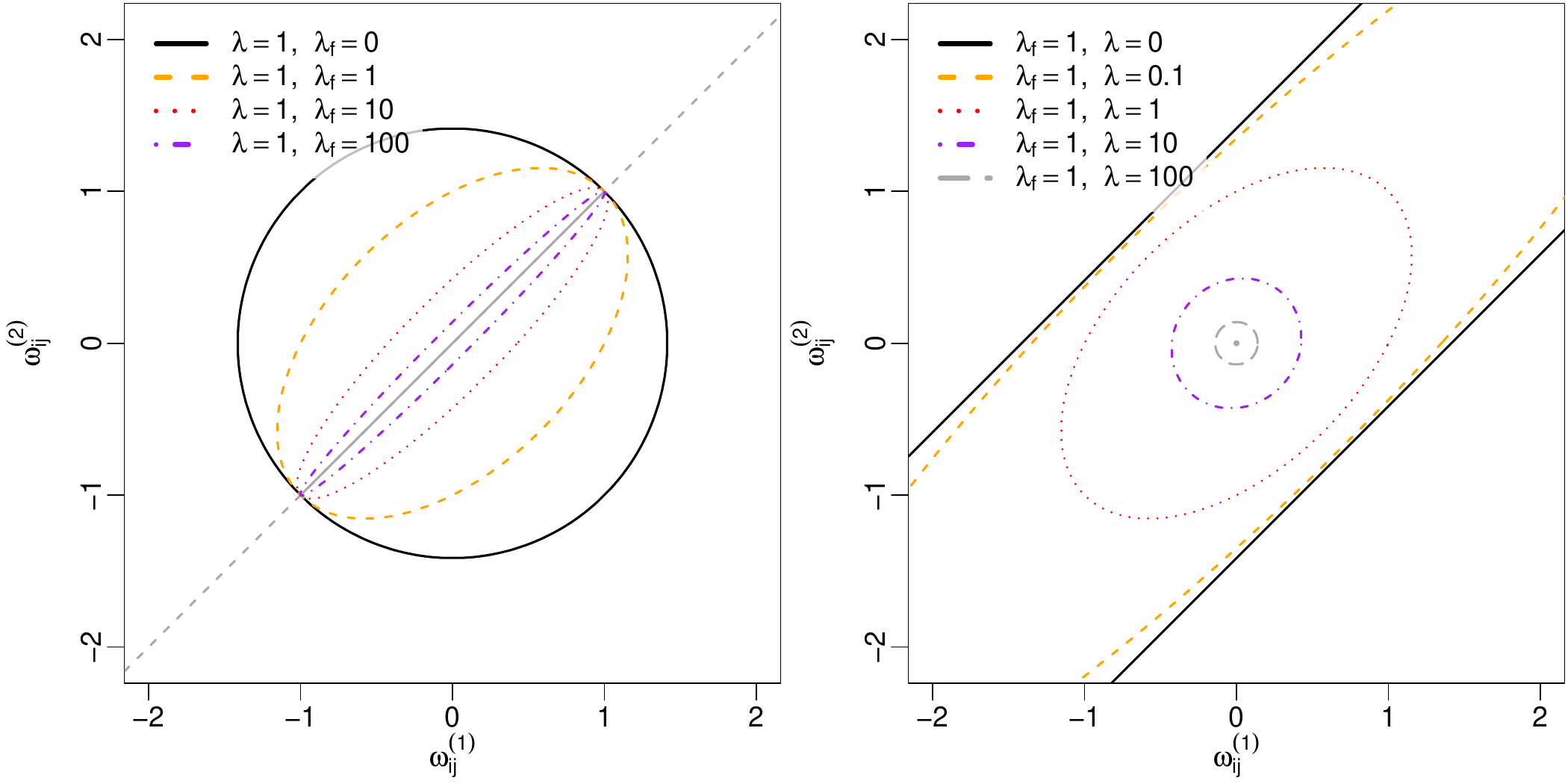}
  \caption{Visualization of the effects of the fused ridge penalty in terms of constraints. The left panel shows the effect of $\lambda_f$ for fixed $\lambda$. Here, $\lambda_f=0$ is the regular ridge penalty. The right panel shows the effect of $\lambda$ while keeping $\lambda_f$ fixed.}
  \label{fig:geom}
\end{figure}

Some intuition behind the fused ridge is provided by pointing to the equivalence of penalized and constrained optimization.
To build this intuition we study the geometric interpretation of the fused ridge penalty in the special case of \eqref{eq:argmax2} with $\vT = \vec{0}$.
In this case $\lambda_{g g} = \lambda$ for all $g$, and $\lambda_{g_1 g_2} = \lambda_f$ for all $g_1 \neq g_2$.
Clearly, the penalty matrix then amounts to $\vLambda = \lambda\vI_G + \lambda_f(\vJ_G-\vI_G)$.
Matters are simplified further by considering $G = 2$ classes and by focusing on a specific entry in the precision matrix, say $(\vOmega_g)_{jj'} = \omega_{jj'}^{(g)}$, for $g=1,2$.
By doing so we ignore the contribution of other precision elements to the penalty.
Now, the fused ridge penalty may be rewritten as:
\begin{footnotesize}
\begin{align*}
  \frac{\lambda}{2}
  \Big(\big\Vert \vOmega_1 \big\Vert_F^2 + \big\Vert \vOmega_2 \big\Vert_F^2 \Big)
  + \frac{\lambda_f}{4} \sum_{g_1 = 1}^2 \sum_{g_2 = 1}^2
  \big\Vert \vOmega_{g_1} - \vOmega_{g_2} \big\Vert_F^2
  &=
  \frac{\lambda}{2}
  \Big(\big\Vert \vOmega_1 \big\Vert_F^2 + \big\Vert \vOmega_2 \big\Vert_F^2 \Big)
  + \frac{\lambda_f}{2} \big\Vert \vOmega_1 - \vOmega_2 \big\Vert_F^2.
\end{align*}\end{footnotesize}\noindent
Subsequently considering only the contribution of the $\omega_{jj'}^{(g)}$ entries implies this expression can be further reduced to:
\begin{align*}
   \frac{\lambda}{2}
   \left[ \big(\omega_{jj'}^{(1)}\big)^2 + \big(\omega_{jj'}^{(2)}\big)^2 \right]
  + \frac{\lambda_f}{2} \big( \omega_{jj'}^{(1)} - \omega_{jj'}^{(2)} \big)^2
  =
  \frac{\lambda + \lambda_f}{2}
  \left[ \big(\omega_{jj'}^{(1)}\big)^2 + \big(\omega_{jj'}^{(2)}\big)^2 \right]
  - \lambda_f \omega_{jj'}^{(1)} \omega_{jj'}^{(2)}.
\end{align*}
It follows immediately that this penalty imposes constraints on the parameters $\omega_{jj'}^{(1)}$ and $\omega_{jj'}^{(2)}$, amounting to the set:
\begin{align}
  \label{eq:constraint}
  \biggl\{
    \big( \omega_{jj'}^{(1)}, \omega_{jj'}^{(2)} \big) \in \bbR^2
    :
    \frac{\lambda + \lambda_f}{2}
    \Bigl[\bigl( \omega_{jj'}^{(1)} \bigr)^2 + \bigl( \omega_{jj'}^{(2)} \bigr)^2\Bigr]
    - \lambda_f  \omega_{jj'}^{(1)}\omega_{jj'}^{(2)} \leq c
  \biggr\},
\end{align}
for some $c\in \bbR_+$.
It implies that the fused ridge penalty can be understood by the implied constraints on the parameters.
Figure \ref{fig:geom} shows the boundary of the set for selected values.

Panel \ref{fig:geom}A reveals the effect of the fused, inter-class penalty parameter $\lambda_f$ (while keeping $\lambda$ fixed).
At $\lambda_f = 0$, the constraint coincides with the regular ridge penalty.
As $\lambda_f$ increases, the ellipsoid shrinks along the minor principal axis $x = -y$ with no shrinkage along $x = y$.
In the limit $\lambda_f \to \infty$ the ellipsoid collapses onto the identity line.
Hence, the parameters $\omega_{jj'}^{(1)}$ and $\omega_{jj'}^{(2)}$ are shrunken towards each other and while their differences vanish, their sum is not affected.
Hence, the fused penalty parameter primarily shrinks the `sum of the parameters', but also fuses them as a bound on their sizes implies a bound on their difference.

Panel \ref{fig:geom}B shows the effect of the intra-class $\lambda$ penalty (while keeping $\lambda_f$ fixed).
When the penalty vanishes for $\lambda \to 0$ the domain becomes a degenerated ellipse (i.e., cylindrical for more than 2 classes) and parameters $\omega_{jj'}^{(1)}$ and $\omega_{jj'}^{(2)}$ may assume any value as long as their difference is less than $\sqrt{2c/\lambda_f}$.
For any $\lambda > 0$, the parameter-constraint is ellipsoidal.
As $\lambda$ increases the ellipsoid is primarily shrunken along the principal axis formed by the identity line and along the orthogonal principal axis $(y = - x)$.
In the limit $\lambda \to \infty$ the ellipsoid collapses onto the point $(0, 0)$.
It is clear that the shape of the domain in \eqref{eq:constraint} is only determined by the ratio of $\lambda$ and $\lambda_{f}$.

The effect of the penalties on the domain of the obtainable estimates can be further understood by noting that the fused ridge penalty \eqref{eq:FR} can be rewritten as
\begin{equation}
  \label{eq:altFR}
  \tilde{\lambda} \sum_{g_1, g_2}
  \sqfnorm[\big]{(\vOmega_{g_1} \!- \vT_{g_1}) + (\vOmega_{g_2} \!- \vT_{g_2})}
  +\tilde{\lambda}_f \sum_{g_1, g_2}
  \sqfnorm[\big]{(\vOmega_{g_1} \!- \vT_{g_1}) - (\vOmega_{g_2} \!- \vT_{g_2})},
\end{equation}
for some penalties $\tilde{\lambda}$ and $\tilde{\lambda}_f$.
The details of this derivation can be found in Section \ref{sec:altFR} below.
The first and second summand of the rewritten penalty \eqref{eq:altFR} respectively shrink the sum and difference of the parameters of the precision matrices.
Their contributions thus coincide with the principal axes along which two penalty parameters shrink the domain of the parameters.

\subsection{Alternative Form for the Fused Ridge Penalty}
\label{sec:altFR}
This section shows that the alternative form \eqref{eq:altFR} for the ridge penalty can be written in the form \eqref{eq:FR}.
We again assume a common ridge penalty  $\lambda_{gg} = \lambda$ and a common fusion penalty $\lambda_{g_1 g_2} = \lambda_f$ for all classes and pairs thereof.
To simplify the notation, let $\vA_g = \vOmega_g \!- \vT_g$. Now,
\begin{align*}
  &f^\mathrm{FR'}\bigl(\{\vOmega_g\}; \tilde{\lambda}, \tilde{\lambda}_f, \{\vT_g\}\bigr)
  \\&=
    \tilde{\lambda} \sum_{g_1, g_2} \big\Vert \OmTpOmT \big\Vert_F^2
  +\tilde{\lambda}_f \sum_{g_1, g_2} \big\Vert \OmTmOmT \big\Vert_F^2
  \\&=
  \tilde{\lambda} \sum_{g_1, g_2} \big\Vert \vA_{g_1} \!+ \vA_{g_2} \big\Vert_F^2
  +\tilde{\lambda}_f \sum_{g_1, g_2} \big\Vert \vA_{g_1} \!- \vA_{g_2} \big\Vert_F^2
  \\&=
  \tilde{\lambda} \sum_{g_1, g_2}
  \Big(
    \big\Vert \vA_{g_1} \big\Vert_F^2
    + \big\Vert\vA_{g_2}\big\Vert_F^2
    + 2\langle\vA_{g_1}, \vA_{g_2}\rangle
  \Big)
  +\tilde{\lambda}_f \sum_{g_1, g_2} \big\Vert \vA_{g_1} \!- \vA_{g_2} \big\Vert_F^2
  \\&=
  \tilde{\lambda} \sum_{g_1, g_2}
  \Big(
    2\big\Vert \vA_{g_1} \big\Vert_F^2
    + 2\big\Vert\vA_{g_2}\big\Vert_F^2
    - \big\Vert \vA_{g_1} - \vA_{g_2} \big\Vert_F^2
  \Big)
  +\tilde{\lambda}_f \sum_{g_1, g_2} \big\Vert \vA_{g_1} \!- \vA_{g_2} \big\Vert_F^2
  \\&=
  4\tilde{\lambda} G \sum_g \big\Vert\vA_g\big\Vert_F^2
  -\tilde{\lambda} \sum_{g_1, g_2} \big\Vert \vA_{g_1} \!- \vA_{g_2} \big\Vert_F^2
  +\tilde{\lambda}_f \sum_{g_1, g_2} \big\Vert \vA_{g_1} \!- \vA_{g_2} \big\Vert_F^2
  \\&=
  4\tilde{\lambda} G \sum_g \big\Vert\vA_g\big\Vert_F^2
  +(\tilde{\lambda}_f - \tilde{\lambda})
  \sum_{g_1, g_2} \big\Vert \vA_{g_1} - \vA_{g_2} \big\Vert_F^2
  \\&=
  4\tilde{\lambda} G \sum_g \big\Vert \OmT{} \big\Vert_F^2
  + (\tilde{\lambda}_f - \tilde{\lambda})
  \sum_{g_1, g_2} \big\Vert \OmTmOmT \big\Vert_F^2.
\end{align*}
Hence, the alternative penalty \eqref{eq:altFR} is also of the form \eqref{eq:FR} and thus
the fused ridge of \eqref{eq:altFR} is equivalent to \eqref{eq:FR} for appropriate choices of the penalties.

\section{Results and Proofs}
Section \ref{sec:Support} contains supporting results from other sources and results in support of Algorithm \ref{alg:fusedridge}.
Section \ref{sec:Proofs} contains proofs of the results stated in the main text as well as additional results conducive in those proofs.

\subsection{Supporting Results}\label{sec:Support}
\begin{lemma}[\citealt{VanWieringen2014}]
\label{Lem:WP}
Amend the log-likelihood \eqref{eq:OLL} with the $\ell_2$-penalty
\begin{equation*}
\frac{\lambda}{2} \sqfnorm[\big]{\vOmega \!- \vT},
\end{equation*}
with $\mathbf{T} \in \calS_+^p$ denoting a fixed symmetric positive semi-definite target matrix, and
where $\lambda \in (0,\infty)$ denotes a penalty parameter.
The zero gradient equation w.r.t. the precision matrix then amounts to
\begin{equation}\label{eq:EMridge}
\hat{\vOmega}^{-1} - (\vS - \lambda\vT) - \lambda\hat{\vOmega} = \vec{0},
\end{equation}
whose solution gives a penalized ML ridge
estimator of the precision matrix:
\begin{equation*}
\hat{\mathbf{\Omega}}(\lambda) =
\left\{\left[\lambda\mathbf{I}_{p} + \frac{1}{4}(\mathbf{S} -
\lambda\mathbf{T})^{2}\right]^{1/2} + \frac{1}{2}(\mathbf{S} -
\lambda\mathbf{T})\right\}^{-1}.
\end{equation*}
\end{lemma}

\begin{lemma}[\citealt{VanWieringen2014}]
\label{lem:NoInverseIdent}
Consider $\hat{\vOmega}(\lambda)$ from Lemma \ref{Lem:WP} and define $[\hat{\vOmega}(\lambda)]^{-1} \equiv \hat{\vSigma}(\lambda)$. The following identity then holds:
\begin{equation*}
\vS - \lambda\vT = \hat{\vSigma}(\lambda) - \lambda\hat{\vOmega}(\lambda).
\end{equation*}
\end{lemma}

\begin{lemma}
\label{lem:concavity}%
Let $\vLambda \in \calS^G$ be a matrix of fixed penalty parameters such that $\vLambda \geq \vec{0}$. Moreover, let $\{\vT_g\} \in \calS_+^p$.
Then if $\diag(\vLambda) > \vec{0}$, the problem of \eqref{eq:argmax0} is strictly concave.
\end{lemma}
\begin{proof}(Proof of Lemma~\ref{lem:concavity})
By $\diag(\vLambda) > \vec{0}$, it is clear that the fused ridge penalty \eqref{eq:FR} is strictly convex as it is a conical combination of strictly convex and convex functions.
Hence, the negative fused ridge penalty is strictly concave.
The log-likelihood of \eqref{eq:loglik} is a conical combination of concave functions and is thus also concave.
Therefore, the penalized log-likelihood is strictly concave.
\end{proof}

\subsection{Proofs and Additional Results}\label{sec:Proofs}
\begin{proof}(Proof of Proposition~\ref{prop:fusedridge})
To find the maximizing argument for a specific class of the general fused ridge penalized log-likelihood problem \eqref{eq:argmax0} we must obtain its first-order derivative w.r.t. that class and solve the resulting zero gradient equation.
To this end we first rewrite the ridge penalty \eqref{eq:FR} into a second alternative form.
Using that $\vLambda = {\vLambda}^\top$, and keeping in mind the cyclic property of the trace as well as properties of $\vOmega_g$ and $\vT_g$ stemming from their symmetry, we may find:
\begin{align}
  &f^\mathrm{FR''}\bigl(\{\vOmega_g\}; \vLambda, \{\vT_g\}\bigr)
  \notag
  \\
  &\qquad=
    \sum_g \frac{\lambda_{gg}}{2}
    \sqfnorm[\big]{ \vOmega_g {-} \vT_g }
  + \sum_{g_1, g_2} \frac{\lambda_{g_1 g_2}}{4}
    \sqfnorm[\big]{ (\vOmega_{g_1} {-} \vT_{g_1})  -
                    (\vOmega_{g_2} {-} \vT_{g_2})  }
  \notag
  \\
  &\qquad=
  \sum_g \frac{\lambda_{g\summed}}{2}
    \tr\left[  (\vOmega_g {-} \vT_g)^\top(\vOmega_g {-} \vT_g) \right]
  - \sum_{\mathclap{\substack{g_1, g_2 \\ g_1 \neq g_2}}}
    \frac{\lambda_{g_1 g_2}}{2}
    \tr\left[  (\vOmega_{g_1} {-} \vT_{g_1})^\top(\vOmega_{g_2} {-} \vT_{g_2}) \right],
    \label{eq:rewrittenFR}
\end{align}
where $\lambda_{g\summed} = \sum_{g'} \lambda_{gg'}$ denotes the sum over the $\nth{g}$ row (or column) of $\vLambda$.
Taking the first-order partial derivative of \eqref{eq:rewrittenFR} w.r.t.\ $\vOmega_{g_0}$ yields:
\begin{align}
  &\frac{\partial}{\partial\vOmega_{g_0}}
  f^\mathrm{FR''}\bigl(\{\vOmega_g\}; \vLambda, \{\vT_g\}\bigr)
  \notag
  \\
  &\qquad=
  \lambda_{g_0\summed}
  \left[2(\vOmega_{g_0} {-} \vT_{g_0}) - (\vOmega_{g_0} {-} \vT_{g_0})\circ \vI_p\right]
  - \sum_{g \neq g_0}
  \lambda_{g g_0}
  \left[2(\vOmega_g {-} \vT_g) - (\vOmega_g {-} \vT_g)\circ\vI_p\right].
  \label{eq:dFR}
\end{align}
The first-order partial derivative of \eqref{eq:loglik} w.r.t.\ $\vOmega_{g_0}$ results in:
\begin{align}
  \frac{\partial}{\partial \vOmega_{g_0}} \calL(\{\vOmega_g\}; \{\vS_g\})
  &=
  \frac{\partial}{\partial \vOmega_{g_0}}
  \sum_g n_g
  \big\{\ln |\vOmega_g|- \tr(\vS_g\vOmega_g) \big\},
  \notag
  \\
  &=
  n_{g_0} \left[2(\vOmega_{g_0}^{-1} \!- \vS_{g_0})
  - (\vOmega_{g_0}^{-1} \!- \vS_{g_0}) \circ \vI_p\right].
  \label{eq:loglikderiv}
\end{align}
Subtracting \eqref{eq:dFR} from \eqref{eq:loglikderiv} yields
\begin{equation}\label{eq:derdiff}
\left[
n_{g_0}(\vOmega_{g_0}^{-1} \!- \vS_{g_0}) -
\lambda_{g_0\summed}(\vOmega_{g_0} {-} \vT_{g_0}) +
\sum_{g \neq g_0}\lambda_{g g_0}(\vOmega_g {-} \vT_g)
\right]\circ(2\vJ_p - \vI_p),
\end{equation}
which, clearly, is $\vec{0}$ only when $
n_{g_0}(\vOmega_{g_0}^{-1} \!- \vS_{g_0}) -
\lambda_{g_0\summed}(\vOmega_{g_0} {-} \vT_{g_0}) +
\sum_{g \neq g_0}\lambda_{g g_0}(\vOmega_g {-} \vT_g) = \vec{0}$.
From \eqref{eq:derdiff} we may then find our (conveniently scaled) zero gradient equation to be:
\begin{equation}\label{eq:estimatingequation}
\hat{\vOmega}_{g_0}^{-1} \!- \vS_{g_0} -
\frac{\lambda_{g_0\summed}}{n_{g_0}}(\hat{\vOmega}_{g_0} {-} \vT_{g_0}) +
\sum_{g \neq g_0}\frac{\lambda_{g g_0}}{n_{g_0}}(\vOmega_g {-} \vT_g) = \vec{0}.
\end{equation}
Now, rewrite \eqref{eq:estimatingequation} to
\begin{equation}
   \hat{\vOmega}_{g_0}^{-1} - \bar{\vS}_{g_0}
  -  \bar{\lambda}_{g_0}(\hat{\vOmega}_{g_0} \!- \bar{\vT}_{g_0})
  = \vec{0},
 \label{eq:barupdate_appendix}
\end{equation}
where
$
 \bar{\vS}_{g_0}
 = \vS_{g_0} -
   \sum_{g \neq g_0}\frac{\lambda_{g g_0}}{n_{g_0}} (\vOmega_g \!- \vT_g)
$,
$\bar{\vT}_{g_0} = \vT_{g_0}$, and $\bar{\lambda}_{g_0} = \lambda_{g_0\summed}/n_{g_0}$.
It can be seen that \eqref{eq:barupdate_appendix} is of the form \eqref{eq:EMridge}.
Lemma \ref{Lem:WP} may then be applied to obtain the solution \eqref{eq:update}.
\end{proof}

\begin{corollary}
\label{prop:fusedridge2}%
Consider the estimator \eqref{eq:update}.
Let $\hvOmega_{g}\bigl(\vLambda, \{\vOmega_{g'}\}_{g' {\neq} g} \bigr)$ be the precision matrix estimate of the \nth{g} class.
Also, let $\diag(\vLambda) > \vec{0}$ and assume that all off-diagonal elements of $\vLambda$ are zero.
Then $\hvOmega_{g}\bigl(\vLambda, \{\vOmega_{g'}\}_{g' {\neq} g} \bigr)$ reduces to the non-fused ridge estimate of class $g$:
\begin{equation}\label{eq:unfusedClass}
\hvOmega_{g}\bigl(\vLambda, \{\vOmega_{g'}\}_{g' {\neq} g} \bigr)
=
\hvOmega_{g}(\lambda_{gg})
=
\left\{
    \left[
      \frac{\lambda_{gg}}{n_g} \vI_p
      + \frac{1}{4}\left(\vS_{g} - \frac{\lambda_{gg}}{n_g}\vT_{g}\right)^2
    \right]^{1/2}
    + \frac{1}{2}\left(\vS_{g} - \frac{\lambda_{gg}}{n_g} \vT_{g}\right)
  \right\}^{-1}.
\end{equation}
\end{corollary}

\begin{proof}(Proof of Corollary~\ref{prop:fusedridge2})
The result follows directly from equations \eqref{eq:update} and \eqref{eq:barupdate} by using that $\sum_{g'\neq g} \lambda_{gg'} = \sum_{g'\neq g} \lambda_{g'g} = 0$ for all $g$.
\end{proof}

\begin{lemma}
\label{lem:bound}%
Let $\{\vT_g\} \in \calS_+^p$ and assume $\lambda_{gg} \in \bbR_{++}$ in addition to $0 \leq \lambda_{g g'} < \infty$ for all $g'\neq g$.
Then
\begin{equation}\nonumber
\lim_{\lambda_{gg} \rightarrow \infty^{-}} \left\| \hvOmega_{g}\bigl(\vLambda, \{\vOmega_{g'}\}_{g' {\neq} g}\bigr) \right\|_F < \infty.
\end{equation}
\end{lemma}

\begin{sloppypar}
\begin{proof}(Proof of Lemma~\ref{lem:bound})
The result is shown through proof by contradiction.
Hence, suppose
\begin{equation*}
\lim_{\lambda_{gg} \rightarrow \infty^{-}} \| \hvOmega_{g}\bigl(\vLambda, \{\vOmega_{g'}\}_{g' {\neq} g}\bigr) \|_F
\end{equation*}
is unbounded.
Let $d[\cdot]_{jj}$ denote the \nth{j} largest eigenvalue.
Then, as
\begin{equation}\nonumber
\left\| \hvOmega_{g}\bigl(\vLambda, \{\vOmega_{g'}\}_{g' {\neq} g}\bigr) \right\|_F =
\left\{\sum_{j=1}^p d\left[ \hvOmega_{g}\bigl(\vLambda, \{\vOmega_{g'}\}_{g' {\neq} g}\bigr) \right]_{jj}^{2}\right\}^{1/2},
\end{equation}
at least one eigenvalue must tend to infinity along with $\lambda_{gg}$.
Assume without loss of generality that this is only the first (and largest) eigenvalue:
\begin{equation}
\lim_{\lambda_{gg} \rightarrow \infty^{-}}
d\left[ \hvOmega_{g}\bigl(\vLambda, \{\vOmega_{g'}\}_{g' {\neq} g}\bigr) \right]_{11} =
\mathcal{O}(\lambda_{gg}^{\gamma}),
\label{eq:EIGval}
\end{equation}
for some $\gamma > 0$.
Now, for any $\lambda_{gg}$, the precision can be written as an eigendecomposition:
\begin{equation}
\hvOmega_{g}\bigl(\vLambda, \{\vOmega_{g'}\}_{g' {\neq} g}\bigr) =
d_{11}\mathbf{v}_{1}\mathbf{v}_1^\top + \sum_{j=2}^p d_{jj} \mathbf{v}_{j} \mathbf{v}_j^\top,
\label{eq:eigDecomp}
\end{equation}
where the dependency of the eigenvalues and eigenvectors on the target matrices and penalty parameters has been suppressed (for notational brevity and clarity).
It is the first summand on the right-hand side that dominates the precision for large $\lambda_{gg}$.
Furthermore, this ridge ML precision estimate of the \nth{g} group satisfies, by \eqref{eq:derdiff}, the following gradient equation:
\begin{equation}\nonumber
n_{g}(\hvOmega_{g}^{-1} \!- \vS_{g}) -
\lambda_{gg}(\hvOmega_{g} {-} \vT_{g}) -
\sum_{g' \neq g}\lambda_{g' g}(\hvOmega_{g} {-} \vT_{g}) +
\sum_{g' \neq g}\lambda_{g' g}(\vOmega_{g'} {-} \vT_{g'}) = \vec{0}.
\end{equation}
We now make three observations:
(i) Item \ref{prop:fusedridge3item1} of Proposition~\ref{prop:fusedridge3} implies that $\hvOmega_{g}\bigl(\vLambda, \{\vOmega_{g'}\}_{g' {\neq} g}\bigr)$ is always positive definite for $\lambda_{gg} \in \bbR_{++}$. Consequently, $\lim_{\lambda_{gg} \rightarrow \infty^{-}} \| \hvOmega_{g}\bigl(\vLambda, \{\vOmega_{g'}\}_{g' {\neq} g}\bigr)^{-1} \|_F < \infty$;
(ii) The target matrices do not depend on $\lambda_{gg}$;
and (iii) The finite $\lambda_{g g'}$ ensure that the norms of $\vOmega_{g'}$ can only exceed the norm of $\hvOmega_g$ by a function (independent of $\lambda_{gg}$) of the constant $\lambda_{gg'}$.
Hence, in the limit, the norms of the $\vOmega_{g'}$ cannot exceed the norm of $\hvOmega_{g}$.
These observations give that, as $\lambda_{gg}$ tends towards infinity, the term $\lambda_{gg}(\hvOmega_{g} {-} \vT_{g})$ will dominate the gradient equation.
In fact, the term $\lambda_{gg}\hvOmega_{g}$ will dominate as, using \eqref{eq:EIGval} and \eqref{eq:eigDecomp}:
\begin{eqnarray*}
\vec{0} & \approx & - \lambda_{gg} ( \hvOmega_g - \vT_g)
\\
& \approx & - \lambda_{gg} d_{11}\mathbf{v}_1 \mathbf{v}_1^\top + \lambda_{gg} \vT
\\
& \approx & - \lambda_{gg}^{1 + \gamma} \mathbf{v}_1 \mathbf{v}_1^\top + \lambda_{gg} \vT
\\
& \approx & - \lambda_{gg}^{1 + \gamma} (\mathbf{v}_1 \mathbf{v}_1^\top + \lambda_{gg}^{-\gamma} \vT)
\\
& \approx & - \lambda_{gg}^{1 + \gamma} \mathbf{v}_1 \mathbf{v}_1^\top.
\end{eqnarray*}
This latter statement is contradictory as it can only be true if the first eigenvalue tends to zero.
This, in turn, contradicts the assumption of unboundedness (in the Frobenius norm) of the precision estimate.
Hence, the fused ridge ML precision estimate must be bounded.
\end{proof}
\end{sloppypar}

\medskip
\begin{proof}(Proof of Proposition~\ref{prop:fusedridge3})
~

(\ref{prop:fusedridge3item1})
Note that \eqref{eq:estimatingequation} for class $g$ may be rewritten to
\begin{equation}\nonumber
\hat{\vOmega}_{g}^{-1} \!- \vS_{g} -
\frac{\lambda_{g\summed}}{n_{g}}
\left\{ \hat{\vOmega}_{g} - \left[ \vT_g + \sum_{g' \neq g} \frac{\lambda_{g g'}}{\lambda_{g\summed}}(\vOmega_{g'} {-} \vT_{g'}) \right] \right\} = \vec{0},
\end{equation}
implying that \eqref{eq:update} can be obtained under the following alternative updating scheme to \eqref{eq:barupdate}:
\begin{equation}\nonumber
  \bar{\vS}_{g} = \vS_{g},
  \quad
  \bar{\vT}_{g} = \vT_{g} + \sum_{g' \neq g}
          \frac{\lambda_{g g'}}{\lambda_{g\summed}}
          (\vOmega_{g'} \!- \vT_{g'}),
  \andwhere
  \bar{\lambda}_{g}
  = \frac{\lambda_{g\summed}}{n_{g}}.
\end{equation}
Now, let $d[\missingarg]_{jj}$ denote the \nth{j} largest eigenvalue.
Then
\begin{align*}
d\left\{[\hat{\vOmega}_g]^{-1}\right\}_{jj} =
d\left[\frac{1}{2}(\vS_g - \bar{\lambda}_{g}\bar{\vT}_{g})\right]_{jj}
+
\sqrt{\left\{d\left[\frac{1}{2}(\vS_g - \bar{\lambda}_{g}\bar{\vT}_{g})\right]_{jj}\right\}^2
+ \bar{\lambda}_{g}}
> 0,
\end{align*}
when $\bar{\lambda}_{g} > 0$.
As $\bar{\lambda}_{g} = \sum_{g'}(\lambda_{g'g}/n_g)$ and as $\lambda_{g'g}$ may be $0$ for all $g'\neq g$, $\hat{\vOmega}_g$ is guaranteed to be positive definite whenever $\lambda_{gg} \in \bbR_{++}$.

(\ref{prop:fusedridge3item2})
Note that $\sum_{g'\neq g} \lambda_{gg'} = \sum_{g'\neq g} \lambda_{g'g} = 0$ implies that $\hat{\vOmega}_g$ reduces to the non-fused class estimate \eqref{eq:unfusedClass} by way of Corollary \ref{prop:fusedridge2}.
The stated right-hand limit is then immediate by using $\lambda_{gg} = 0$ in \eqref{eq:unfusedClass}.
Under the distributional assumptions this limit exists with probability 1 when $p \leq n_g$.

(\ref{prop:fusedridge3item3})
Consider the zero gradient equation \eqref{eq:estimatingequation} for the \nth{g} class.
Multiply it by $n_g/\lambda_{g\summed}$ to factor out the dominant term:
\begin{equation}
   \label{eq:estimatingequation2}
   \frac{n_g}{\lambda_{g\summed}}\hat{\vOmega}_g^{-1} \!- \frac{n_{g}}{\lambda_{g\summed}}\vS_g
  - (\hat{\vOmega}_g \!- \vT_{g})
  + \sum_{g' \neq g}
  \frac{\lambda_{g'g}}{\lambda_{g\summed}} (\vOmega_{g'} \!- \vT_{g'})
  = \vec{0}.
\end{equation}
When $\lambda_{gg} \to \infty^{-}$, $\lambda_{g\summed} = \sum_{g'} \lambda_{gg'} \to \infty^{-}$, implying that the first two terms of \eqref{eq:estimatingequation2} vanish.
Under the assumption that $\lambda_{gg'} < \infty$ for all $g' \neq g$ we have that $\lambda_{g'g}/\lambda_{g\summed} \to 0$ when $\lambda_{gg} \to \infty^{-}$ for all $g' \neq g$.
Thus, all terms of the sum also vanish as Lemma \ref{lem:bound} implies that the $\vOmega_{g'}$ are all bounded.
Hence, when $\lambda_{gg} \to \infty^{-}$ and $\lambda_{gg'} < \infty$ for all $g' \neq g$, the zero gradient equation reduces to
$\hat{\vOmega}_g \!- \vT_g = \vec{0}$, implying the stated left-hand limit.

(\ref{prop:fusedridge3item4})
The proof strategy follows the proof of item \ref{prop:fusedridge3item3}.
Multiply the zero gradient equation \eqref{eq:estimatingequation} for the \nth{g_1} class with $n_{g_1}/\lambda_{g_1 g_2}$ to obtain:
\begin{equation}
   \label{eq:estimatingequation3}
   \frac{n_{g_1}}{\lambda_{g_1 g_2}}\hat{\vOmega}_{g_1}^{-1} \!-
     \frac{n_{g_1}}{\lambda_{g_1 g_2}}\vS_{g_1}
  - \frac{\lambda_{g_1 \summed}}{\lambda_{g_1 g_2}}(\hat{\vOmega}_{g_1} \!- \vT_{g_1})
  + \sum_{g' \neq g_1}
  \frac{\lambda_{g'g_1}}{\lambda_{g_1 g_2}} (\vOmega_{g'} \!- \vT_{g'})
  = \vec{0}.
\end{equation}
The first two terms are immediately seen to vanish when $\lambda_{g_1 g_2} \to \infty^{-}$.
Under the assumption that all penalties except $\lambda_{g_1 g_2}$ are finite, we have that
$\lambda_{g_1 \summed}/\lambda_{g_1 g_2} \to 1$ for $\lambda_{g_1 g_2} \to \infty^{-}$.
Similarly, all elements of the sum term in \eqref{eq:estimatingequation3} vanish except the element where $g' = g_2$.
Hence, when $\lambda_{g_1 g_2} \to \infty^{-}$ and when $\lambda_{g_1' g_2'} < \infty$ for all $\{g_1',g_2'\} \neq \{g_1,g_2\}$, the zero gradient equation for class $g_1$ reduces to:
\begin{equation}\label{eq:zg.g1}
  - (\hat{\vOmega}_{g_1} \!- \vT_{g_1})
  + (\vOmega_{g_2} \!- \vT_{g_2})
  = \vec{0}.
\end{equation}
Conversely, by multiplying the zero gradient equation \eqref{eq:estimatingequation} for the \nth{g_2} class with $n_{g_2}/\lambda_{g_1 g_2}$ one obtains, through the same development as above, that the zero gradient equation for class $g_2$ reduces to the $\hat{\vOmega}_{g_2}$-analogy of equation \eqref{eq:zg.g1}.
The result \eqref{eq:zg.g1} then immediately implies the stated limiting result.
\end{proof}

\begin{corollary}
\label{cor:SpecialT}%
Consider item \ref{prop:fusedridge3item4} of Proposition \ref{prop:fusedridge3}.
When, in addition, $\vT_{g_1} = \vT_{g_2}$, we have that
\begin{equation*}
  \lim\limits_{\lambda_{g_1 g_2} \to \infty^-}(\hvOmega_{g_1} - \vT_{g_1})
  = \lim\limits_{\lambda_{g_1 g_2} \to \infty^-} (\hvOmega_{g_2} - \vT_{g_2})
  \qquad \Longrightarrow \qquad \hat{\vOmega}_{g_1} = \hat{\vOmega}_{g_2}.
\end{equation*}
\end{corollary}

\begin{proof}(Proof of Corollary~\ref{cor:SpecialT})
The implication follows directly by using $\vT_{g_1} = \vT_{g_2}$ in \eqref{eq:zg.g1}.
\end{proof}

\medskip
\begin{proof}(Proof of Proposition~\ref{prop:InvLess})
The result follows directly from Proposition \ref{prop:fusedridge} and Lemma \ref{lem:NoInverseIdent}.
\end{proof}

\medskip
\begin{proof}(Proof of Proposition~\ref{prop:PosRealm})
Note that line \ref{lst:Initial} of  Algorithm \ref{alg:fusedridge} implies that the initializing estimates are positive definite.
Moreover, regardless of the value of the fused penalties (in the feasible domain), the estimate in line \ref{lst:UpdateStep} of  Algorithm \ref{alg:fusedridge} is positive definite as a consequence of Proposition~\ref{prop:fusedridge3}.
\end{proof}

\vskip 0.2in
\bibliography{references_manual}

\begin{thebibliography}{60}
\providecommand{\natexlab}[1]{#1}
\providecommand{\url}[1]{\texttt{#1}}
\expandafter\ifx\csname urlstyle\endcsname\relax
  \providecommand{\doi}[1]{doi: #1}\else
  \providecommand{\doi}{doi: \begingroup \urlstyle{rm}\Url}\fi

\bibitem[Alizadeh et~al.(2000)Alizadeh, Eisen, Davis, Ma, Lossos, Rosenwald,
  Boldrick, Sabet, Tran, Yu, Powell, Yang, Marti, Moore, Hudson, Lu, Lewis,
  Tibshirani, Sherlock, Chan, Greiner, Weisenburger, Armitage, Warnke, Levy,
  Wilson, Grever, Byrd, Botstein, Brown, and Staudt]{Alizadeh2000}
A.~A. Alizadeh, M.~B. Eisen, R.~E. Davis, C.~Ma, I.~S. Lossos, A.~Rosenwald,
  J.~C. Boldrick, H.~Sabet, T.~Tran, X.~Yu, J.~I. Powell, L.~Yang, G.~E. Marti,
  T.~Moore, J.~Hudson, L.~Lu, D.~B. Lewis, R.~Tibshirani, G.~Sherlock, W.~C.
  Chan, T.~C. Greiner, D.~D. Weisenburger, J.~O. Armitage, R.~Warnke, R.~Levy,
  W.~Wilson, M.~R. Grever, J.~C. Byrd, D.~Botstein, P.~O. Brown, and L.~M.
  Staudt.
\newblock Distinct types of diffuse large {B}-cell lymphoma identified by gene
  expression profiling.
\newblock \emph{Nature}, 403\penalty0 (6769):\penalty0 503--511, 2000.

\bibitem[Banerjee et~al.(2008)Banerjee, El~Ghaoui, and
  D'Aspremont]{Banerjee2008}
O.~Banerjee, L.~El~Ghaoui, and A.~D'Aspremont.
\newblock Model selection through sparse maximum likelihood estimation for
  multivariate {G}aussian or binary data.
\newblock \emph{The Journal of Machine Learning Research}, 9:\penalty0
  485--516, 2008.

\bibitem[Barab{\'a}si(2009)]{Barabasi2009}
A.~L. Barab{\'a}si.
\newblock Scale-free networks: A decade and beyond.
\newblock \emph{Science}, 325\penalty0 (5939):\penalty0 412--413, 2009.

\bibitem[Barab{\'a}si and Albert(1999)]{Barabasi1999}
A.~L. Barab{\'a}si and R.~Albert.
\newblock Emergence of scaling in random networks.
\newblock \emph{Science}, 286\penalty0 (5439):\penalty0 509--512, 1999.

\bibitem[Barrett et~al.(2013)Barrett, Wilhite, Ledoux, Evangelista, Kim,
  Tomashevsky, Marshall, Phillippy, Sherman, Holko, Yefanov, Lee, Zhang,
  Robertson, Serova, Davis, and Soboleva]{Barrett2013}
T.~Barrett, S.~E. Wilhite, P.~Ledoux, C.~Evangelista, I.~F. Kim,
  M.~Tomashevsky, K.~A. Marshall, K.~H. Phillippy, P.~M. Sherman, M.~Holko,
  A.~Yefanov, H.~Lee, N.~Zhang, C.~L. Robertson, N.~Serova, S~Davis, and
  A.~Soboleva.
\newblock {NCBI GEO: A}rchive for functional genomics data sets--update.
\newblock \emph{Nucleic Acids Research}, 41\penalty0 (D1):\penalty0 D991--D995,
  2013.

\bibitem[Bera and Bilias(2001)]{BeraBil01}
A.~K. Bera and Y.~Bilias.
\newblock {Rao's} score, {Neyman's} $c(\alpha)$ and {Silvey's} {LM} tests: An
  essay on historical developments and some new results.
\newblock \emph{Journal of Statistical Planning and Inference}, 97\penalty0
  (1):\penalty0 9--44, 2001.

\bibitem[Bid{\`e}re et~al.(2009)Bid{\`e}re, Ngo, Lee, Collins, Zheng, Wan,
  Davis, Lenz, Anderson, Arnoult, Vazquez, Sakai, Zhang, Meng, Veenstra,
  Staudt, and Lenardo]{Bidere2009}
N.~Bid{\`e}re, V.~N. Ngo, J.~Lee, C.~Collins, L.~Zheng, F.~Wan, R.~E. Davis,
  G.~Lenz, D.~E. Anderson, D.~Arnoult, A.~Vazquez, K.~Sakai, J.~Zhang, Z.~Meng,
  T.~D. Veenstra, L.~M. Staudt, and M.~J. Lenardo.
\newblock Casein kinase 1$\alpha$ governs antigen-receptor-induced
  {NF}-$\kappa${B} activation and human lymphoma cell survival.
\newblock \emph{Nature}, 458\penalty0 (7234):\penalty0 92--96, 2009.

\bibitem[Bilgrau and Falgreen(2014)]{DLBCLdata}
A.~E. Bilgrau and S.~Falgreen.
\newblock \emph{\texttt{DLBCLdata}: Automated and Reproducible Download and
  Preprocessing of DLBCL Data}, 2014.
\newblock URL \url{http://github.com/AEBilgrau/DLBCLdata}.
\newblock {\R} package version 0.9.

\bibitem[Bilgrau et~al.(2018)Bilgrau, Br{\o}ndum, Eriksen, Dybk{\ae}r, and
  B{\o}gsted]{Bilgrau2015b}
A.~E. Bilgrau, R.~F. Br{\o}ndum, P.~S. Eriksen, K.~Dybk{\ae}r, and
  M.~B{\o}gsted.
\newblock Estimating a common covariance matrix for network meta-analysis of
  gene expression datasets in diffuse large {B-cell} lymphoma.
\newblock \emph{The Annals of Applied Statistics}, 12\penalty0 (3):\penalty0
  1894--1913, 2018.

\bibitem[Boyle et~al.(2017)Boyle, Li, and Pritchard]{Omnigenic}
E.~A. Boyle, Y.~I. Li, and J.~K. Pritchard.
\newblock An expanded view of complex traits: From polygenic to omnigenic.
\newblock \emph{Cell}, 169:\penalty0 1177--1186, 2017.

\bibitem[Browning et~al.(1997)Browning, Sizing, Lawton, Bourdon, Rennert,
  Majeau, Ambrose, Hession, Miatkowski, Griffiths, ek~A., W., D., and
  S.]{Browning1997}
J.~L. Browning, I.~D. Sizing, P.~Lawton, P.~R. Bourdon, P.~D. Rennert, G.~R.
  Majeau, C.~M. Ambrose, C.~Hession, K.~Miatkowski, D.~A. Griffiths, Ngam
  ek~A., Meier W., Benjamin~C. D., and Hochman~P. S.
\newblock Characterization of lymphotoxin-$\alpha \beta$ complexes on the
  surface of mouse lymphocytes.
\newblock \emph{The Journal of Immunology}, 159\penalty0 (7):\penalty0
  3288--3298, 1997.

\bibitem[Cai(2017)]{Cai}
T.~T. Cai.
\newblock Global testing and large-scale multiple testing for high-dimensional
  covariance structures.
\newblock \emph{Annual Review of Statistics and Its Application}, 4:\penalty0
  423--446, 2017.

\bibitem[Care et~al.(2013)Care, Barrans, Worrillow, Jack, Westhead, and
  Tooze]{Care2013}
M.~A. Care, S.~Barrans, L.~Worrillow, A.~Jack, D.~R. Westhead, and R.~M. Tooze.
\newblock A microarray platform-independent classification tool for cell of
  origin class allows comparative analysis of gene expression in diffuse large
  {B}-cell lymphoma.
\newblock \emph{PLoS One}, 8\penalty0 (2):\penalty0 e55895, 2013.

\bibitem[Dai et~al.(2005)Dai, Wang, Boyd, Kostov, Athey, Jones, Bunney, Myers,
  Speed, Akil, Watson, and Meng]{Dai2005}
M.~Dai, P.~Wang, A.~D. Boyd, G.~Kostov, B.~Athey, E.~G. Jones, W.~E. Bunney,
  R.~M. Myers, T.~P. Speed, H.~Akil, S.~J. Watson, and F.~Meng.
\newblock Evolving gene/transcript definitions significantly alter the
  interpretation of {GeneChip} data.
\newblock \emph{Nucleic Acids Research}, 33\penalty0 (20):\penalty0 e175, 2005.

\bibitem[Danaher et~al.(2014)Danaher, Wang, and Witten]{Danaher2013}
P.~Danaher, P.~Wang, and D.~M. Witten.
\newblock The joint graphical lasso for inverse covariance estimation across
  multiple classes.
\newblock \emph{Journal of the Royal Statistical Society, Series B},
  76\penalty0 (2):\penalty0 373--397, 2014.

\bibitem[Dybk\ae{}r et~al.(2015)Dybk\ae{}r, B\o{}gsted, Falgreen, B\o{}dker,
  Kjeldsen, Schmitz, Bilgrau, Xu-Monette, Li, Bergkvist, Laursen,
  Rodrigo-Domingo, Marques, Rasmussen, Nyegaard, Gaihede, M{\o}ller, Samworth,
  Shah, Johansen, El-Galaly, Young, and Johnsen]{DykaerBoegsted2015}
K.~Dybk\ae{}r, M.~B\o{}gsted, S.~Falgreen, J.~S. B\o{}dker, M.~K. Kjeldsen,
  A.~Schmitz, A.~E. Bilgrau, Z.~Y. Xu-Monette, L.~Li, K.~S. Bergkvist, M.~B.
  Laursen, M.~Rodrigo-Domingo, S.~C. Marques, S.~B. Rasmussen, M.~Nyegaard,
  M.~Gaihede, M.~B. M{\o}ller, R.~J. Samworth, R.~D. Shah, P.~Johansen, T.~C.
  El-Galaly, K.~H. Young, and H.~E. Johnsen.
\newblock A diffuse large {B}-cell lymphoma classification system that
  associates normal {B}-cell subset phenotypes with prognosis.
\newblock \emph{Journal Of Clinical Oncology}, 33\penalty0 (12):\penalty0
  1379--1388, 2015.

\bibitem[Eddelbuettel(2013)]{Rcpp2013}
D.~Eddelbuettel.
\newblock \emph{Seamless {\R} and \texttt{C++} Integration with \texttt{Rcpp}}.
\newblock Springer-Verlag, New York, 2013.

\bibitem[Eddelbuettel and Fran\c{c}ois(2011)]{Eddelbuettel2011}
D.~Eddelbuettel and R.~Fran\c{c}ois.
\newblock \texttt{Rcpp}: Seamless {\R} and \texttt{C++} integration.
\newblock \emph{Journal of Statistical Software}, 40\penalty0 (8), 2011.

\bibitem[Efron(2005)]{Efron2005}
B.~Efron.
\newblock Local false discovery rates.
\newblock Technical report, Stanford University Division of Biostatistics, 03
  2005.

\bibitem[Efron et~al.(2001)Efron, Tibshirani, Storey, and Tusher]{EfronLocFDR}
B.~Efron, R.~Tibshirani, J.~D. Storey, and V.~Tusher.
\newblock Empirical {B}ayes analysis of a microarray experiment.
\newblock \emph{Journal of the American Statistical Association}, 96:\penalty0
  1151--1160, 2001.

\bibitem[Erd\"{o}s and R\'{e}nyi(1959)]{RandomGraph}
P.~Erd\"{o}s and A.~R\'{e}nyi.
\newblock On random graphs {I}.
\newblock \emph{Publicationes Mathematicae}, 6:\penalty0 290--297, 1959.

\bibitem[Fran\c{c}ois et~al.(2012)Fran\c{c}ois, Eddelbuettel, and
  Bates]{RcppArmadillo}
R.~Fran\c{c}ois, D.~Eddelbuettel, and D.~Bates.
\newblock \emph{\texttt{RcppArmadillo}: \texttt{Rcpp} Integration for Armadillo
  Templated Linear Algebra Library}, 2012.
\newblock URL \url{http://CRAN.R-project.org/package=RcppArmadillo}.
\newblock {R}~package version~0.3.6.1.

\bibitem[Friedman et~al.(2008)Friedman, Hastie, and Tibshirani]{Friedman2008}
J.~Friedman, T.~Hastie, and R.~Tibshirani.
\newblock Sparse inverse covariance estimation with the graphical lasso.
\newblock \emph{Biostatistics}, 9\penalty0 (3):\penalty0 432--41, 2008.

\bibitem[Gautier et~al.(2004)Gautier, Cope, Bolstad, and Irizarry]{Gautier2004}
L.~Gautier, L.~Cope, B.~M. Bolstad, and R.~A. Irizarry.
\newblock \texttt{affy}---analysis of {Affymetrix GeneChip} data at the probe
  level.
\newblock \emph{Bioinformatics}, 20\penalty0 (3):\penalty0 307--315, 2004.

\bibitem[Guo et~al.(2011)Guo, Levina, Michailidis, and Zhu]{GLMZ2011}
Y.~Guo, E.~Levina, G.~Michailidis, and J.~Zhu.
\newblock Joint estimation of multiple graphical models.
\newblock \emph{Biometrika}, 98\penalty0 (1):\penalty0 1--15, 2011.

\bibitem[Ha et~al.(2015)Ha, Baladandayuthapani, and Do]{DINGO}
M.~J. Ha, V.~Baladandayuthapani, and K.~A. Do.
\newblock {DINGO}: differential network analysis in genomics.
\newblock \emph{Bioinformatics}, 31:\penalty0 3413--3420, 2015.

\bibitem[Jones and West(2005)]{Jones2005}
B.~Jones and M.~West.
\newblock Covariance decomposition in undirected {Gaussian} graphical models.
\newblock \emph{Biometrika}, 92:\penalty0 779--786, 2005.

\bibitem[Kanehisa and Goto(2000)]{Kanehisa2000}
M.~Kanehisa and S.~Goto.
\newblock {KEGG: Kyoto Encyclopedia of Genes and Genomes}.
\newblock \emph{Nucleic Acids Research}, 28\penalty0 (1):\penalty0 27--30,
  2000.

\bibitem[Lauritzen(1996)]{Lauritz96}
S.~L. Lauritzen.
\newblock \emph{Graphical Models}.
\newblock Clarendon Press, Oxford, 1996.

\bibitem[Lu and Zhang(2006)]{Lu2006}
X.~Lu and X.~Zhang.
\newblock The effect of {GeneChip} gene definitions on the microarray study of
  cancers.
\newblock \emph{Bioessays}, 28\penalty0 (7):\penalty0 739--46, 2006.

\bibitem[Maurya(2016)]{Maurya}
A.~Maurya.
\newblock A well-conditioned and sparse estimation of covariance and inverse
  covariance matrices using a joint penalty.
\newblock \emph{Journal of Machine Learning Research}, 17:\penalty0 345--372,
  2016.

\bibitem[Mei et~al.(2011)Mei, Zhang, and Cao]{Mei2011}
S.~Mei, X.~Zhang, and M.~Cao.
\newblock \emph{Power Grid Complexity}.
\newblock Tsinghua University Press, Beijing and Springer-Verlag Berlin, 2011.

\bibitem[Mersmann(2014)]{Mersmann2014}
O.~Mersmann.
\newblock \emph{\texttt{microbenchmark}: Accurate Timing Functions}, 2014.
\newblock URL \url{http://CRAN.R-project.org/package=microbenchmark}.
\newblock {\R} package version 1.4-2.

\bibitem[Newman(2010)]{Newman10}
M.~E.~J. Newman.
\newblock \emph{Networks: An Introduction}.
\newblock Oxford University Press, Oxford, 2010.

\bibitem[Nowakowski et~al.(2015)Nowakowski, LaPlant, Macon, Reeder, Foran,
  Nelson, Thompson, Rivera, Inwards, Micallef, Johnston, Porrata, Ansell,
  Gascoyne, Habermann, and Witzig]{Nowakowski2015}
G.~S. Nowakowski, B.~LaPlant, W.~R. Macon, C.~B. Reeder, J.~M. Foran, G.~D.
  Nelson, C.~A. Thompson, C.~E. Rivera, D.~J. Inwards, I.~N. Micallef, P.~B.
  Johnston, L.~F. Porrata, S.~M. Ansell, R.~D. Gascoyne, T.~M. Habermann, and
  T.~E. Witzig.
\newblock Lenalidomide combined with {R-CHOP} overcomes negative prognostic
  impact of non-germinal center {B}-cell phenotype in newly diagnosed diffuse
  large {B}-cell lymphoma: A phase {II} study.
\newblock \emph{Journal of Clinical Oncology}, 33\penalty0 (3):\penalty0
  251--257, 2015.

\bibitem[Peeters et~al.(2019)Peeters, Bilgrau, and van Wieringen]{rags}
C.~F.~W. Peeters, A.~E. Bilgrau, and W.~N. van Wieringen.
\newblock \emph{\texttt{rags2ridges}: Ridge Estimation of Precision Matrices
  from High-Dimensional Data}, 2019.
\newblock URL \url{https://CRAN.R-project.org/package=rags2ridges}.
\newblock {\R} package version 2.1.1.

\bibitem[Peterson et~al.(2015)Peterson, Stingo, and Vannucci]{PSV2015}
C.~Peterson, F.~C. Stingo, and M.~Vannucci.
\newblock Bayesian inference of multiple {G}aussian graphical models.
\newblock \emph{Journal of the American Statistical Association}, 110\penalty0
  (509):\penalty0 159--174, 2015.

\bibitem[Price et~al.(2015)Price, Geyer, and Rothman]{Price2014}
B.~S. Price, C.~J. Geyer, and A.~J. Rothman.
\newblock Ridge fusion in statistical learning.
\newblock \emph{Journal of Computational and Graphical Statistics}, 24\penalty0
  (2):\penalty0 439--454, 2015.

\bibitem[{{\R} Core Team}(2012)]{R}
{{\R} Core Team}.
\newblock \emph{{\R}: A Language and Environment for Statistical Computing}.
\newblock {\R} Foundation for Statistical Computing, Vienna, Austria, 2012.
\newblock URL \url{http://www.R-project.org/}.

\bibitem[Roschewski et~al.(2014)Roschewski, Staudt, and Wilson]{Roschewski2014}
M.~Roschewski, L.~M. Staudt, and W.~H. Wilson.
\newblock Diffuse large {B}-cell lymphoma-treatment approaches in the molecular
  era.
\newblock \emph{Nature Reviews Clinical Oncology}, 11\penalty0 (1):\penalty0
  12--23, 2014.

\bibitem[Rothman(2012)]{Rothman12}
A.~Rothman.
\newblock Positive definite estimators of large covariance matrices.
\newblock \emph{Biometrika}, 99:\penalty0 733--740, 2012.

\bibitem[Ruan et~al.(2011)Ruan, Martin, Furman, Lee, Cheung, Vose, LaCasce,
  Morrison, Elstrom, Ely, Chadburn, Cesarman, Coleman, and Leonard]{Ruan2011}
J.~Ruan, P.~Martin, R.~R. Furman, S.~M. Lee, K.~Cheung, J.~M. Vose, A.~LaCasce,
  J.~Morrison, R.~Elstrom, S.~Ely, A.~Chadburn, E.~Cesarman, M.~Coleman, and
  J.~P. Leonard.
\newblock Bortezomib plus {CHOP}-rituximab for previously untreated diffuse
  large {B}-cell lymphoma and mantle cell lymphoma.
\newblock \emph{Journal of Clinical Oncology}, 29\penalty0 (6):\penalty0
  690--697, 2011.

\bibitem[Saegusa and Shojaie(2016)]{SS16}
T.~Saegusa and A.~Shojaie.
\newblock Joint estimation of precision matrices in heterogeneous populations.
\newblock \emph{Electronic Journal of Statistics}, 10:\penalty0 1341--1392,
  2016.

\bibitem[Sandberg and Larsson(2007)]{Sandberg2007}
R.~Sandberg and O.~Larsson.
\newblock Improved precision and accuracy for microarrays using updated probe
  set definitions.
\newblock \emph{BMC Bioinformatics}, 8\penalty0 (1):\penalty0 48, 2007.

\bibitem[Sanderson(2010)]{Sanderson2010}
C.~Sanderson.
\newblock \emph{\texttt{Armadillo}: An Open Source \texttt{C++} Linear Algebra
  Library for Fast Prototyping and Computationally Intensive Experiments.}
\newblock Technical Report, NICTA, 2010.
\newblock URL \url{http://arma.sourceforge.net}.

\bibitem[Sch\"{a}fer and Strimmer(2005{\natexlab{a}})]{SS05}
J.~Sch\"{a}fer and K.~Strimmer.
\newblock A shrinkage approach to large-scale covariance matrix estimation and
  implications for functional genomics.
\newblock \emph{Statistical Applications in Genetics and Molecular Biology},
  4:\penalty0 art. 32, 2005{\natexlab{a}}.

\bibitem[Sch\"{a}fer and Strimmer(2005{\natexlab{b}})]{SS05a}
J.~Sch\"{a}fer and K.~Strimmer.
\newblock An empirical bayes approach to inferring large-scale gene association
  networks.
\newblock \emph{Bioinformatics}, 21:\penalty0 754--764, 2005{\natexlab{b}}.

\bibitem[Schuetz et~al.(2012)Schuetz, Johnson, Morin, Scott, Tan, Ben-Nierah,
  Boyle, Slack, Marra, Connors, Brooks-Wilson, and Gascoyne]{Schuetz2012}
J.~M. Schuetz, N.~A. Johnson, R.~D. Morin, D.~W. Scott, K.~Tan, S~Ben-Nierah,
  M~Boyle, G.~W. Slack, M.~A. Marra, J.~M. Connors, A.~R. Brooks-Wilson, and
  R.~D. Gascoyne.
\newblock {BCL2} mutations in diffuse large {B}-cell lymphoma.
\newblock \emph{Leukemia}, 26\penalty0 (6):\penalty0 1383--90, 2012.

\bibitem[St\"{a}dler and Mukherjee(2017)]{SM17}
N.~St\"{a}dler and S.~Mukherjee.
\newblock Two-sample testing in high-dimensions.
\newblock \emph{Journal of the Royal Statistical Society, Series B},
  79:\penalty0 225--246, 2017.

\bibitem[{The Non-Hodgkin's Lymphoma Classification
  Project}(1997)]{Project1997}
{The Non-Hodgkin's Lymphoma Classification Project}.
\newblock A clinical evaluation of the international lymphoma study group
  classification of non-{Hodgkin's} lymphoma.
\newblock \emph{Blood}, 89\penalty0 (11):\penalty0 3909--3918, 1997.

\bibitem[van Wieringen and Peeters(2016)]{VanWieringen2014}
W.~N. van Wieringen and C.~F.~W. Peeters.
\newblock Ridge estimation of inverse covariance matrices from high-dimensional
  data.
\newblock \emph{Computational Statistics \& Data Analysis}, 103:\penalty0
  284--303, 2016.

\bibitem[Watts and Strogatz(1998)]{Watts1998}
D.~J. Watts and S.~H. Strogatz.
\newblock Collective dynamics of `small-world' networks.
\newblock \emph{Nature}, 393\penalty0 (6684):\penalty0 440--442, 1998.

\bibitem[Williams-Abbott et~al.(1997)Williams-Abbott, Walter, Cheung, Goh,
  Porter, and Ware]{WilliamsAbbott1997}
L.~Williams-Abbott, B.~N. Walter, T.~C. Cheung, C.~R. Goh, A.~G. Porter, and
  C.~F. Ware.
\newblock The lymphotoxin-$\alpha$ (lt$\alpha$) subunit is essential for the
  assembly, but not for the receptor specificity, of the membrane-anchored
  lt$\alpha 1 \beta 2$ heterotrimeric ligand.
\newblock \emph{The Journal of Biological Chemistry}, 271\penalty0
  (31):\penalty0 19451--19456, 1997.

\bibitem[Witten and Tibshirani(2009)]{Wit09}
D.~M. Witten and R.~Tibshirani.
\newblock Covariance-regularized regression and classification for
  high-dimensional problems.
\newblock \emph{Journal of the Royal Statistical Society, Series B},
  71:\penalty0 615--636, 2009.

\bibitem[Xia et~al.(2015)Xia, Cai, and Cai]{XCC15}
Y.~Xia, T.~Cai, and T.~T. Cai.
\newblock Testing differential networks with applications to the detection of
  gene-by-gene interactions.
\newblock \emph{Biometrika}, 102:\penalty0 247--266, 2015.

\bibitem[Yang et~al.(2012)Yang, Shaffer, Emre, Ceribelli, Zhang, Wright, Xiao,
  Powell, Platig, Kohlhammer, M., Zhao, Yang, Xu, Buggy, Balasubramanian,
  Mathews, Shinn, Guha, Ferrer, Thomas, Waldmann, and Staudt]{Yang2012}
Y.~Yang, A.~L. Shaffer, N.~C.~T. Emre, M.~Ceribelli, M.~Zhang, G.~Wright,
  W.~Xiao, J.~Powell, J.~Platig, H.~Kohlhammer, Young~R. M., H.~Zhao, Y.~Yang,
  W.~Xu, J.~J. Buggy, S.~Balasubramanian, L.~A. Mathews, P.~Shinn, R.~Guha,
  M.~Ferrer, C.~Thomas, T.~A. Waldmann, and L.~M. Staudt.
\newblock Exploiting synthetic lethality for the therapy of {ABC} diffuse large
  {B} cell lymphoma.
\newblock \emph{Cancer cell}, 21\penalty0 (6):\penalty0 723--737, 2012.

\bibitem[Yuan and Lin(2007)]{YL2007}
M.~Yuan and Y.~Lin.
\newblock Model selection and estimation in the {G}aussian graphical model.
\newblock \emph{Biometrika}, 94:\penalty0 19--35, 2007.

\bibitem[Yuan(2008)]{Yuan2008b}
Y.~Yuan.
\newblock Efficient computation of $\ell_1$ regularized estimates in {G}aussian
  graphical models.
\newblock \emph{Journal of Computational and Graphical Statistics},
  17:\penalty0 809--826, 2008.

\bibitem[Zhang and Wiemann(2009)]{Zhang2015}
J.~D. Zhang and S.~Wiemann.
\newblock \texttt{KEGGgraph}: A graph approach to {KEGG} pathway in {\R} and
  {B}ioconductor.
\newblock \emph{Bioinformatics}, 25\penalty0 (11):\penalty0 1470--1471, 2009.

\bibitem[Zhao et~al.(2014)Zhao, Cai, and Li]{Zhao}
S.~D. Zhao, T.~T. Cai, and H.~Li.
\newblock Direct estimation of differential networks.
\newblock \emph{Biometrika}, 101:\penalty0 253--268, 2014.

\end{thebibliography}

\end{document}